%% file: tesis_1.tex
\documentclass[12pt,a4paper,dipdefs]{book}
\usepackage[english, activeacute]{babel}
\usepackage{rotating}
\usepackage{epsfig}
\usepackage{psfig}
\usepackage{amsmath}
\usepackage{amssymb}
\usepackage{mathrsfs}
\usepackage{amscd}
\usepackage{leqno}
\usepackage{lscape}
\usepackage{fancyheadings}
\usepackage{amscd}
\usepackage{indentfirst} 
\begin{document}
\include{titulo}
\newpage
\pagenumbering{roman}

\tableofcontents

\listoffigures

\listoftables
\newpage

\pagenumbering{arabic}


\pagestyle{fancyplain}




\lhead[\fancyplain{}{\bfseries\thepage}]{\fancyplain{}{\bfseries\rightmark}}
\rhead[\fancyplain{}{\bfseries\leftmark}]{\fancyplain{}{\bfseries\thepage}}
\cfoot{}
\setcounter{page}{0}
\thispagestyle{empty}

\chapter{Motivation and  physics}

{\it  In this chapter we give an overview of the $^{146}$Gd region and the reasons for studying the p-h multiplets and double-phonon states in this nucleus, as well as a summary of previous work on $^{146}$Gd. A general overview of the nuclear vibrational modes and their importance, focusing on octupole modes, is given.} 
  
\vspace*{0.6cm}

\section{Abstract}

In this work I have studied the $^{144}$Sm($\alpha$,2n) fusion-evaporation reaction with a beam of 26.3 MeV $\alpha$-particles at the Institute for Nuclear Physics (IKP) of the University of Cologne (Germany) in order to identify the double octupole states and two-particle configuration states in the spherical even-even nucleus $^{146}$Gd. The target was surrounded by a compact array of nine individual Ge detectors at 90, $\pm$45 and $\pm$35 degrees to the beam direction (five of them had anti-Compton shields), and by a EUROBALL CLUSTER detector placed at 90 degrees to act as a non-orthogonal $\gamma$-ray polarimeter. The experiment provided excellent data on $\gamma$-$\gamma$ coincidences as well as information on the $\gamma$-ray anisotropies and their $\gamma$-ray polarizations. 
A total of 35 new $\gamma$-rays have been identified corresponding to 28 new 
states (some of them with firm spin and parity assignment), this togeter with
previous experiment makes a total of 44 new levels, 
as well as 31 new $\gamma$-rays corresponding to 26 previously known levels. In addition, 3 previously known $\gamma$-rays were seen for the first time in an in-beam experiment. Amongst the new levels, new candidates for the two-particle configuration states have been found as well as for the (3$^{-}$$\times$2$^{+}$) and (3$^{-}$$\times$3$^{-}$) two-phonon multiplets. A very important result is the unequivocal identification of the 6$^{+}$ member of the two-phonon octupole in $^{146}$Gd by identifying the {\it E}3 branching to the one-phonon 3$^{-}$ state. This results present the first conclusive observation of a 6$^{+}$$\rightarrow$3$^{-}$$\rightarrow$0$^{+}$ double {\it E}3 cascade in the decay of a two-phonon octupole state.

\section{Introduction}
The present work concerns the study of particle-hole multiplets and possible double-phonon states in $^{146}$Gd, a spherical nucleus which presents characteristics of a doubly-magic nucleus.\
The experimental investigation of the above mentioned states demands the identification of yrast and above-yrast states in the nucleus. While the former are relatively easy to study experimentally, the latter present serious difficulties.\
In the first chapter, we will explain the motivation to perform such studies. We will summarise previous knowledge concerning this nucleus and the possible means to populate above-yrast states, and we will give an overview of the vibrational modes and center our attention on the octupole modes. We will also give the reasons why we think that the new experimental developments allow us to reach this goal now.
In chapter 2 we will discuss the experiment and the analysis methods where directional angular distributions and polarization analysis methods will be described in detail.\
In chapter 3 we will show the experimental results from the present work.\
In chapter 4 we will discuss the results concerning the particle-hole states and make multiplet assignments to the firmly established levels.\
In chapter 5 the two-phonon octupole states are explored in depth and multiplet candidates are proposed.\

\section{The $^{146}$Gd region}
\begin{figure}[h]
\begin{center}
\includegraphics[height=11.cm,angle=0.]{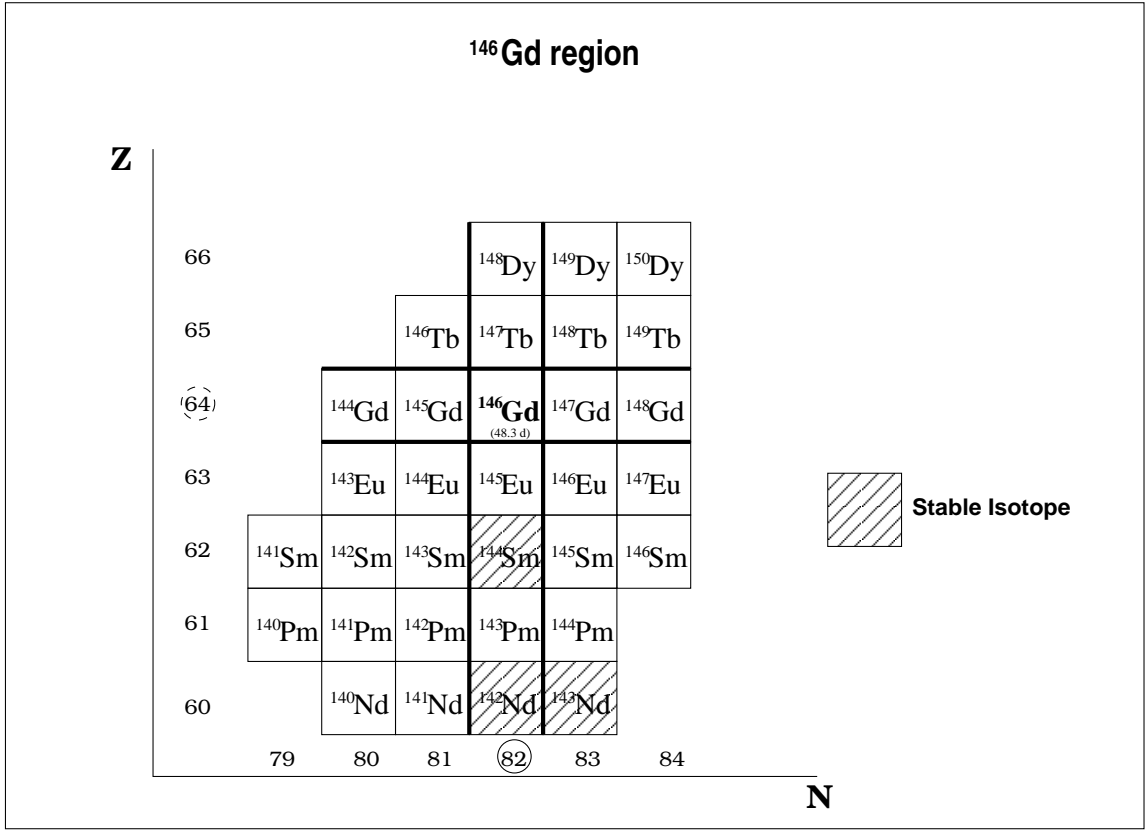}
\end{center}
\caption{The $^{146}$Gd region in the nuclidic chart.}
\label{fig:chart}
\end{figure}

\begin{figure}[h]
\begin{center}
\includegraphics[height=12.cm,angle=0.]{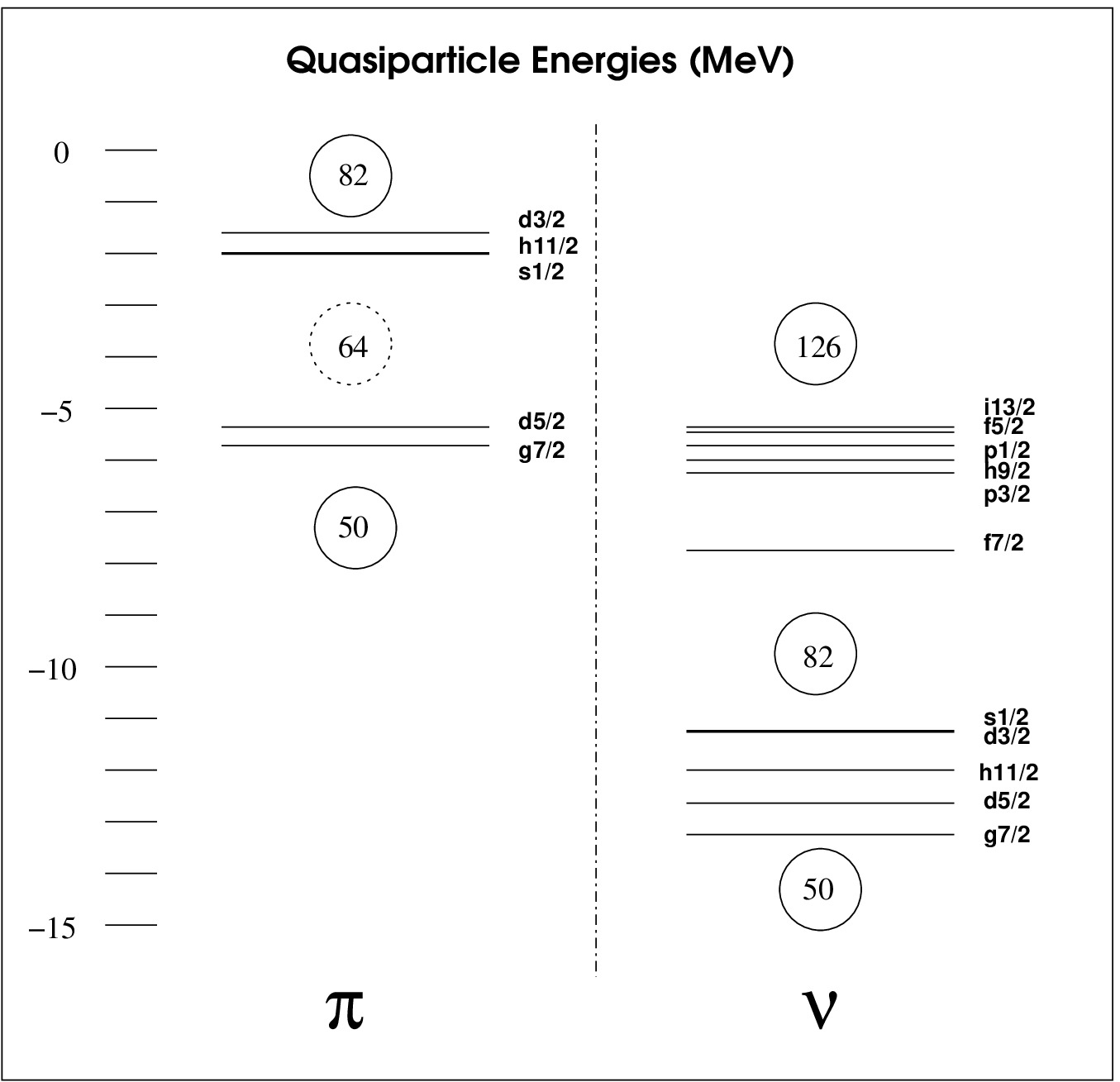}
\end{center}
\caption{Quasiparticle experimental values of the energy for protons and neutrons in the N=82 and Z=64 region.}
\label{fig:gaps}
\end{figure}

The gadolinium isotopes are situated in the periodic table in the rare earth or lanthanide region. $^{146}$Gd is an unstable nucleus, but is not very far away from stability (see Figure~[\ref{fig:chart}]). In previous work~\cite{KLEIN79,NAGAI} on $^{146}$Gd and its neighbouring nuclei, it was established that the gap ($\sim$2.4 MeV) in the proton single-particle spectrum at Z=64 and the well-known magic character at N=82 give to $^{146}$Gd many of the features of a doubly closed shell nucleus (see Figure~[\ref{fig:gaps}]). In addition, $^{146}$Gd is the only even-even nucleus besides $^{208}$Pb that exhibits a 3$^{-}$ first excited state. As we will see later, this fact favours the identification of the two-phonon states on this nucleus. An additional advantage of studying $^{146}$Gd is that it is easily accessible by fusion-evaporation reactions, which allows spectroscopic studies of yrast and above-yrast states.\

\

The energy splittings of p-h multiplets in $^{146}$Gd are of particular interest because they give us information about the nucleon-nucleon residual interaction and provide fundamental data for shell model calculations in this region. These data are specially important for understanding the yrast and near-yrast states of more complex structures in the $^{146}$Gd neighbouring nuclei, where very frequently the particle-hole excitations across the Z=64 shell gap contribute. In addition, $^{146}$Gd presents an advantage over $^{208}$Pb because the proton p-h states in $^{146}$Gd are lower in energy than the neutron p-h excitations; this makes the characterization of the states easier due to the fact that there are fewer p-h states at that low excitation energy. \\

The existence of two-phonon octupole multiplets in doubly-magic nuclei was predicted long ago (see for instance~\cite{ALDER56}). Experimentally these states have been sought for a long time in $^{208}$Pb, and very solid candidates for these excitations have been found recently in~\cite{MINFANG}. These states have also been investigated in $^{146}$Gd. Closely related states were identified~\cite{KLEIN82,LUNARDI} in $^{147}$Gd($\nu${\it f}$_{7/2}$$\times$3$^{-}$$\times$3$^{-}$) and in $^{148}$Gd($\nu$$^{2}$$\times$3$^{-}$$\times$3$^{-}$). In the fusion-evaporation study of~\cite{YATES} two possible candidates for the (3$^{-}$$\times$3$^{-}$) states were proposed in $^{146}$Gd. The confirmation of the existence of these states and the identification of the other multiplet members is one of the main goals of the present investigation.\\

\section{Shell model and particle-hole states}

The occurrence of the magic numbers has been one of the strongest motivations for the formulation of the shell model. At these proton and neutron numbers, effects analogous to the electron shell closure in atoms are observed. The main characteristic at these numbers is that the nucleus is especially stable. The shell model is based on the assumption of an average potential (built up by the action of all the nucleons) in which the nucleons can move independently.\

In the doubly magic nuclei, where both protons and neutrons fill the shells defined by the magic numbers, nucleons are strongly bounded and the nucleus is very stable against excitations. In these nuclei the first excitations are either of vibrational (see next section) or of particle-hole character. In a particle-hole excitation a pair of protons or neutrons coupled to 0$^{+}$ by the pairing force is broken, and one of the nucleons is promoted to an empty shell above the energy gap between the shells. The excitation energy depends, in a first approximation, on the energy gap and the pairing force. But the particle-hole nucleons can suffer what it is called a ``residual interaction'' that changes the previously defined energy of the multiplet, and also splits in energy the members of the multiplet depending on the {\it j} coupling of the particle and the hole employed to build the final spin J. The expected splittings are proportional to

\begin{equation}
V_{SDI}=A_{T}(-1)^{2(n_{1}+n_{2})}\frac{(2j_{1}+1)(2j_{2}+1)}{2}(-1)^{2(j_{2}+l_{2})}[1-(-1)^{J+T}]\left(\begin{array}{ccc}j_{1}&j_{2}&J\\\frac{1}{2}&\frac{-1}{2}&0\end{array}\right)^{2}\times\\ \nonumber 
\end{equation}\
\begin{equation}
\times[1+(-1)^{T}]\left(\begin{array}{ccc}j_{1}&j_{2}&J\\\frac{1}{2}&\frac{1}{2}&-1\end{array}\right)^{2}
\end{equation}\
 



where A$_{T}$ = A'$_{T}$C(R$_{0}$).\

Here A$_{T}$ is defined as the product of the strength A'$_{T}$ and the value of the radial integral C(R$_{0}$). A typical estimation of A$_{T}$ in the $^{146}$Gd region is A$_{T}$=25000/A, where A is the atomic mass number. In Chapter 4 we will calculate particle-hole multiplets below 4 MeV in the $^{146}$Gd nucleus and their residual interactions. As mentioned earlier, $^{146}$Gd has many of the features of a doubly closed shell nucleus. Consequently, and looking at Figure~[\ref{fig:gaps}], one expects the first excitations to come from the promotion of a proton from the d$_{5/2}$ and g$_{7/2}$ levels to the s$_{1/2}$, h$_{11/2}$ and d$_{3/2}$ levels. The next possibility is to promote more than one particle and then create two-particle-hole configurations.\

Although in doubly magic nuclei, such as $^{208}$Pb, the first excited states are either of vibrational or of particle-hole character, in $^{146}$Gd the first excited state is a mixture. On the one hand is a very collective state, decaying by a 37 W.u. {\it E}3 transition, but on the other hand, its wave function has a dominating component from the $\pi$h$_{11/2}$$\pi$d$^{-1}$$_{5/2}$ particle-hole excitation across the Z=64 gap within the 50 to 82 major shells. This contribution was estimated to be 50$\%$ by Conci et al.~\cite{CONCI} from QRPA calculations and, thus the level preserves its particle-hole nature. This is very important in our discussion since it will perturb the energy of some of the levels due to the Pauli principle.\\

\section{Vibrational states}

Two types of collective nuclear motions appear when describing the macroscopic properties of nuclei. These motions are \textbf {vibrations} (for spherical and almost spherical nuclei) and \textbf {rotations} (for deformed nuclei), which are based on the ``liquid drop'' model. Since $^{146}$Gd can be considered as a doubly closed-shell nucleus, the model that better describes the system is the vibrational model. In this section we will describe this model in detail.\

In complex systems such as nuclei, which are composed of many particles, it is possible to describe the excitation spectra in terms of elementary excitation modes corresponding to the different fluctuations around equilibrium. These fluctuations depend on the internal structure of the system and could be considered approximately independent. The elementary modes may be associated with excitations of individual particles or they may represent collective vibrations of the nucleus shape.\

The possibility of collective shape oscillations in the nucleus is strongly suggested by the fact that some nuclei are found to have non-spherical equilibrium shapes whereas others, such as closed-shell nuclei, have an equilibrium with spherical shape. Thus, we expect to find intermediate cases in which the shape presents rather large fluctuations away from the equilibrium shape.\\

The vibrational model \footnote{Originally proposed by Bohr and Mottelson and later developed by Faessler and Greiner.} describes collective movements of the nucleus assuming that the nucleus behaves as a liquid drop. In order to describe this model we will assume that the nucleus has a spherical shape of radius {\it R}$_{0}$ in its ground state, which represents the equilibrium state. The nucleus can be considered as an homogeneous fluid with shape fluctuations about equilibrium described by the surface coordinates or amplitudes {\it $\alpha$}$_{\lambda\mu}$.\

\begin{equation}
R({\it \theta},{\it \phi}) = R_{0} \bigg\{ 1 + \sum^{\infty}_{\lambda=0} \sum^{\lambda}_{\mu=-\lambda} {\it \alpha}_{\lambda \mu} ({\it t}){\it Y}_{\lambda \mu}({\it \theta},{\it \phi}) \bigg\}
\end{equation}\

Each vibrational mode is given by $\lambda$ and described by the 2$\lambda$+1 amplitudes ({\it $\alpha$}$_{\lambda\mu}$, $\mu$= -$\lambda$,...,$\lambda$). These amplitudes describe the expansion of the shape fluctuations in spherical harmonics and they are not independent and present rotational invariance. The states corresponding to each vibrational mode have angular momentum {\it J} = $\lambda${\it $\hbar$} and parity P = (-1)$^{\lambda}$. Thus, there exist infinite vibrational movements. The lowest vibrational modes could be associated with the $\lambda$ value and defined by their corresponding spherical harmonic and are expected to have density variations with no radial nodes and may be referred to as shape oscillations. Below is an overview of the lowest vibrational modes.

\begin{itemize}
\item[$\bullet$] $\lambda$=1, dipole mode. The dipole mode is the first vibrational mode that presents changes in the nuclear shape. The isoscalar and the isovector components induce different behaviours.
\begin{itemize}
\item[-] isoscalar (T=0). There occurs a change in the center of mass, but the nucleus structure does not change.
\item[-] isovector (T=1). The neutrons and protons move out of phase. this represents the so-called giant dipole resonance, {\it J}$^{\pi}$=1$^{-}$, studied since 1940.
\end{itemize}
\item[$\bullet$] $\lambda$=2, quadrupole mode. This is the fundamental mode in the vibrational model, since it is the first that induces non-spherical shape oscillations in the nucleus. The nucleus oscillates between prolate and oblate forms passing through the spherical equilibrium shape. 
\item[$\bullet$] $\lambda$=3, octupole mode. This vibrational mode is much more complex than the previous modes and the vibrating nucleus undergoes pear-shaped distortions, with the "stem end" and the "blossom end" exchanging places periodically.
\end{itemize}

In the present discussion we avoided to discuss the density vibrations since they are of no importance in nuclear structure at the energies relevant to this work.\

As we have seen before, the amplitudes {\it $\alpha$}$_{\lambda\mu}$ are more appropriate to describe the collective excitations of the nucleus than using the individual positions of the nucleons. The Hamiltonian for a vibrational mode of $\lambda$-order can be written in terms of the amplitudes as\

\begin{equation}
H_{\lambda}=\frac{1}{2}D_{\lambda}\sum_{\mu}\left|\frac{d{\it \alpha}_{\lambda \mu}}{d{\it t}}\right|^{2}+\frac{1}{2}C_{\lambda}\sum_{\mu}\left|{\it \alpha}_{\lambda \mu}\right|^{2}
\end{equation}\

where

\begin{equation}
D_{\lambda}=\frac{\rho R^{5}_{0}}{\lambda}
\end{equation}\

and 

\begin{equation}
C_{\lambda}=\frac{1}{4\pi}(\lambda - 1)(\lambda + 2)\alpha_{S}A^{2/3}-\frac{5}{2\pi}\frac{\lambda - 1}{2\lambda + 1}\alpha_{C}\frac{Z(Z-1)}{A^{1/3}}
\end{equation}\

The nuclear radius at equilibrium, R$_{0}$, can be approximated as R$_{0}$=1.2$\times$A$^{1/3}$ fm, $\rho$ is its mass density, and $\alpha_{S}$ and $\alpha_{C}$ contain the surface and Coulomb energy terms, respectively, of the liquid-drop model ($\alpha_{S}$= 18.3 MeV and $\alpha_{C}$= 0.7 MeV). The first term of the Hamiltonian is the kinetic energy of the harmonic oscillator and the quantity {\it D}$_{\lambda}$ is referred to as the mass parameter. The second term is the potential energy of deformation and the coefficient {\it C}$_{\lambda}$ is referred to as the restoring force parameter~\cite{BOHMOT2}. If the vibrational modes are not coupled, since the Hamiltonian is independent of time or constant in movement, the derivative with respect to time gives the equation for the system

\begin{equation}
D_{\lambda}\frac{d^{2}{\it \alpha}_{\lambda \mu}}{d{\it t}^{2}}+C_{\lambda}{\it \alpha}_{\lambda \mu} = 0
\end{equation}\

which is the equation of a harmonic oscillator of frequency

\begin{equation}
\omega_{\lambda} = \sqrt{\frac{C_{\lambda}}{D_{\lambda}}}
\end{equation}\

From the last relation it can be easily appreciated that the oscillator frequency derives from the shape properties of the nucleus.\\

When quantizing the oscillator, the vibrational states become defined by three quantum numbers, $\mid${\it N};$\lambda$$\mu$$\rangle$, where {\it N} is the number of vibrational energy quanta of multipolarity $\lambda$, called \textbf {phonons}. The phonon is a boson of spin {\it J} = $\lambda${\it $\hbar$} and parity $\pi$=(-1)$^{\lambda}$, while $\mu$ is the $\lambda$ projection. The state that corresponds to the vacuum is $\mid$0;00$\rangle$, and the state corresponding to 1 phonon is obtained after applying the creation operator $\beta^{\dag}_{\lambda\mu}$ to the vacuum state. The creation and annihilation operators result from the amplitude quantization, considered as operators and properly normalized, and fulfil the commutation rules of the creation and annihilation boson operators. 

\begin{equation}
[\beta^{\dag}_{\lambda\mu},\beta^{\dag}_{\lambda\mu'}]=0\; ; \;[\beta_{\lambda\mu},\beta_{\lambda\mu'}]=0 \; ; \; [\beta_{\lambda\mu},\beta^{\dag}_{\lambda\mu'}]=\delta_{\mu\mu'}
\end{equation}\

Thus, such a system of bosons can be treated in terms of the operators  $\beta^{\dag}_{\lambda\mu}$ and $\beta_{\lambda\mu}$ that create and annihilate a quantum of excitation. The vibrational Hamiltonian  corresponding to the mode $\lambda$ is\

\begin{equation}
H_{\lambda} = \hbar\omega_{\lambda}\sum_{\mu}\bigg(\eta_{\lambda\mu} + \frac{1}{2}\bigg)
\end{equation}

where the number of quanta in the projection $\mu$ of multipolarity $\lambda$ is given by the operator $\eta_{\lambda\mu}$, which is  $\eta_{\lambda\mu}$ = $\beta^{\dag}_{\lambda\mu}$$\beta_{\lambda\mu}$. If we sum the projections, 
\begin{equation}
 N_{\lambda}=\sum^{\lambda}_{\mu=-\lambda}\eta_{\lambda\mu} 
\end{equation}

we obtain the number of phonons of multipolarity $\lambda$. The energy of the vibrational states is given by the expression

\begin{equation}
E_{\lambda} = \hbar\omega_{\lambda}\sum_{\mu}\bigg(n_{\lambda\mu} + \frac{1}{2}\bigg)
\end{equation}

It is easy to observe that the levels are equally spaced by $\hbar\omega_{\lambda}$, with E$_{\lambda}$=N$_{\lambda}$$\hbar\omega_{\lambda}$ the energy and N$_{\lambda}$ the number of phonons of multipolarity $\lambda$. In  Table~[\ref{tab:tabla_phonones}] we have the first three quadrupole ($\lambda$=2) and octupole ($\lambda$=3) phonons with spin and parity assignments of their multiplet members.\

\begin{table}[thbp!]
\begin{center}
\begin{tabular}{|c|c|c|c|}
\hline
$\lambda$ & Phonon & Energy & Multiplet members\\
\hline
 & 0 & 0$\hbar\omega_{2}$ & 0$^{+}$\\ 
2 & 1 & $\hbar\omega_{2}$ & 2$^{+}$\\ 
 & 2 & 2$\hbar\omega_{2}$ & 0$^{+}$,2$^{+}$,4$^{+}$\\ 
\hline
 & 0 & 0$\hbar\omega_{3}$ & 0$^{+}$\\ 
3 & 1 & $\hbar\omega_{3}$ & 3$^{-}$\\ 
 & 2 & 2$\hbar\omega_{3}$ & 0$^{+}$,2$^{+}$,4$^{+}$,6$^{+}$\\ 
\hline
\end{tabular}
\end{center}
\caption{Lowest three quadrupole and octupole phonons, with spin and parity assignments of their multiplet members}
\label{tab:tabla_phonones}
\end{table}

An important consequence (as can be seen from this table) is that the two-phonon states should occur at twice the energy of the one-phonon state for both vibrational modes. The importance of the phonon states resides in their role in the collectivity of the nucleus. Vibrational excitations in nuclei have been studied for many years. While there are examples of excitations up to the three-phonon quadrupole states in even-even nuclei ($^{118}$Cd)~\cite{HAMILTON}, even in the case of two-phonon octupole states the information is sparse. Since studying multi-phonon octupole states is one of the aims of the present work, we will present a deeper overview of our knowledge of octupole states in the next section.

\section{Octupole states}

Many studies have been carried out to identify one and two-phonon octupole excitations in nuclei. In particular, in the rare earth region, extensive studies have been carried out using fusion-evaporation reactions. Two-phonon octupole excitations have been identified coupled to one or two particle excitations~\cite{LACH,PIIP1990,ZOLTS,BIZ}. These cases were relatively easily identified because the double {\it E}3 cascades lied along the yrast line and possible competing lower-multiplicity de-excitations do not occur easily. These nuclei are $^{147}$Gd and the N=84 isotones $^{144}$Nd, $^{146}$Sm and $^{148}$Gd. A more difficult quest has been to identify the two-phonon octupole quartet members (0$^{+}$,2$^{+}$,4$^{+}$,6$^{+}$) in spherical even-even nuclei expected to occur at twice the energy of the 3$^{-}$ one-phonon state but clearly above the yrast line. Three regions of nuclei~\cite{SPEAR} where the octupole multiphonon excitations might be expected with large {\it E}3 strengths (more than 30 W.u.) in their B({\it E}3;3$^{-}$$\rightarrow$0$^{+}$) transitions have been identified. These regions are near $^{96}$Zr, $^{146}$Gd and $^{208}$Pb. But only two known nuclei exhibit the 3$^{-}$ phonon as the first excited state: $^{146}$Gd and $^{208}$Pb. In this nuclei only the 0$^{+}$ and the 6$^{+}$ members of the two-phonon multiplet can decay by an {\it E}3 transition to the 3$^{-}$ state. Their expected strengths are, in first approximation, twice the B({\it E}3;3$^{-}$$\rightarrow$0$^{+}$). In the $^{146}$Gd case, the estimated strength is 57 W.u.~\cite{LUNARDI}. \ 

\begin{figure}[h!]
\begin{center}
\includegraphics[width=14.cm,height=8.cm,angle=0.]{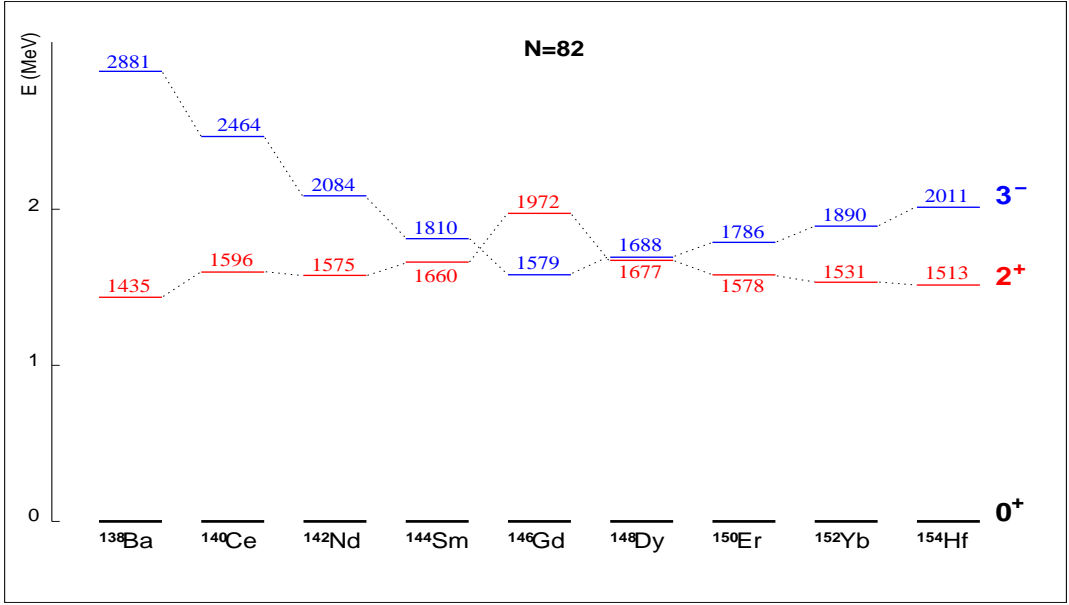}
\end{center}
\caption{Energies of the lowest 2$^{+}$ and 3$^{-}$ states in even-A N=82 nuclei.}
\label{fig:2_3}
\end{figure}

For many years studies of two-phonon octupole vibrations were focused on the $^{208}$Pb case and its neighbourhood, but in the last 25 years they have been extended to nuclei around $^{146}$Gd. Historically, $^{146}$Gd became of special interest after the establishment of the {\it J$^{\pi}$}= 3$^{-}$ ~\cite{KLEIN783-} for the first excited state, characterizing it as the second even-even nucleus, besides $^{208}$Pb, showing this feature. Furthermore, it was found that the unusually large {\it E}3 strength to the ground state, comparable with that found in $^{208}$Pb, indicated strong octupole collectivity in this nucleus. The fact that the first 2$^{+}$ state is about 300 keV higher than in any other N=82 nucleus (see Figure~[\ref{fig:2_3}]), was interpreted as spectroscopic evidence for a pronounced energy gap in the single-particle spectrum at Z=64 between the 2d$_{5/2}$ and the 1h$_{11/2}$ proton orbitals.

Apart from their similarities in terms of the octupole vibrational mode, it is expected to be easier to identify multiphonon excited states in $^{146}$Gd than in $^{208}$Pb, because its one-phonon octupole state occurs about 1 MeV lower in energy. This implies a lower density of states in the energy range where the two-phonon states are expected, so it would be easier to distinguish them from the particle-hole states that will lie in the same region.\

Another interesting aspect of these two nuclei is the different microscopic structures of their cores. The 3$^{-}$ state of $^{208}$Pb is composed of contributions from both the proton and neutron particle-hole excitations across the Z=82 and N=126 shells, respectively, while in $^{146}$Gd, the 3$^{-}$ state is dominated by the proton component from the $\pi$h$_{11/2}$$\pi$d$^{-1}$$_{5/2}$ particle-hole excitation across the Z=64 gap within the 50 to 82 major shells. This contribution was estimated to be 50$\%$ by Conci et al.~\cite{CONCI} from QRPA calculations. This behaviour is evident from the systematic variation of the energy of the 3$^{-}$ state in the N=82 isotones (see Figure~[\ref{fig:2_3}]). The lowest excitation value corresponds to the $^{146}$Gd nucleus, whereas it increases when depleting the $\pi$d$_{5/2}$ shell at lower Z and when filling the $\pi$h$_{11/2}$ at higher Z nuclei.

\section{Previous knowledge of $^{146}$Gd}

The $^{146}$Gd region presents clear advantages for the study of two-phonon states in even-even nuclei. Its placement in the nuclidic chart, with eight neutrons less than the lightest stable Gd isotope, makes possible the study of the proton p-h multiplets by different means.
\begin{enumerate}
\item The instability of the $^{146}$Gd nucleus limits its study and constrains the number of techniques which can be employed to study it. However, it makes $^{146}$Gd accessible to in-beam $\gamma$-ray measurements, following suitably chosen fusion-evaporation reactions, where little angular momentum is transferred to the compound nucleus. Yrast states up to $\sim$9 MeV have been identified previously by $\boldsymbol {in}$-$\boldsymbol {beam}$ $\gamma$-ray measurements~\cite{KLEIN79,BRODA} using fusion-evaporation reactions. The lowest 0$^{+}$ and 2$^{+}$  excited states are also known from other in-beam experiments~\cite{JULIN,OGAWA,YATES87} explicitly designed to enhance the population of these states. Twenty years ago, an ($\alpha$,2n) fusion-evaporation experiment~\cite{YATES} was used to study non-yrast states and to search for the double-octupole excitations in $^{146}$Gd. In this experiment two germanium detectors were used to record $\gamma$-$\gamma$ coincidences and $\gamma$-ray angular distributions. This study led to a substantial extension of the $^{146}$Gd level scheme.\

Before the present work, a similar fusion-evaporation measurement~\cite{TESINA} was performed to improve the nucleus knowledge of $^{146}$Gd. A $^{144}$Sm($\alpha$,2n) experiment at the IKP (University of Cologne) using 26.5 MeV $\alpha$-particles impinging on a self-supporting Sm metal foil 10.0 mg/cm$^{2}$ thick and enriched to 97.6$\%$ in $^{144}$Sm was carried out. The sensitivity of that experiment was about 10 times higher than in~\cite{YATES}. The experimental set-up consisted of one EUROBALL Cluster detector consisting of seven encapsulated coaxial germanium detectors in a common cryostat that was placed in front of the target, and five tapered germanium detectors placed at $\sim$142 degrees with respect to the beam direction. In that work, a total of 21 new $\gamma$-rays were identified corresponding to the decay of 16 new states and 19 $\gamma$-rays corresponding to 13 known levels. Also, 7 $\gamma$-rays were seen for the first time in an in-beam experiment. Unfortunately, in that experiment no information on the $\gamma$-ray angular distributions could be extracted and thus spin assignments were mainly based on the level decay pattern after a careful energy matching inspection. For this reason we chose to repeat the experiment and extract angular distribution and linear polarization information on the transitions in $^{146}$Gd.
\item A $\boldsymbol {\beta}$-$\boldsymbol {decay}$ experiment~\cite{STYCZEN} to study the decay of 23-{\it s} $^{146}$Tb (J$^{\pi}$=5$^{-}$), which proceeds by allowed Gamow-Teller transitions to specific p-h  excitations in $^{146}$Gd, provided information about the location of the neutron p-h states at energies higher than 3.4 MeV. A second $\beta$-decay experiment to study the decay of the 1$^{+}$ isomer with T$_{1/2}$=8 s populated states with J$^{\pi}$=0$^{+}$ and 2$^{+}$~\cite{KLEINAR}. 
\item Most of the nuclei near $^{146}$Gd are unstable. However, $^{148}$Gd lives long enough (74.6 a) to allow the construction of a radioactive target. Twenty years ago such a target was made. The neutron pairing vibrational state at an energy slightly greater than 3 MeV and an associated 2$^{+}$ state were identified in a $^{148}$Gd(p,t)$^{146}$Gd $\boldsymbol {transfer}$ reaction by Flynn et al.~\cite{FLYNN} in 1983. Additional $^{146}$Gd levels were also observed in this experiment and the angular distributions obtained were compared with distorted-wave Born approximation (DWBA) calculations in order to obtain information on the $\Delta$L transferred in the reaction and consequently on the spin of the populated states in $^{146}$Gd. Six firm L-transfer values were obtained from this comparison.\
Six years later, a similar experiment was performed by Mann et al.~\cite{MANN}. Comparing the angular distributions with DWBA calculations they obtained about eleven firm {\it L}-transfer values and approximate {\it L} values for about fifty more excited states. 
\item Two $\boldsymbol {conversion}$-$\boldsymbol {electron}$ experiments have been performed~\cite{YATES,JULIN,YATES87} to search for high-energy 0$^{+}$ states in $^{146}$Gd. In the former, the second 0$^{+}$ state in $^{146}$Gd was identified through observation of the {\it E}0(0$^{+}_{2}$$\rightarrow$0$^{+}_{1}$) transition. In the later, three new {\it E}0 transitions were identified. One de-excited the two-neutron pairing vibrational state in $^{146}$Gd, but it was not clear whether either of the others could be associated with the 0$^{+}$ member of the (3$^{-}$$\times$3$^{-}$) two-phonon octupole multiplet.
\end{enumerate}

As mentioned previously, there are several limitations on the available reactions that allow us to study p-h multiplets in $^{146}$Gd. We find problems if we want to do single-particle transfer reactions due to the short half-lives of the $^{145}$Gd, $^{145}$Eu, $^{147}$Tb and $^{147}$Gd nuclei. These kinds of studies could give us the most straightforward information about the p-h multiplets. The above described two-nucleon transfer reaction or multinucleon transfer reactions will suffer from low energy resolution and moreover, they will not populate particle-hole states in $^{146}$Gd. In conclusion, it seems that the only kinds of experiments that can allow us to add to our present knowledge of proton p-h multiplets are fusion-evaporation reactions with low angular momentum input. This requirement can be met by reactions such as ($^{3}$He,n) or ($\alpha$,2n) where the incident particle is very light and not more than two particles are evaporated. As was noted earlier, the ($^{3}$He,n) reaction on $^{144}$Sm~\cite{JULIN,OGAWA} was used to enhance explicitly the 0$^{+}$ and the 2$^{+}$ states but this reaction has a slightly positive {\it Q}-value. Taking into account the Coulomb barrier, the reaction is only possible at energies far above the threshold energy, which leads to complications because other reaction channels appear and dominate. On the contrary, the ($\alpha$,2n) reaction has a negative {\it Q}-value and it has been demonstrated~\cite{OGAWA} that non-yrast states are populated. A study of the optimum bombarding energy for the population of the non-yrast states has been made by Yates et al.~\cite{YATES}, where a bombarding energy of about 26 MeV was found to be the optimal. In this work yrast and above-yrast states were observed in $^{146}$Gd and many of them were interpreted as members of two-nucleon multiplets in $^{146}$Gd. In the present work we will re-investigate the same reaction study with improved sensitivity.\

Attempts to locate the two-phonon octupole states in $^{146}$Gd have had limited success, because in addition to the {\it E}3 transition to the one-phonon octupole state, the two-phonon states can decay through low multipolarity transitions which make its identification difficult. In Yates et al.~\cite{YATES}, three 6$^{+}$ and three 4$^{+}$ states in the expected energy range for the two-phonon states were found. But 6$^{+}$ and 4$^{+}$ states from other configurations are expected in the same region of excitation energy making firm configuration assignments difficult.\ 

\chapter{The experiment}

{\it  In this chapter the experimental details of the measurements will be presented. A description of the electronics used for the data acquisition will be given as well. Later, we will explain how the energy and efficiency calibrations were done. Finally, we will see in depth the different information which could be extracted from our detectors set-up: directional angular distributions and directional linear polarization.} 
  
\vspace*{0.6cm}
\section{The experimental setup}
\subsection{The Tandem accelerator}

\begin{figure}[h]
\begin{center}
\includegraphics[width=15.cm,height=11.5cm,angle=0.]{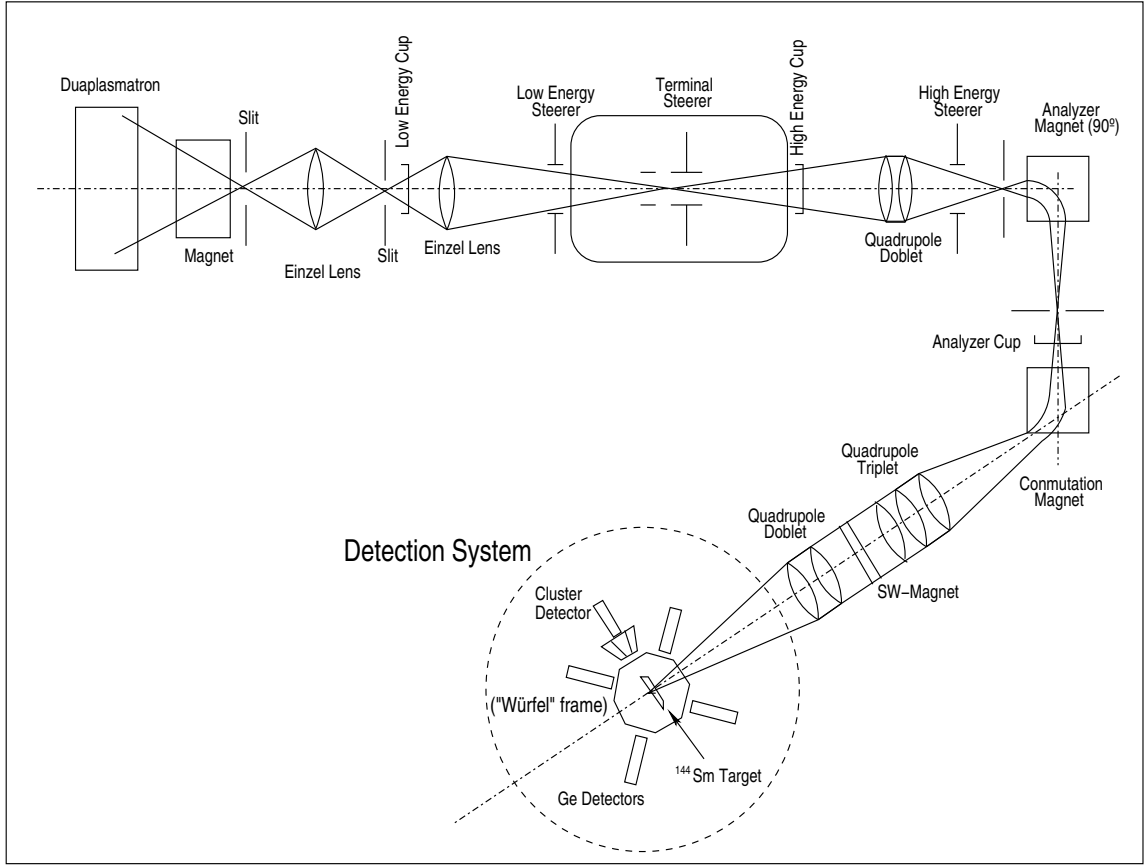}
\end{center}
\caption{Schematic overview of the Tandem accelerator at the Institute for Nuclear Physics (IKP),University of Cologne (Germany) and the detection setup used in the experiment.}
\label{fig:tandem_plan}
\end{figure}

   All the measurements, $\gamma$-singles and $\gamma$-$\gamma$ coincidences, presented in this work were made at the Institute for Nuclear Physics (IKP) of the University of Cologne (Germany) with an $\alpha$-particle beam produced at the FN Tandem Van de Graaff Accelerator. This accelerator has a working voltage of up to 11 MV and three different ion sources. In our case, the voltage of the accelerator was 8.75 MV and a duoplasmatron source was used. An overview of the tandem accelerator and the beam line elements is shown in Figure~[\ref{fig:tandem_plan}].\\

\subsection{Excitation function of the $^{146}$Gd + $\alpha$ reaction }\

Depending on the energy of the beam, different reaction exit channels will be favoured compared to others. Since we are interested in the study of low-lying non-yrast levels in $^{146}$Gd, we have to select an energy which maximizes the population of this kind of states.\
 
The optimum bombarding energy for the population of non-yrast states in $^{146}$Gd was known from the measurements by Yates et al.~\cite{YATES}. These data indicated 26.3 MeV as an optimum bombarding energy for an ($\alpha$,2n$\gamma$) study of non-yrast levels in $^{146}$Gd. At this energy, the maximum in the excitation functions for $\gamma$-rays de-exciting non-yrast levels has been attained, while the competing ($\alpha$,n) exit channel is not so strong to obscure the lines of interest. Any further increase in bombarding energy will lead to a greater yield of well-known and strongly dominant yrast transitions. At this bombarding energy, the maximum excitation energy for the $^{146}$Gd is about 5 MeV, and this will permit us to study the levels lying in the region where the two-phonon states are predicted to be (about twice the one-phonon energy of 1.58 MeV).\

\subsection{Detection system }

As mentioned earlier, we have studied the $^{144}$Sm($\alpha$,2n) fusion-evaporation reaction using 26.3 MeV $\alpha$-particles impinging on 3.0 mg/cm$^{2}$ thick target enriched to 97.6\% in $^{144}$Sm and supported by a 0.5 mg/cm$^{2}$ thick Au backing made at the Laboratori Nazionale di Legnaro (Italy). The beam was stopped with a Ta beam dump placed in the beam pipe about one meter behind the target.\\

\begin{figure}[h!]
\begin{center}
\includegraphics[width=15.cm,height=8.cm,angle=0.]{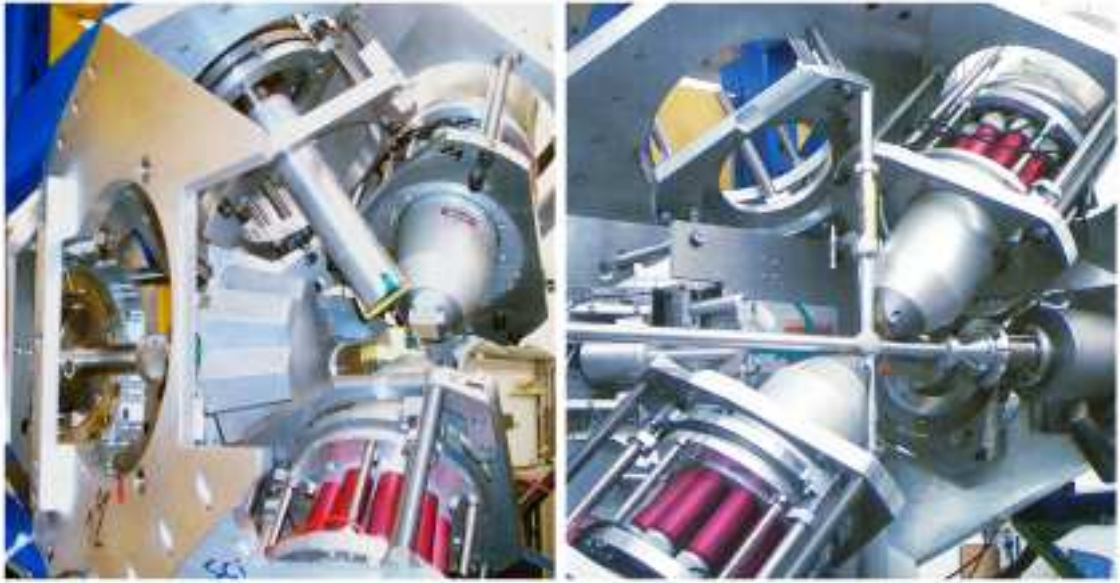}
\end{center}
\caption{Picture of the detection system mounted in the W\"urfel frame designed at the IKP institute. The W\"urfel frame was opened in order to visualize the detectors. In the left picture, the CLUSTER detector which is beside the other Ge detectors can be seen, while in the right picture, we show the rest of the detectors and the beam tube.}
\label{fig:latdetec2a}
\end{figure}\

\begin{figure}[h!]
\begin{center}
\includegraphics[width=13.cm,height=10.cm,angle=0.]{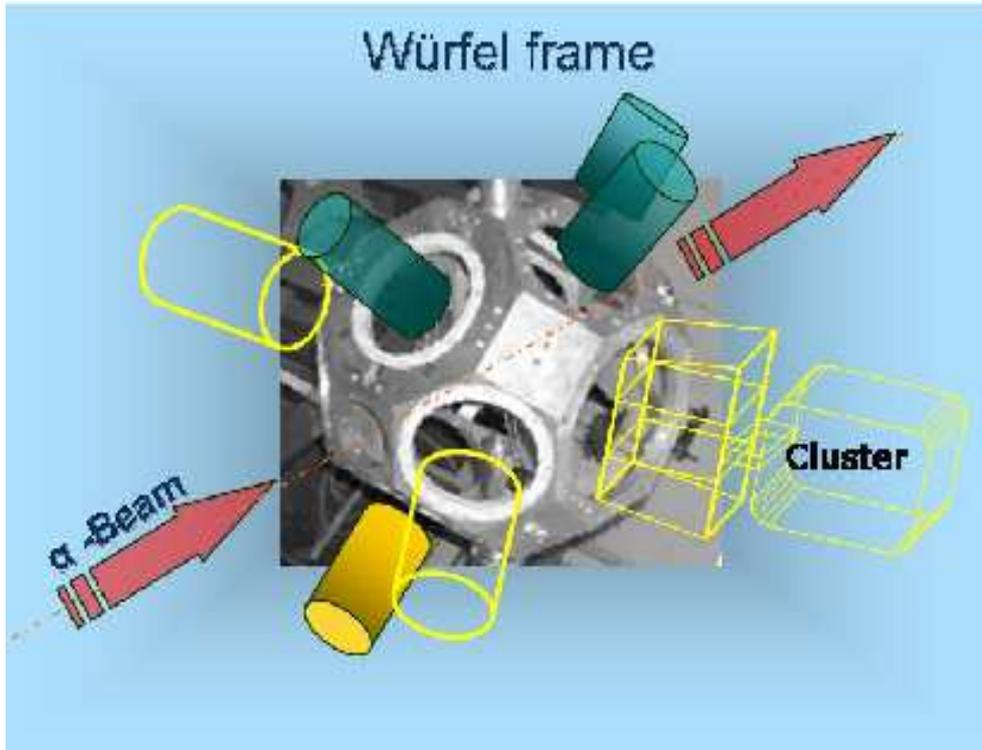}
\end{center}
\caption{Schematic diagram of the detection setup consisting of ten germanium detectors placed in a W\"urfel frame designed at the IKP institute.}
\label{fig:latdetec1}
\end{figure}\

The average energy of the $\alpha$-particles when they react in our target is 26 MeV, and the average energy of the recoiling $^{146}$Gd nuclei is 6.3 MeV. At this energy the mean range of the recoiling $^{146}$Gd nuclei is 1.2 mg/cm$^{2}$ before it is fully stopped, and the corresponding time interval is of the order of 1 ps. This means that gammas de-exciting levels with half-lives of the order of 1 ps will be observed with both shifted and stopped components. If the half-life is clearly longer, the peak will be observed only at the stopped position and in the cases where the half-life is much shorter the peak observed will be fully shifted due to the Doppler effect. This effect could be checked with the case of the 2$^{+}$ 1972.0 keV state, whose half-life is shorter than 0.32 ps  ~\cite{TESINA}. In this case, we see both peaks, the stopped and the shifted. Thus, the presence or absence of the Doppler effect for a transition in our spectra will tell us about the half-life of the de-exciting level (neglecting the side feeding).\\

\begin{figure}[h!]
\begin{center}
\includegraphics[width=12.cm,height=10.cm,angle=0.]{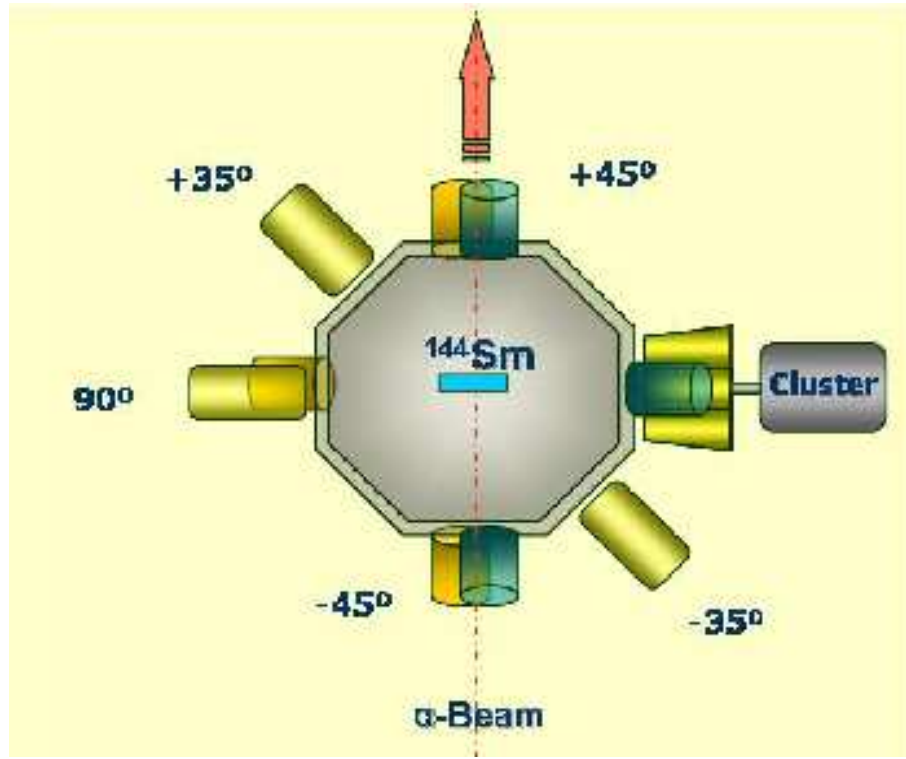}
\end{center}
\caption{Schematic diagram of the detection setup showing the different angles of the detectors.}
\label{fig:latdetec2}
\end{figure}

In order to construct the $^{146}$Gd level scheme we have used the germanium detectors to measure $\gamma$-$\gamma$ coincidences. The detection set-up consisted of ten germanium detectors placed in a W\"urfel frame designed at the IKP (see Figure~[\ref{fig:latdetec2a}], Figure~[\ref{fig:latdetec1}] and Figure~[\ref{fig:latdetec2}]). Distances between the detectors and the target are shown in Table~[\ref{tab:tabla_det_1}]. Such a configuration permitted us to have detectors placed at five different angles with respect to the beam direction (see also Table~[\ref{tab:tabla_det_1}]). As we will see later, these five angles made an angular distribution measurement possible. In addition, one of the detectors was a EUROBALL Cluster detector comprised of seven encapsulated coaxial germanium detectors in a common cryostat that was placed at 90 degrees with respect to the beam direction. This detector, placed at that angle, also permitted us to obtain a linear polarization measurement. Half of the ten germanium detectors had relative efficiencies (compared to the corresponding efficiency of a 3''$\times$3'' NaI(Tl) crystal at a source-detector distance of 25 cm) of 50\%, while the other half had 30\%.\\

The current registered in the Faraday cup during the experiment was of the order of 2.5 nA measured occasionally at the analyzer cup.\

The sensitivity of the present set-up was about a factor of 10 higher in coincidences in comparison with the experiment carried out by Yates et al.~\cite{YATES} where only two low-efficiency Ge(Li) detectors were used.\\\\\\

\

\begin{table}[thbp!]
\begin{center}
\begin{tabular}{|c|c|c|c|c|c|c|c|}
\hline
&&&\small{Relative}&&\footnotesize{Anti-Compton}&\small{Distance}&\\
Detector&$\theta$&$\phi$&\small{efficiency}&Absorbers&\small{shield}&\small{to target}&ADC \\
\hline
\hline
\footnotesize{CLUSTER}&90&0&50\%&\small{none}&no&14.4 cm&16k\\
\hline
Ge 1&45&90&50\%&\small{1 mm Cu}&yes&8.4 cm&16k\\
\hline
Ge 2&145&0&30\%&\small{1 mm Cu, 1 mm Pb}&no&9.3 cm&16k\\
\hline
Ge 3&90&305&30\%&\small{1 mm Pb, 2 mm Al}&no&8.8 cm&16k\\
\hline
Ge 4&135&270&50\%&\small{1 mm Cu}&yes&8.4 cm&16k\\
\hline
Ge 5&135&90&50\%&\small{1 mm Cu}&yes&8.4 cm&16k\\
\hline
Ge 6&45&270&50\%&\small{1 mm Cu}&yes&8.4 cm&8k\\
\hline
Ge 7&35&180&30\%&\small{2 mm Cu, Pb arround}&no&10.2 cm&8k\\
\hline
Ge 8&90&125&30\%&\small{2 mm Cu, 1 mm Al}&no&9.8 cm&8k\\
\hline
Ge 9&90&180&30\%&\small{1.25 mm Cu}&yes&15.2 cm&8k\\
\hline
\end{tabular}
\end{center}
\caption{Settings and characteristics of the detectors used in the experiment placed in the W\"urfel frame. Note that the CLUSTER is treated as one detector since an {\it addback} was done in the data sorting for build up the matrices.}
\label{tab:tabla_det_1}
\end{table}\

\section{Data and calibrations }

Data were recorded using a data acquisition system developed at the University of Cologne, which allowed us to record spectra in two different ways: a direct spectrum for each detector, where all the signals coming from the detector are continuously recorded without any restriction (singles), and coincidence events, where the requirement to validate an event was that at least two detectors fired with a time difference smaller than 300 ns. This condition was fixed electronically as we will describe it later.\

A total of 132 runs of one hour each were accumulated during the experiment. At some times between two reaction runs, a $^{226}$Ra source was put close to the target position to make energy calibrations. Immediately after the last run, we made measurements of different durations of the activation in the target. This kind of measurement allowed us to identify peaks in the spectra originating from the isotropic decay of nuclei produced in the target. After this measurement, $^{226}$Ra and $^{133}$Ba sources were placed at the target position in order to make efficiency calibrations for all the detectors.\

During the running period, minor instabilities in the electronics might led to slight gain shifts in the spectra. As a consequence, corrections to the spectra became necessary. First we corrected all the runs and all detectors, and later we proceeded with the energy calibration. Thus, for each detector we ``shifted'' with a linear function all the runs to match the closest in time to a $^{226}$Ra energy calibration run, and we added all of them together to get the total singles spectrum  for each detector. This procedure was followed for correcting the gain-shift when building the $\gamma$-$\gamma$ matrices as well.\

We repeated the same procedure for the $^{226}$Ra and $^{133}$Ba efficiency calibration runs, but in this case shifting all the runs of accumulated statistics to the first in time by a linear gain-shift correction and finally adding them. The small gap of time from the last run of reaction data to the first efficiency calibration run makes any gain-shift correction for the latter neglectible.\

The energy calibration has been obtained by fitting the highest peaks of the $^{226}$Ra spectra and making a linear regression fit to all of them with the nominal values of the energies taken from~\cite{CALIBTABLE}. These energy calibrations obtained for the different detectors were applied to their corresponding total singles spectra and were also used to sort the data from the coincidences to build up the different matrices.\\

\section{Efficiency }

The intensities of the gamma rays of interest can be extracted from the area of each peak, fitted with a gaussian and after background subtraction, corrected for the detector efficiency. In order to obtain the intensity, we need to apply the relation:\\

\begin{equation}
Intensity = \frac{Area}{Efficiency}
\end{equation}\\

Consequently, we have to determine the efficiency curve. This is done by integrating the peaks of the known $^{226}$Ra and $^{133}$Ba decay sources, located at the target position, and dividing the areas by the known absolute intensity of the source as given in~\cite{CALIBTABLE} and in the specifications of the two sources. We should point out that in the present experiment we calculated absolute efficiencies.\\

In order to get the efficiency calibration we proceeded in a similar way as for the energy calibration. For each detector we shifted by a linear gain-shift correction all the runs to the first and added them. Then we fitted the highest photopeaks with gaussian functions with a tail on the left side after the background subtraction to obtain the areas. Thus, the efficiencies were calculated since the intensities of the calibration sources, when they were produced, are known.\

\begin{figure}[h!]
\begin{center}
\includegraphics[width=14.cm,height=9.2cm,angle=0.]{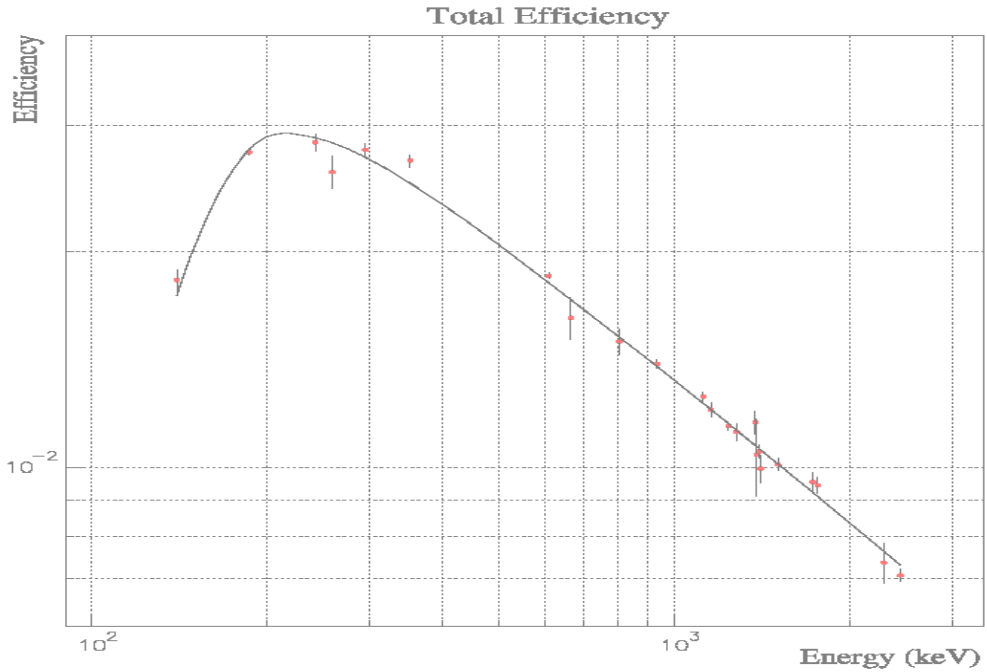}
\end{center}
\caption{Absolute efficiency of the detection system. The experimental points with the error bars and the fit to the points with the J$\ddot{a}$ckel function as a continuous curve are shown.}
\label{fig:totaleff}
\end{figure}

After efficiency values were calculated for the highest photopeaks of both sources, $^{226}$Ra and $^{133}$Ba (see Figure~[\ref{fig:totaleff}]), we made a fit with the function proposed by J$\ddot{a}$ckel et al.~\cite{JACK87} for the case of germanium detectors:

\begin{equation}
\ln \epsilon (E_{\gamma}) + 25 = (b_{1} + b_{2}x + b_{3}x^{2}) \frac{2}{\pi} \arctan (\exp(b_{4} + b_{5}x + b_{6}x^{2}))
\end{equation}

\begin{equation}
\textrm{where } x = \ln E_{\gamma}
\end{equation}\\

 Thus, we obtained 10 individual efficiency curves that are of crucial importance for the $\gamma$-ray intensities and angular distribution determinations. The germanium detector used to determine the intensities was the detector ``Ge 4'' placed at 135 degrees with respect to the beam direction, which was the detector with the best energy resolution of those closest to 126 degrees where the transition intensities can be compared unaffected by their angular distribution (see explanation in section 2.5.). For the use of the Cluster as a polarimeter, two efficiency curves have been determined for the scatterer-analyzer pairs at 30 and 90 degrees, respectively, using the $^{226}$Ra and the $^{133}$Ba sources and following the same procedure used for sorting the $\gamma$-$\gamma$ coincidences.\

 Differences in efficiencies between these two scatterer-analyzer pairs are mainly due to the number of detectors that form them ( the ``30 degrees'' data include twice the number of pairs as the ``90 degrees'') since the intrinsic efficiencies of the Cluster capsules are similar and the solid angle they covered was the same.\\

\section{Electronics and sorts }

\begin{figure}[h!]
\begin{center}
\includegraphics[width=15.cm,height=9.cm,angle=0.]{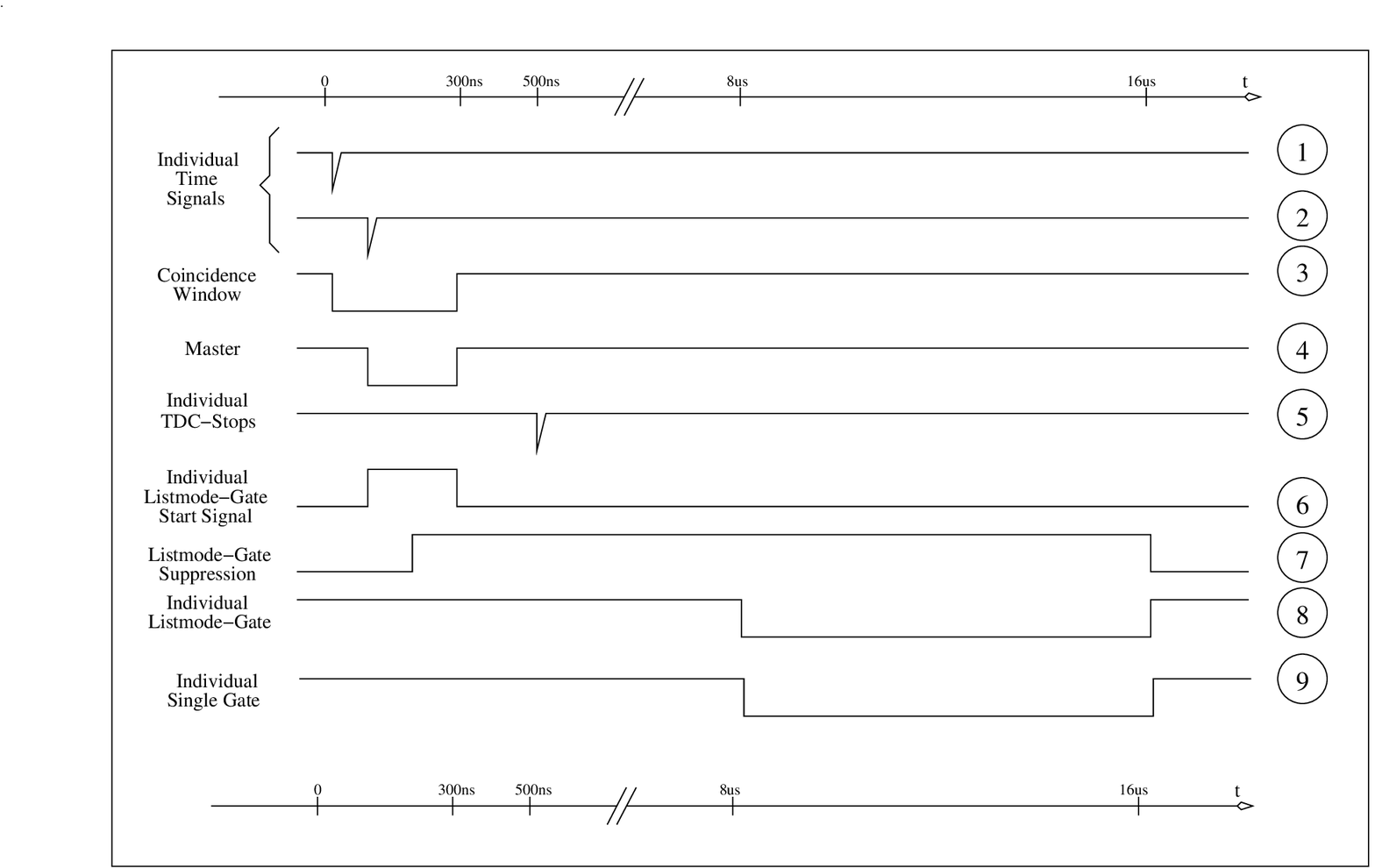}
\end{center}
\caption{Visual signal scheme used in the experiment.}
\label{fig:time_scheme}
\end{figure}

The electronics and data acquisition system used in this experiment were developed at the University of Cologne and allow us to acquire data in singles and coincidence mode simultaneously. In the first case, all signals registered by a detector are continuously stored without any restriction, but in the $\gamma$-$\gamma$ coincidence spectra, only signals produced with a time difference lower than a predetermined time window are accepted and stored in list-mode. In our experiment, this {\it coincidence window } was 300 ns, so we accepted all the $\gamma$-ray coincidences that occurred with a maximum difference in time of 300 ns (see Figure~[\ref{fig:time_scheme}]). The data adquisition system was FERA based. The FERA controller was developed and optimized at the University of Cologne to reduce the dead time of the system~\cite{HARALD}.\\\\

\begin{figure}[h!]
\begin{center}
\includegraphics[width=15.cm,height=9.cm,angle=0.]{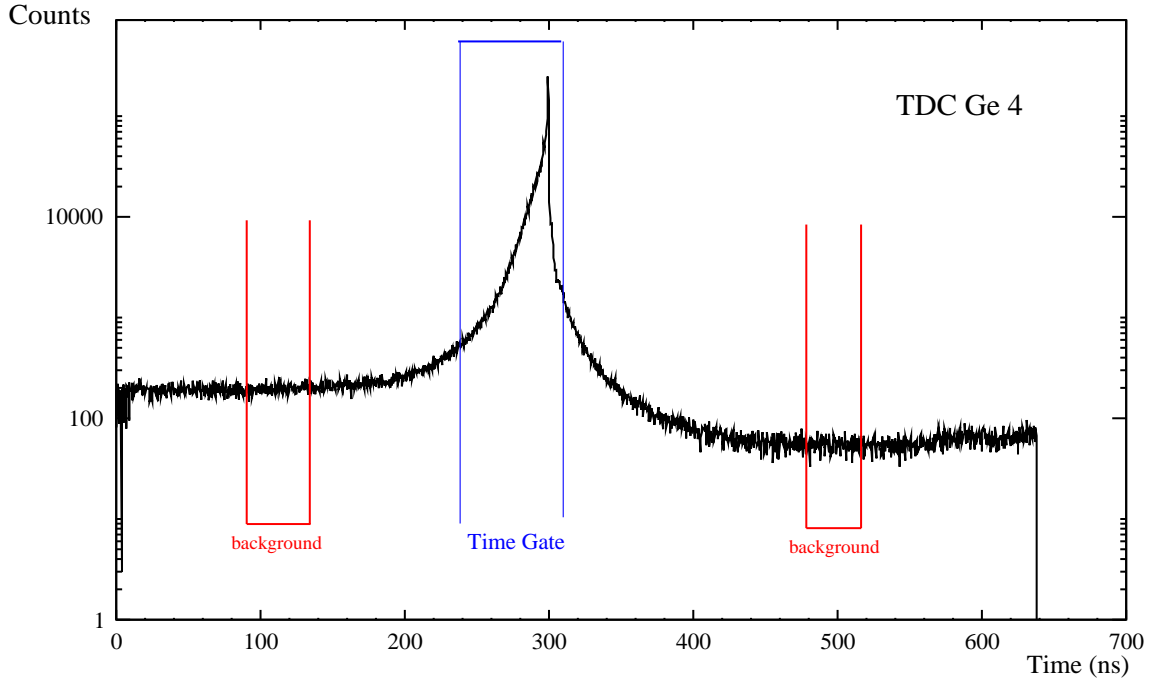}
\end{center}
\caption{TDC spectra of detector ``Ge 4''. In this figure a typical example of the time window selection for sorting the data to obtain the $\gamma$-$\gamma$ coincidences matrices is also shown.}
\label{fig:tdc10}
\end{figure}

\begin{figure}[h]
\begin{center}
\includegraphics[width=15.cm,height=17.5cm,angle=0.]{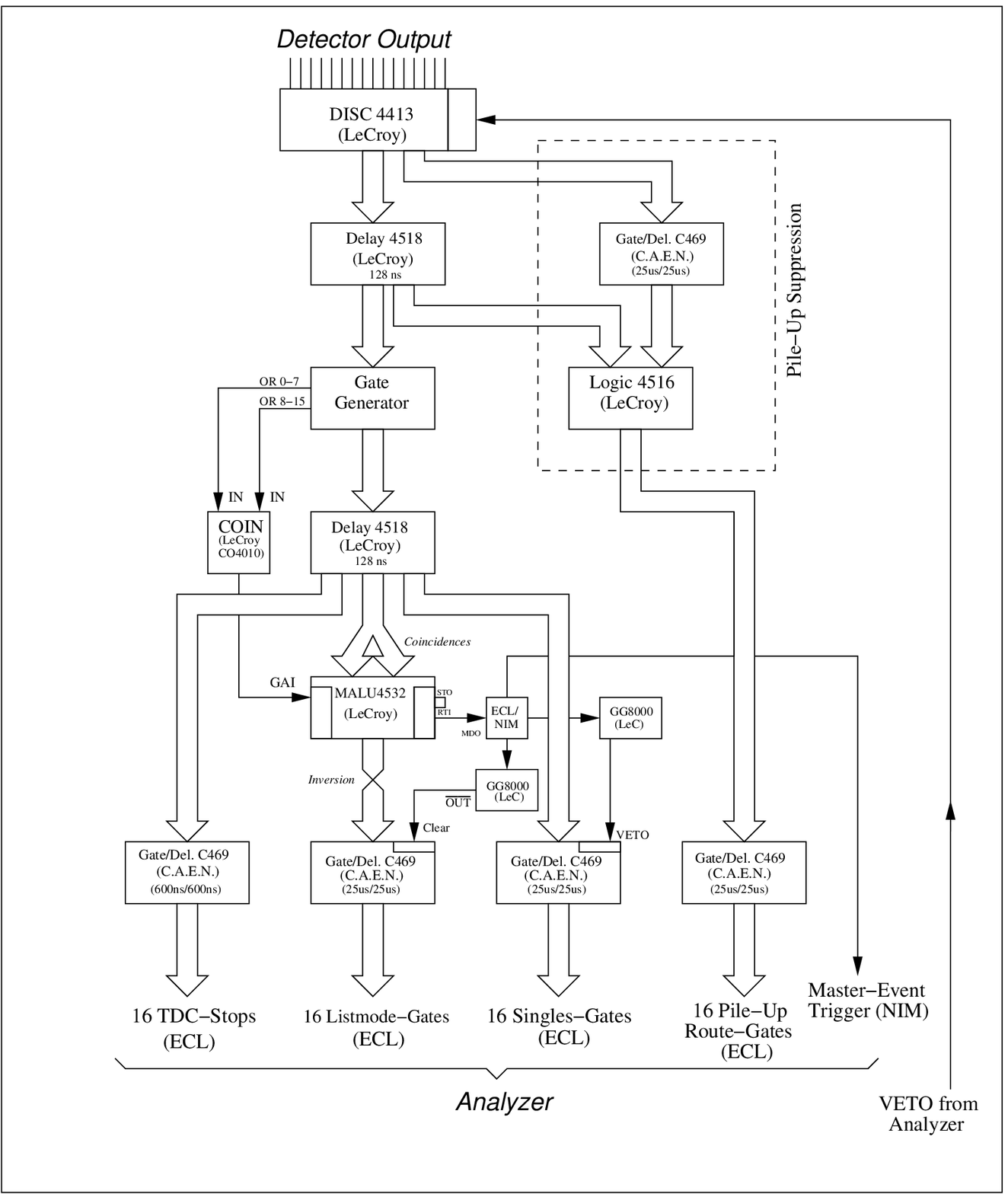}
\end{center}
\caption{Electronics scheme used in the experiment.}
\label{fig:electronics}
\end{figure}

The system had a slow analogue branch and a fast timing circuit for each detector. Every energy signal was amplified through a spectroscopic amplifier (ORTEC 671 and 572 modules). A timing filter amplifier (ORTEC TFA 474 module) and a constant fraction discriminator (ORTEC CDF 584 and 473A modules) were used to produce the individual timing signals. The logic of the fast timing circuit is given in Figure~[\ref{fig:electronics}], while the delays and gates involved can be better seen in Figure~[\ref{fig:time_scheme}]. Every time a coincidence of two germanium detectors (time signals $\textcircled{1}$ and $\textcircled{2}$) occurred within the 300 ns coincidence window $\textcircled{3}$, a master signal associated with the second detector was generated. This master signal generates, after 8 $\mu$s of delay, the gate $\textcircled{8}$ for the ADCs which will then start to convert the energy signals (individual list-mode gate), and starts $\textcircled{6}$, the fast TDC modules (individual list-mode gate start signal). The individual TDC-Stop signal $\textcircled{5}$ is associated with the start detector after a delay of 500 ns. Finally, the time information is given as 16 TDC spectra, each associated with one detector, where the time of this detector is recorded every time a master signal has been created and this detector has fired. In Figure~[\ref{fig:tdc10}] the TDC projection of ``Ge 4'' is given. The sharp peak at $\sim$300 ns corresponds to the case when the master signal was given by detector ``Ge 4''.\\\\\\\\\\

\

The counting rates throughout the experiment were between 7 kHz and 12 kHz for each individual detector; meanwhile for the master trigger ($\gamma$-$\gamma$ coincidences), it was $\sim$11 kHz.\

The energy signals from all the detectors were sent into either 8k or 16k ADCs and the time signals to 2k TDCs. The electronic scheme is shown in Figure~[\ref{fig:electronics}].\

A minimum multiplicity of 2 was required to validate the $\gamma$-$\gamma$ coincidences. Using these $\gamma$-$\gamma$ coincidence events, we constructed a total matrix of 8k$\otimes$8k channels with all the coincidences between detectors. We built also 4k$\otimes$4k matrices with all possible combinations of groups of detectors depending on the angles at which they were placed. All these matrices can be added in such a way that one obtains matrices with all the detectors against an angle (see Table~[\ref{tab:tabla_matrices}] where all the matrices are shown). The two matrices we obtained were used in order to extract angular distributions for transitions not clean enough in the singles spectra whose information has to be extracted from gating in gammas on coincidence with them. They are also very valuable for identifying transitions that present a Doppler shift, since one can observe and compare the spectra at different angles.\

The matrices have been built up by sorting the list-mode data recorded using prompt and delayed gating conditions, i.e., we made a prompt matrix that included all coincidence events inside a time window of $\sim$ 50 ns (see Figure~[\ref{fig:tdc10}]), and a delayed matrix which included the ``random'' events in a time window with the same width as the prompt matrix. Once we have these two matrices, we subtract the delayed matrix from the prompt matrix and we obtain the matrices listed in the table. A graphical explanation is provided in Figure~[\ref{fig:tdc10}].\\

\begin{table}[thbp!]
\begin{center}
\begin{tabular}{|c|}
\hline
Total $\gamma$-$\gamma$ coincidence matrix (8k$\otimes$8k)\\
\hline
\\
(all detectors)$\otimes$(all detectors)\\
\\
\hline
\hline
Angular correlation matrices (4k$\otimes$4k)\\
\hline
\\
(90 degrees)$\otimes$(90 degrees)\\
(90 degrees)$\otimes$(45 degrees)\\
(90 degrees)$\otimes$(35 degrees)\\
(45 degrees)$\otimes$(45 degrees)\\
(45 degrees)$\otimes$(35 degrees)\\
(35 degrees)$\otimes$(35 degrees)\\
\\
\hline
\hline
Angular distribution coincidence matrices (4k$\otimes$4k)\\
\hline
\\
(all detectors)$\otimes$(90 degrees)\\
(all detectors)$\otimes$(40 degrees)\\
\\
\hline
\end{tabular}
\end{center}
\caption{All $\gamma$-$\gamma$ coincidence matrices constructed in the data sort. The last two were used for the directional angular distribution analysis.}
\label{tab:tabla_matrices}
\end{table}

\section{Directional angular distributions }

The level scheme can be constructed from the $\gamma$-$\gamma$ coincidence matrix, which gives us information about the energy of the levels and how they are fed and de-excite, but the spin and parities are still unknown. This is the reason why a directional angular distribution measurement becomes crucial.\

The way to assign spins and parities to the levels of the nucleus is to know the character (electric or magnetic) and multipolarity of the transitions connecting the levels. If one makes use of conservation of parity, measures the character and multipolarity of the transition to a level, and also knows the spin and parity of the final level, then the spin and parity of the initial level can be often established:

\begin{equation}
\pi_{i}·\pi_{f}= \Delta\pi_{\gamma}
\end{equation}\

An easier way is to make two complementary measurements: a linear polarization measurement of the $\gamma$ radiation and a directional angular distribution measurement. From both data one can often determine the multipolarity and nature of the transition. The former will be presented more in detail in the next section while the later will be explained in the next lines.\

Nuclei in excited states formed in nuclear reactions are in general oriented with respect to the beam direction. The degree of orientation depends on the formation process and, therefore, is subject to the reaction mechanism. In general, the angular momentum {\it j} of a state has 2{\it j}+1 components {\it m} ({\it m}=-{\it j},...,{\it j}) along a quantization axis. In an experiment we find a number of substates {\it m} with respect to a suitable symmetry axis as the quantization axis (the direction of the projectiles in nuclear reactions), that are characterized from the statistical point of view by the population parameter {\it P(m)}.\\

Let us now consider the angular momentum conservation in a fusion-evaporation process like our $^{144}$Sm($\alpha$,2n)$^{146}$Gd reaction. In such a reaction when we deal with a heavy target nucleus, as in the case of $^{144}$Sm, the angular momentum transferred in the reaction to the compound nucleus must be dissipated by the emission of neutrons and the succeeding $\gamma$-ray cascade. With an impinging $\alpha$-particle energy of 26.3 MeV, the energy of the first evaporated neutron is $\sim$ 1.1 MeV and $\sim$ 1.0 MeV for the second. These low-energy neutrons are quite inefficient in taking away angular momentum. Therefore, the total angular momentum dissipated by neutron emission is not expected ever to come close to the average angular momentum brought in by the incident alpha particle.\

After the emission of the last neutron, the angular momentum as well as the excitation energy must be dissipated by gamma-rays and internal conversion electrons. Since the gamma de-excitation probability goes as E$^{(2\lambda + 1)}$, where $\lambda$ is the multipolarity of the transition and lower multipolarities are always faster than higher multipolarities, the $\gamma$-ray de-excitation tends to proceed through $\gamma$-rays of high energy and low multipolarity. Consequently, the first gamma de-excitation reaches either the yrast line or states not far away from it. The {\it yrast line} is formed by the states with lowest energy for each spin.\\

In our case, an even-even target with spin 0 is bombarded with $\alpha$-particles. The large angular momentum transferred to the compound nucleus acts only in the direction perpendicular to the beam. In other words, the compound nucleus state is completely  aligned to the beam direction, and the population parameters are simply

\begin{equation}
{\it P(m)} \left\{\begin{array}{ll}
1 & \textrm{for {\it m}=0}\\
0 & \textrm{for {\it m}$\neq$0}
\end{array}\right.
\end{equation}\\

This aligned compound state emits first neutrons and later $\gamma$-rays of high energy until reaching the low-lying states in the final nucleus. These first cascades are very fragmented in intensity, and it is only when reaching relatively low excitation energy and, therefore, a region of low density of states, that the subsequent gamma emission reaches the energy regime of interest. After reaching these states, the original orientation formed in the collision is retained to a considerable extent if the angular momenta transferred by the projectile are large, since the angular momenta carried off by neutrons and early $\gamma$-rays are too small to induce a serious change of orientation. Furthermore, neutrons and $\gamma$-rays tend to feed levels of decreasing spin because of the higher density of lower-spin states in general. Such stretch-type emission retains the orientation to the maximum extent. The directional angular distribution of gamma radiation emitted from an axially symmetric oriented source is\

\begin{equation}
{\it W}({\it \theta}) = \frac{{\it d}\Omega}{4\pi} \sum_{{\it \lambda=even}} {\it B_{\lambda}}({\it I_{i}) A_{\lambda}}{\it P}_{\it \lambda}(cos {\it \theta}) = \frac{{\it d}\Omega}{4\pi} \sum_{{\it \lambda=even}}{\it a_{\lambda}}{\it P}_{\it \lambda}(\cos {\it \theta})
\end{equation}\\

where {\it P}$_{\it \lambda}$(cos {\it $\theta$}) are the Legendre polynomials and the coefficients {\it a$_{\lambda}$} depend on the nuclear orientation degree, on the spins of the levels between which the transition occurs, and on the multipole order of the transition, but not on its electric or magnetic character.\  

Therefore, one can obtain an angular distribution function by measuring the intensities of a transition at different angles with respect to the beam direction and determine its multipolarity. As was described in section 2.1.3, we had detectors placed at five different angles (35, 45, 90, 135, 145 degrees) that, due to the symmetry of the angular distribution functions around 90 degrees, could be considered as three different angles at 35, 45 and 90 degrees. The intensities of the gammas are obtained from the areas of the photopeaks in the histogrammed spectra, after correction for the detector efficiency. In order to extract the area of the photopeak, a fit is done with a Gaussian function with an exponential tail on the left of the photopeak. We have followed this procedure for the intense photopeaks in the spectra obtained at the different angles, if the lines were not obscured by contaminating photopeaks. But for the cases with poor intensities or where contaminants are present, it became necessary to build up coincidence matrices. After examining the data, evaluating the statistics of our experiment, and considering that the real angular difference (taking into account the solid angles subtended by the detectors) between 35 and 45 degrees was small, we added both of these angles matrices and treated the result as a group of detectors at an effective angle of 40 degrees. The matrices constructed for extracting the angular information are listed in Table~[\ref{tab:tabla_matrices}].\\

The procedure for extracting the angular information from these matrices is to put a gate on a transition in coincidence with the one we want to determine its angular distribution in the {\it all detectors} projection and obtain its intensity, corrected by its corresponding efficiency, in the 90 and 40 degrees projections. It must be noticed that the gating conditions (region and background subtractions) should be the same at both angles to compare the intensities.\
  
The ideal procedure is to fit the directional angular distribution equation (until second order) with the intensities obtained at different angles

\begin{equation}
{\it W}({\it \theta}) = 1 + {\it A}_{2}^{\it max}{\it P}_{2}(\cos {\it \theta}) + {\it A}_{4}^{\it max}{\it P}_{4}(\cos {\it \theta})
\end{equation}\

where {\it A}$_{n}^{\it max}$ are the angular distribution coefficients for completely aligned nuclei. This is not what happens in reality, where the nuclei are partially aligned and the function may be replaced by

\begin{equation}
{\it W}({\it \theta}) = 1 + {\it A}_{2}{\it P}_{2}(\cos {\it \theta}) + {\it A}_{4}{\it P}_{4}(\cos {\it \theta})
\end{equation}\

where{\it A}$_{n}$= $\alpha_{n}${\it A}$_{n}^{\it max}$.\

$\alpha_{n}$'s are the attenuation coefficients, which depend on the angular momentum {\it J} of the transition and the distribution of the nuclear state over its {\it m} substates. The attenuation coefficients can be extracted from tables of Mateosian and Sunyar~\cite{MATSUN}. From Yamazaki's work ~\cite{YAMAZAKI}, the partial alignment may be represented by a Gaussian distribution of {\it m}-states characterized by a parameter $\sigma$, which is the half-width of the assumed Gaussian distribution. With this assumption $\alpha_{n}$ can be expressed as a function of {\it J} and $\sigma$/{\it J}. To obtain the $\sigma$/{\it J} value for our experiment we can look for a clear case and compare the {\it A}$_{n}^{\it max}$ value obtained from the tables with the experimental value.\

But we can rewrite the previous expression in terms of measured intensities

\begin{equation}
{\it I}_{\gamma}({\it \theta}) = {\it I}_{\gamma} [1 + {\it a}_{2}{\it \mu}_{2}{\it P}_{2}(\cos {\it \theta}) + {\it a}_{4}{\it \mu}_{4}{\it P}_{4}(\cos {\it \theta})]
\end{equation}\

where {\it a}$_{n}$ are the angular distribution coefficients and {\it $\mu$}$_{n}$ are the geometrical attenuation coefficients. It should be remarked that the geometrical attenuation coefficient is independent of the attenuation coefficient due to the partial alignment of the nuclei. These geometrical attenuation coefficients are introduced because of the size of the detectors (the real solid angle subtended by the detector attenuates effectively the measured angular distribution). A complete study of these geometrical attenuation coefficients for axial and planar detectors at different distances detector-target could be found in ~\cite{CAM69}. From that work, one can see that the attenuation for coaxial detectors can be taken as 1 for distances to the target larger than 10 cm. Consequently, we assumed this value to be one. As we mentioned in section 2.3, the intensities of transitions with different multipolarities can be compared at 126$^{o}$ since at that angle the Legendre polynomial P$_{2}$(cos$\theta$) is zero and thus, I$_{\gamma}$($\theta$=126$^{o}$)$\simeq$I$_{\gamma}$. For this reason, we have chosen the detector ``Ge 4'' for determining the intensities (it had the better energy resolution and the closest angle to 126$^{o}$).\

 We should mention here that, since we are constrained to only two angles (40 and 90 degrees) for determine angular distributions, we can extract only a value for {\it a}$_{2}$ and not for {\it a}$_{4}$. However, the most important information is contained in the sign of {\it a}$_{2}$ which, when combined with the polarization information, will tell us about the character and multipolarity of the transition as can be seen in Table ~[\ref{tab:tabla_signos}]. This sign could be easily obtained just by calculating the ratio between the intensities at 40 and 90 degrees.

\begin{equation}
{\it a}_{2} \left\{\begin{array}{ll}
>0 & \textrm {  for $\frac {W(40)}{W(90)}>1$}\\\\
<0 & \textrm {  for $\frac {W(40)}{W(90)}<1$}
\end{array}\right.
\end{equation}\\\\

\begin{table}[thbp!]
\begin{center}
\begin{tabular}{|c|c|c|}
\hline
Nature and&Angular&Polarization\\
Multipolarity&distribution sign&sign\\
\hline
\hline
{\it E}1 ({\it J}$\rightarrow${\it J}+1)&\large{-}&\small{+}\\
\hline
{\it M}1 ({\it J}$\rightarrow${\it J}+1)&\large{-}&\large{-}\\
\hline
{\it E}1 ({\it J}$\rightarrow${\it J}-1)&\large{-}&\small{+}\\
\hline
{\it M}1 ({\it J}$\rightarrow${\it J}-1)&\large{-}&\large{-}\\
\hline
{\it E}1 ({\it J}$\rightarrow${\it J})&\small{+}&\large{-}\\
\hline
{\it M}1 ({\it J}$\rightarrow${\it J})&\small{+}&\small{+}\\
\hline
{\it M}1/{\it E}2&{\it f}($\delta$)&{\it f}($\delta$)\\
\hline
{\it E}2&\small{+}&\small{+}\\
\hline
{\it M}2&\small{+}&\large{-}\\
\hline
{\it M}2/{\it E}3&{\it f}($\delta$)&{\it f}($\delta$)\\
\hline
{\it E}3&\small{+}&\small{+}\\
\hline
\end{tabular}
\end{center}
\caption{Angular distribution and polarization signs corresponding to the most common transition multipolarities. Mixed transition signs depend on the ratio of the amplitudes of the mixed-multipole transition ($\delta$).}
\label{tab:tabla_signos}
\end{table}

\section{Directional linear polarization}

\begin{figure}[h]
\begin{center}
\includegraphics[width=11.cm,height=15.cm,angle=270.]{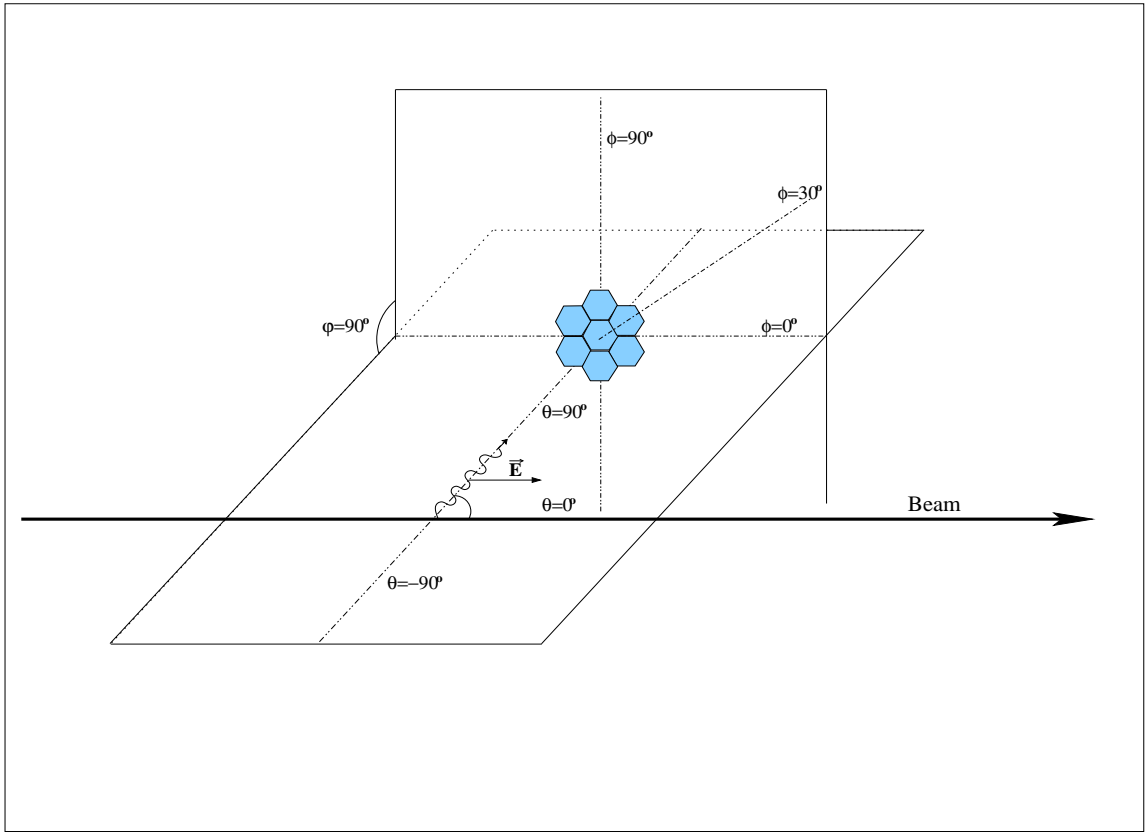}
\end{center}
\caption{The figure shows the polarimeter working principle. A $\gamma$-ray is emitted at $\theta$=90$^{o}$ with respect to the beam direction. When the $\gamma$-ray is Compton scattered with an angle $\varphi$=90$^{o}$ respect to its initial direction, the most probable direction of emission of the scattered $\gamma$-ray corresponds to the plane defined by both $\gamma$-rays and perpendicular to the electric field of the incident $\gamma$-ray ($\phi$=90$^{o}$). The minimum probability corresponds to $\phi$=0$^{o}$. In the figure, the electric field is parallel to the beam direction as in the case of a pure {\it E}1 transition with {\it m}=0.}
\label{fig:ploplane}
\end{figure}

As described in section 2.1.3, a CLUSTER detector was placed at 90 degrees with respect to the beam direction. This placement permitted us to use this detector as a Compton polarimeter. Compton polarimeters are used in $\gamma$-ray spectroscopy to determine the degree of linear polarization of photons emitted in the de-excitation of nuclear states whose spins are oriented with respect to a given direction. In other words, measuring the linear polarization (the direction of the electric field vector) with respect to the beam-detector plane (see Figure~[\ref{fig:ploplane}]) permits a distinction between the electric and magnetic multipoles. This information, as we have seen in the previous section, cannot be extracted from the angular distribution of the $\gamma$-rays.\\

\begin{figure}[h]
\begin{center}
\includegraphics[width=13.cm,height=9.cm,angle=0.]{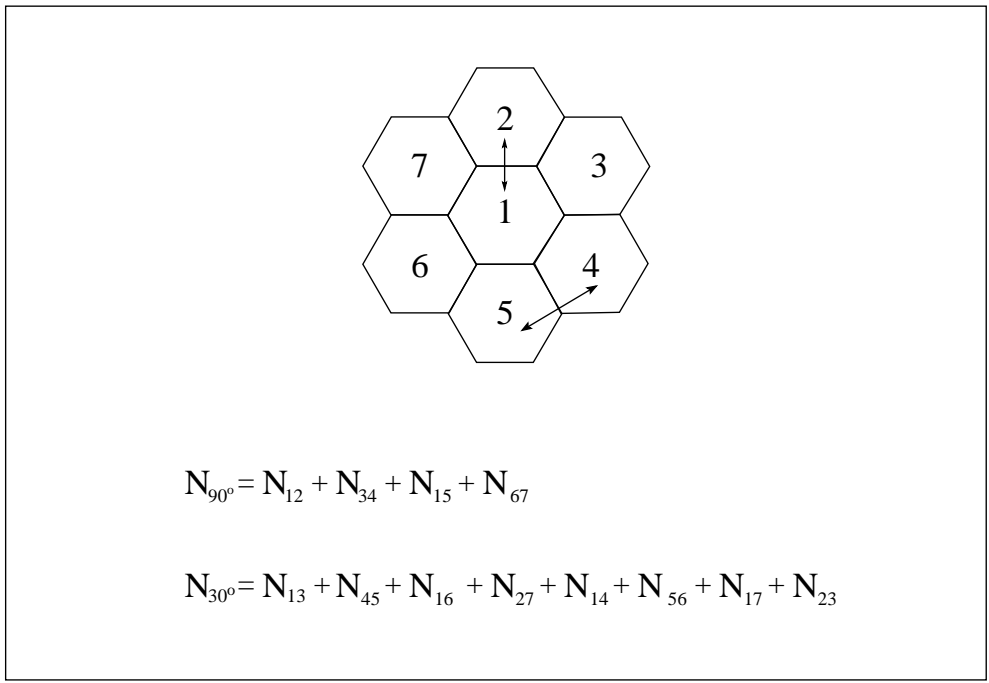}
\end{center}
\caption{Schematic front view of the CLUSTER detector showing the two different combinations of scatterer-analyzer pairs used for the polarization measurement. Note that 30$^{o}$ and 150$^{o}$ are equivalent. N$_{\it ij}$ represents the coincidence counting rate of capsules {\it i} and {\it j}.}
\label{fig:polarization}
\end{figure}

Compton polarimeters are based on the asymmetry of the Compton dispersion probability for linearly polarized photons. This dispersion probability has a maximum for directions perpendiculars to the polarization. Thus, a usual measurement is to compare coincidence counting rates of two detectors (one used as the scatterer and the second as the analyzer) between perpendicular and parallel directions to the polarization reference plane. This can be done either by having a pair of detectors and rotating one of them or by having detectors at both directions during the experiment. In many cases, including the present experiment, it is difficult to move or to place detectors in such a way. Our CLUSTER detector consists of seven closely-packed individually-canned Ge crystals of tapered hexagonal shape. This detector can be used as Compton polarimeter because we can combine the CLUSTER detector capsules and make many scatterer-analyzer pairs. For instance, we can use the central capsule as scatterer and the rest of surrounding capsules as analyzers. In Figure~[\ref{fig:polarization}] we can see all the different scatterer-analyzer pairs that can be made within a CLUSTER detector and how they can be grouped in two polarization direction angles: 30 and 90 degrees. We have four scatterer-analyzer pairs at 90 degrees and eight at 30. The use of this kind of detectors as Compton polarimeter has been studied by Garcia-Raffi et al.~\cite{RAFFI95}.\\

\

As explained in the previous section, in our reaction there is an alignment of the spins in a plane perpendicular to the beam direction. The non-isotropic $\gamma$-ray distribution is not the only consequence of this, as the emitted $\gamma$-rays are linearly polarized. The linear polarization angular distribution of gamma radiation emitted from an axially symmetric oriented source~\cite{STE75} is \

\begin{equation}
{\it W}({\it \theta},{\it \Phi}) = \frac{{\it d}\Omega}{8\pi} \sum_{{\it \lambda=even}} {\it B_{\lambda}}({\it I_{i})\times \bigg[A_{\lambda}}{\it P}_{\it \lambda}(\cos {\it \theta}) + 2A_{\lambda ,2} \bigg( \frac{(\lambda - 2)!}{(\lambda + 2)!}\bigg)^{1/2}{\it P}_{\it \lambda}^{(2)}(\cos {\it \theta})\cos ({2 \it \Phi})\bigg]
\label{eq:ang1}
\end{equation}\

where ${\it \theta}$ is the angle between the photon and the beam, ${\it \Phi}$
the angle between the plane defined by the incident beam and the photon emission direction and the polarimeter axis, ${\it P}_{\it \lambda}^{(2)}$ are the generalized second-order Legrendre polynomials, ${\it B_{\lambda}}({\it I_{i}})$ are the orientation parameters for the initial spin ${\it I_{i}}$, and $A_{\lambda}, A_{\lambda ,2}$ are the angular distribution coefficients.\

With a polarimeter we are sensitive to the degree of linear polarization, thus we define a ${\it \theta}$ angle and measure the Compton scattering to different ${\it \Phi}$ angles. The ${\it \theta}$ angle typically is 90 degrees since the degree of polarization is maximum at that angle. This is almost our case, since the central capsule of the CLUSTER is placed at ${\it \theta}$=90$^{o}$ and the external capsules are at ${\it \theta}$=68$^{o}$ and ${\it \theta}$=112$^{o}$. Also, the ${\it \Phi}$ typical angles are 0 and 90 degrees but with the CLUSTER, as its seen in Figure~[\ref{fig:polarization}], the ${\it \Phi}$ angles are 30 and 90 degrees.\

The excited oriented nuclei emit $\gamma$ radiation with the electric vector (direction of polarization) either parallel or perpendicular to the reference plane defined by the direction of emission and the direction defining the orientation. The degree of polarization of a radiation is defined by the parallel and perpendicular intensities as\

\begin{equation}
{\it P} = \frac {{\it I}_{\parallel} - {\it I}_{\perp}}{{\it I}_{\parallel} + {\it I}_{\perp}}
\label{eq:pol1}
\end{equation}\\

The linear polarization distribution depends on the parity (electric or magnetic character) of the electromagnetic radiation towards the second member of equation~[\ref{eq:ang1}]. Thus, a polarization measurements gives us information about the parity of the electromagnetic radiation and, therefore, about the parities of the initial and final states.\

As it has been already mentioned, a polarimeter is based on the fact that the Compton dispersion depends on the degree of polarization of the radiation. This probability is given by the Klein-Nishina formulae which, integrated over all polarization directions of the scattered photon, has the form\

\begin{equation}
\bigg( \frac{{\it d}\sigma}{{\it d}\Omega} \bigg)_{e} ({\it \theta},{\it \Phi})= \frac{{\it r}_{0}^{2}}{4} \bigg( \frac {{\it E}'}{{\it E}_{0}} \bigg)^{2} \bigg(\frac{{\it E}_{0}}{{\it E}'} + \frac {{\it E}'}{{\it E}_{0}} - 2\sin^{2}{\it \theta}\cos^{2}{\it \Phi} \bigg)
\end{equation}\\

\begin{equation}
\textrm{where  }   {\it E}' = \frac{{\it E}_{0}}{1+\alpha(1- \cos {\it \theta})} \textrm{ , and } \alpha ={\it E}_{0}/m_{e}c^{2}
\end{equation}

Here, {\it d}$\sigma$ is the Compton scattering cross-section in {\it d}$\Omega$, {\it r}$_{0}$ is the classical electron radius, {\it E}$_{0}$ is the incident photon energy and, $\theta$ and $\Phi$ are the polar angles defined relative to the incident momentum and the plane of polarization, respectively.\
If we take $\theta$=90$^{o}$ and have angles of $\Phi$ and $\Phi$', the coincidence counting rate for each scatterer-analyzer combination, normalized by the efficiencies ($\epsilon$), can be expressed as\

\begin{equation}
\frac{N_{\Phi}}{\epsilon_{\Phi}} = I_{\parallel}\sigma_{\Phi} + I_{\perp}\sigma_{90-\Phi}
\end{equation}

\begin{equation}
\frac{N_{\Phi'}}{\epsilon_{\Phi'}} = I_{\parallel}\sigma_{\Phi'} + I_{\perp}\sigma_{90-\Phi'}
\end{equation}\\

where $\sigma$$_{\Phi}$ $\equiv$ $\frac{{\it d}\sigma}{{\it d}\Omega}({90^{o}},{\it \Phi}$).\\\\

Then we define the asymmetry of the counting rate between both directions as\

\begin{equation}
{\it A} = \frac {N_{\Phi'}/\epsilon_{\Phi'}-N_{\Phi}/\epsilon_{\Phi}}{N_{\Phi'}/\epsilon_{\Phi'}+N_{\Phi}/\epsilon_{\Phi}}
\end{equation}\\

The relation between the asymmetry ({\it A}) and the degree of polarization ({\it P}) can be written as\

\begin{equation}
{\it A} = \frac {{\it QP}}{1+{\it \alpha QP}}
\end{equation}\

\begin{equation}
\textrm{where  }   \alpha = \frac {\sin^{2}\Phi' - \cos^{2}\Phi}{\cos^{2}\Phi - \cos^{2}\Phi'}
\end{equation}\\

and {\it Q} is the polarimeter sensitivity that depends on the energy by the relation\

\begin{equation}
{\it Q} = (\cos^{2}{\Phi}-\cos^{2}{\Phi'}) \frac {{\it a}}{{\it a}^{2}-{\it a}+1}
\end{equation}\

\begin{equation}
\textrm{where } {\it a}=\frac{1}{1+{\it E}_{0}/{\it m}_{e}}
\end{equation}\\

It should be noticed that the absolute value of the asymmetry depends on the sign of the polarization. The polarimeter sensitivity {\it Q} can be determined experimentally from the measured asymmetries for transitions of known {\it P} by the relation\

\begin{equation}
{\it Q} = \frac {{\it A}}{1-{\it \alpha A}} \frac{1}{{\it P}}
\end{equation}

\begin{figure}[h!]
\begin{center}
\includegraphics[width=11.cm,height=16.cm,angle=-90.]{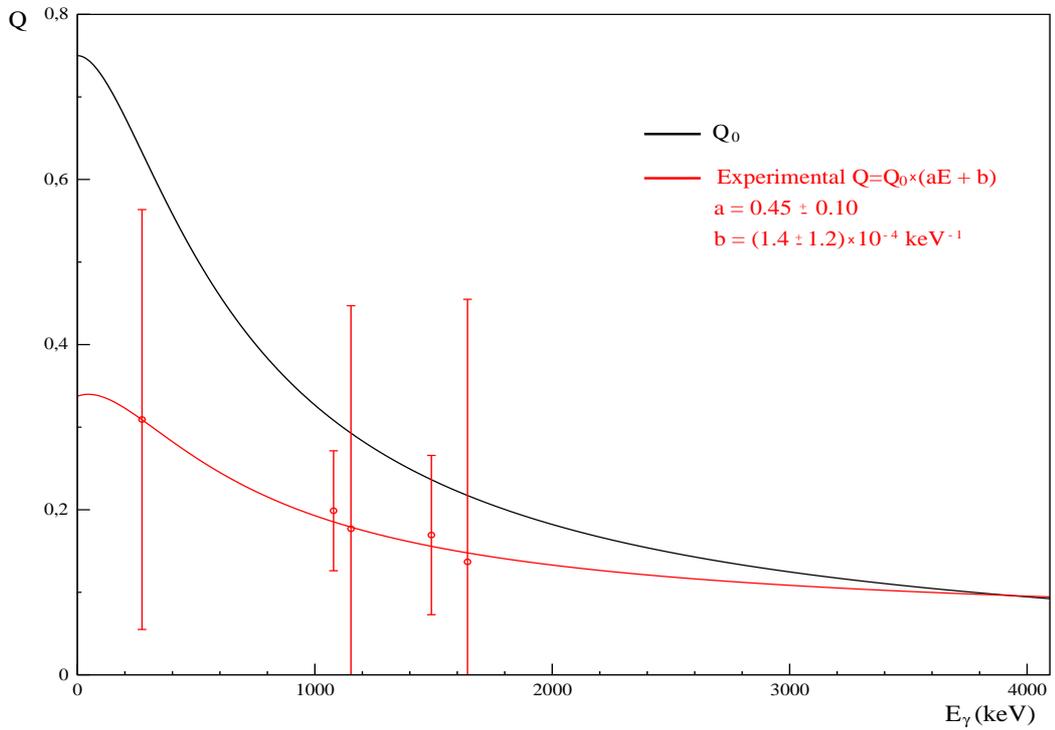}
\caption{Figure shows the theoretical sensitivity Q$_{0}$ for a point-like-detectors polarimeter and the fitted experimental polarimeter sensitivity Q.}
\label{fig:QQo_plot}
\end{center}
\end{figure}

A Compton polarimeter is characterized by its sensitivity which, in the case of an ideal polarimeter composed of point-like detectors, can be calculated with expression~[\ref{eq:pol1}]. But this is not reality, and we have to deal with extended detectors and find the experiment relationship between polarization and sensitivity. Since we have angular distributions of the $\gamma$-rays, we can obtain the distribution coefficients of the most intense $\gamma$-rays and calculate their theoretical polarization. This argument is valid only for pure electric transitions and it is expressed as

\begin{equation}
{\it P} = \pm \frac { \frac{3}{2}{\it a}_{2} + \frac{5}{8}{\it a}_{4} }{ 1 + \frac{1}{2}{\it a}_{2} + \frac{3}{8}{\it a}_{4} }
\end{equation}\

where  the sign + corresponds to pure {\it E}2 transitions and the sign - to pure {\it E}1 transitions. {\it a}$_{2}$ and {\it a}$_{4}$ are the angular distribution coefficients defined in the previous section.\

It is difficult to find clean transitions in $^{146}$Gd that cover the full energy range so we had to use transitions from $^{147}$Gd, which is also produced in our reaction. In this way we obtained different {\it Q} values from asymmetry and polarization ratios. In Figure~[\ref{fig:QQo_plot}] we can see the results and also appreciate that in addition to a reduction in {\it Q} there is a dependence on energy. In order to fit the Q dependence on energy we used the expression 

\begin{equation}
{\it Q} = {\it Q}_{0}({\it aE}+{\it b})
\end{equation}\

,as proposed in~\cite{KIM75} and~\cite{RIK85}, and obtained the {\it a} and {\it b} parameters values. This relation takes into account that we are integrating the Klein-Nishina cross-section over a certain {\it $\theta$} and {\it $\Phi$} interval that depends on energy.\

Once we have this sensitivity function, it is easy to calculate the polarization from the measured asymmetries of the transitions and combine these values with the angular distribution results to determine the spins and parities of the transitions (see Table~[\ref{tab:tabla_signos}]).


\chapter{The level scheme analysis}

{\it  In this chapter the procedures for the analysis of the $\gamma$-ray data and construction of the $^{146}$Gd level scheme will be described. The experimental results will be described and experimental spectra will be shown.}
  
\vspace*{0.6cm}
\section{The $^{146}$Gd level scheme}

  As mentioned in Chapter 1, the latest published work about the $^{146}$Gd level scheme based on in-beam fusion-evaporation experiments was published by Yates et al.~\cite{YATES}. This experiment was very similar to the present one in terms of the reaction used, but the detection efficiency was significantly lower (two 20$\%$ Ge(HP) detectors in close geometry at $\pm$120$^{o}$).  In the present set-up we have repeated this experiment with a modern array of large volume Ge detectors at 90,$\pm$ 45 and $\pm$ 35 degrees. Five of them had anti-Compton shields. The beam energy was similar to the one used in the previous experiment. We expect to confirm the previous results and observe more levels, and thus obtain better level assignments. In addition, a EUROBALL CLUSTER detector was placed at 90$^{o}$ to act as a non-orthogonal $\gamma$-ray Compton polarimeter~\cite{RAFFI95}, as described in Chapter 2.\\

\begin{figure}[!htbp]
\begin{center}
\includegraphics[width=13.8cm,angle=0.]{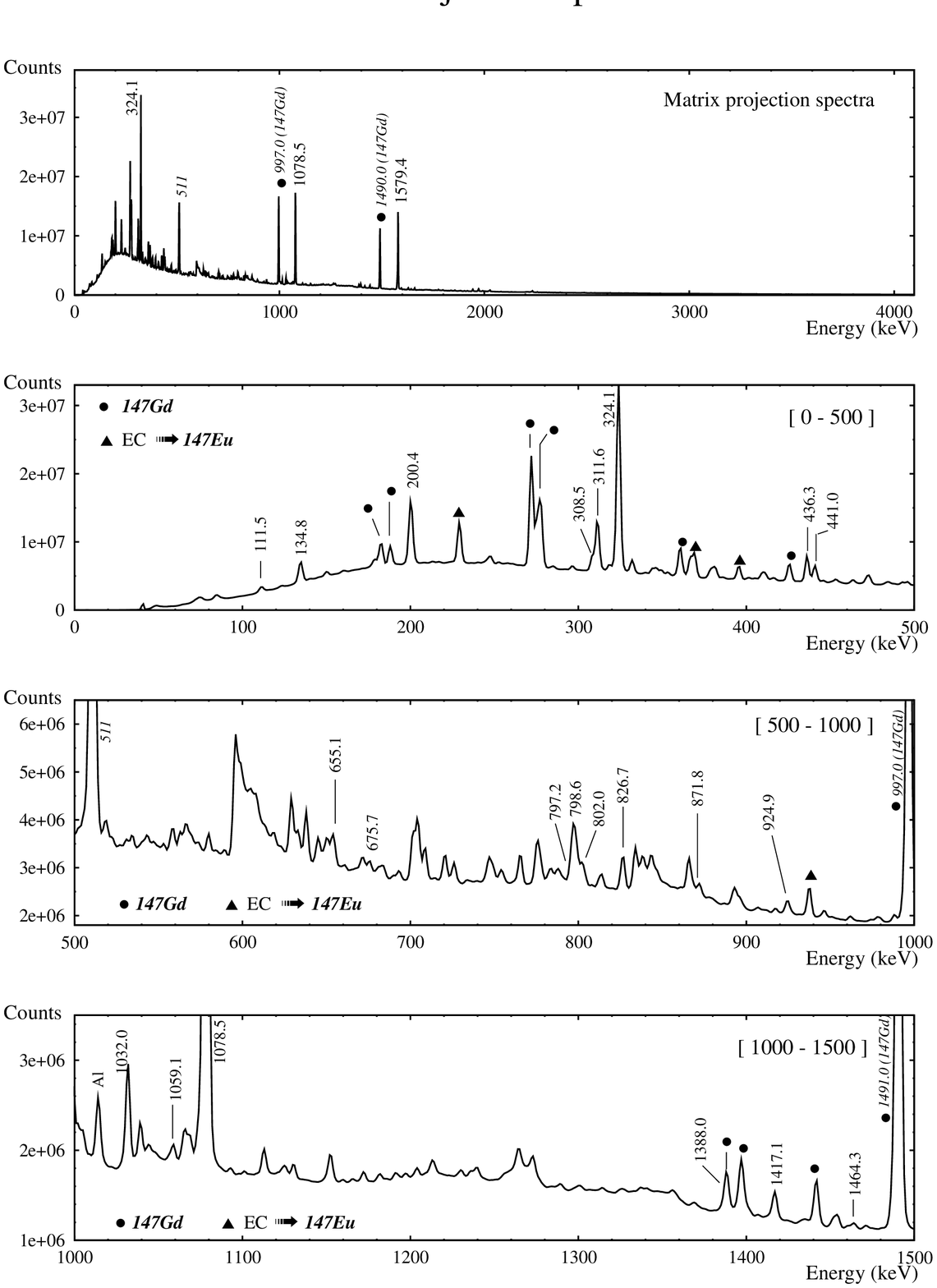}
\end{center}
\caption{The panels show the matrix projection spectrum in ranges of 500 keV. Only the most intense peaks are labeled.}
\label{fig:projection_a}
\end{figure}

\begin{figure}[!htbp]
\begin{center}
\includegraphics[width=13.8cm,angle=0.]{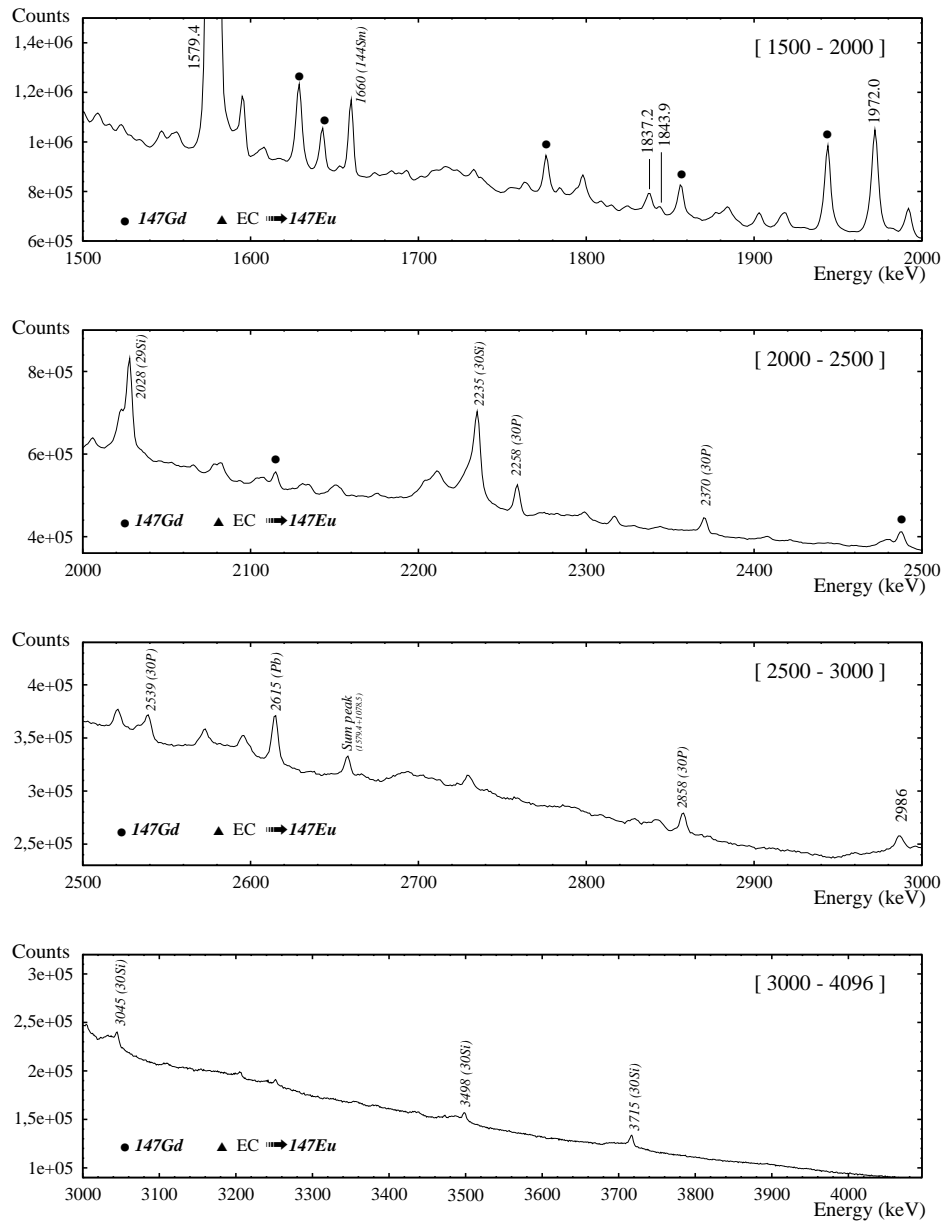}
\end{center}
\caption{The panels show the matrix projection spectrum in ranges of 500 keV (except the last panel). Only the most intense peaks are labeled.}
\label{fig:projection_b}
\end{figure}

  The construction of the level scheme was mainly based on the analysis of the $\gamma$-$\gamma$ coincidence matrices. The construction of these matrices was explained in Chapter 2. A first inspection of the projection (see Figures~[\ref{fig:projection_a}] and~[\ref{fig:projection_b}] in pages 54-55) shows that in the present experiment two primary reaction channels were open: the ($\alpha$,2n) channel populating excited states in $^{146}$Gd and the ($\alpha$,n) channel populating excited states in $^{147}$Gd. The ($\alpha$,p) channel populating excited states in $^{147}$Eu is also open, but the cross section is not as large as the two mentioned above. In addition to these channels, another two appeared due to the $\alpha$-particles impinging in the target frame made of aluminium. These two channels are the ($\alpha$,n) channel populating excited states in $^{30}$P and the ($\alpha$,p) channel populating excited states in $^{30}$Si. Since there is considerable knowledge in literature about these nuclei, it was relatively easy to identify the most intense peaks in the projection and, by putting gates on these transitions, we could easily identify to which nuclei they belonged.\


  All the gates are placed on one of the projections of the coincidence matrix. It does not matter which projection is used since the coincidence matrix is symmetrical and, consequently, the two projections are identical. The way in which the gates have been placed was to select the region of interest for the gate and also select a background on both sides of the peak to subtract from the gate, trying to avoid peaks in the background region. We made use of the angular distribution coincidence matrices to identify peaks that presented Doppler shifts, since one can compare the peaks at 90 degrees, where there is no Doppler effect, with the peaks at $\pm$40 degrees, where the shift appears. The Doppler effect appears when the emitting nucleus is moving.\

  The first step in the analysis was to check that our data confirmed the level scheme known from previous work (~\cite{YATES} and~\cite{TESINA} ). After gating on all previously known gammas, the reported states were confirmed. In this process, many new transitions appeared which were then individually examined placing new gates. This procedure allowed us to place most of the new transitions in $^{146}$Gd, although sometimes intensity arguments were used to decide the gamma de-excitation sequence. The intensities were obtained by integrating the peaks in the singles spectra when possible. For that analysis, a standard fit of the peaks with a Gaussian peak-shape, minimising the $\chi$$^{2}$, was used. We considered a linear background subtraction. This was possible in general for $\gamma$-rays de-exciting yrast levels. For the rest of the peaks the intensities were obtained from the gated spectra. In this case, the intensities were normalized using at least one peak observed in the gated spectrum with known intensity from the singles spectra. In most of the cases, this reference peak was one of the yrast transitions. The integrated values from the fits were corrected by the array or corresponding detector efficiency to obtain the $\gamma$-ray intensity.\\

In a preparatory $^{144}$Sm($\alpha$,2n) experiment~\cite{TESINA}, a total of 21 new $\gamma$-ray transitions from 16 new levels were identified, as well as 19 new $\gamma$-rays corresponding to 13 previously known levels. Also, 7 $\gamma$-rays were seen for the first time in an in-beam experiment.\

In the present work, a total of 35 new $\gamma$-rays from $^{146}$Gd have 
been identified corresponding to 28 new states (together with the tesina this
makes 44 new levels). Also, 31 new $\gamma$-rays, corresponding to 26 previously known levels, were identified; 3 $\gamma$-rays were seen for the first time in an in-beam experiment. In Table~[\ref{tab:tabla_gammas_11}] the results of our analysis are presented by level energy, while in Table~[\ref{tab:tabla_gammas_21}] they are presented ordered by $\gamma$-ray energy. All the energies and intensities are listed together with their uncertainties. The intensity values are related to the 1579.4 keV $\gamma$-ray, which was chosen as the reference peak since it is the most intense peak in the projection and, at the same time, it was not contaminated by any other gamma-ray transition. Angular anisotropy and polarization data were extracted where possible,and the level spins and parities were deduced from the data. For transparency, we have included in the tables comments if the transitions were observed in previous experiments.\

\begin{figure}[h!]
\begin{center}
\includegraphics[width=11.cm,height=13.cm,angle=-90.]{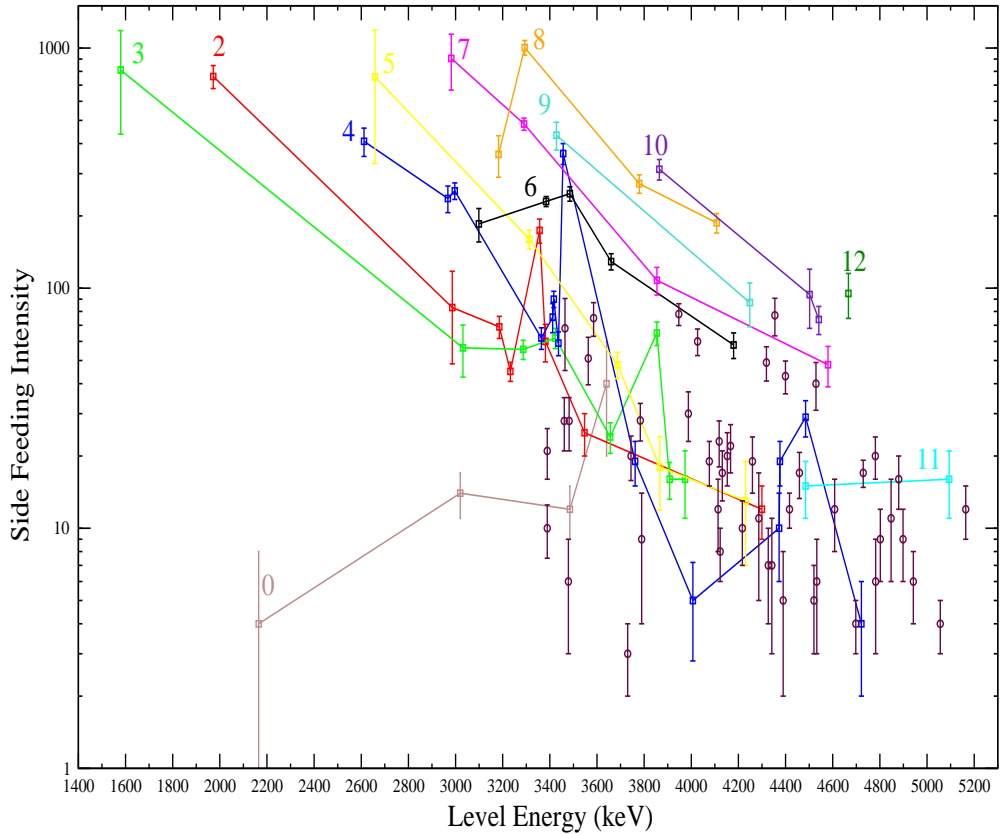}
\end{center}
\caption{Relative level population in $^{146}$Gd observed in the $^{144}$Sm($\alpha$,2n) reaction at {\it E}$_{\alpha}$=26.3 MeV. Squares refer to firmly assigned levels while circles represent levels not firmly assigned. Solid lines connect firmly assigned levels with the same J.}
\label{fig:side_feeding}
\end{figure}

Also, the level population (commonly known as level side-feeding) may help in the level assignment task. While not a strong argument, it is sometimes helpful discarding level assignments. We can see in Figure~[\ref{fig:side_feeding}] that there is a decrease in the level population with increasing excitation energy.\

From the $\gamma$-$\gamma$ coincidence analysis, the analysis of the gamma-ray intensities, and the information on angular distributions (obtained from the 40$^{o}$/90$^{o}$ anisotropy) and on polarizations, we have constructed the level scheme presented at the end of the chapter.\\\\\\\\

\begin{landscape}
\begin{table}[thbp!]
\begin{center}
\begin{tabular}{|c|c|c|c|c|c|c|c|}
\hline
\hline
Level&Transition&Rel. $\gamma$-ray&(40$^{o}$/90$^{o}$)&&&&\\
Energy(keV)&Energy (keV)&Intensity&Anisotropy&\small{Polarization}&\small{Multipolarity}&I${_i}^{\pi}$$\rightarrow$I${_f}^{\pi}$&Comments\\
\hline
\hline
1579.4 (1)&1579.4 (1)&10000&1.63 (10)&0.55 (20)&E3&3$^{-}$$\rightarrow$ 0$^{+}$&\small{a,b,c,d,e,f,g}\\
\hline
1972.0 (1)&1972.0 (1)&1095 (77)&1.25 (12)&&E2&2$^{+}$$\rightarrow$ 0$^{+}$&\small{$\divideontimes$,b,c,d,e,f,g}\\
\hline
2164.7 (1)&192.7 (1)&4 (1)&0.69 (24)&&E2&0$^{+}$$\rightarrow$ 2$^{+}$&\small{c,d,e,f,g}\\
\hline
2611.5 (1)&639.6 (1)&20 (2)&1.65 (23)&&E2&4$^{+}$$\rightarrow$ 2$^{+}$&\small{d,e,f,g}\\
&1032.0 (1)&585 (50)&0.85 (10)&0.35 (22)&E1&4$^{+}$$\rightarrow$ 3$^{-}$&\small{b,d,e,f,g}\\
\hline
2657.9 (1)&1078.5 (1)&7124 (365)&1.47 (11)&0.61 (21)&E2&5$^{-}$$\rightarrow$ 3$^{-}$&\small{a,b,d,e,f,g}\\
\hline
\hline
\end{tabular}
\end{center}
\caption{Transitions in $^{146}$Gd observed in the $^{144}$Sm($\alpha$,2n) reaction at {\it E}$_{\alpha}$=26.3 MeV.}
\label{tab:tabla_gammas_11}
\end{table}
\end{landscape}

\begin{landscape}
\begin{table}[thbp!]
\begin{center}
\begin{tabular}{|c|c|c|c|c|c|c|c|}
\hline
\hline
Level&Transition&Rel. $\gamma$-ray&(40$^{o}$/90$^{o}$)&&&&\\
Energy(keV)&Energy (keV)&Intensity&Anisotropy&\small{Polarization}&\small{Multipolarity}&I${_i}^{\pi}$$\rightarrow$I${_f}^{\pi}$&Comments\\
\hline
\hline
3287.2 (1)&675.7 (1)&46 (5)&0.89 (10)&0.26 (23)&M1&3$^{+}$$\rightarrow$ 4$^{+}$&\small{b,f,g}\\
&1315.2 (2)&9 (1)&0.96 (15)&&M1&3$^{+}$$\rightarrow$ 2$^{+}$&\small{g}\\
\hline
3290.5 (3)&308.5 (3)&448 (28)&1.44 (9)&0.49 (19)&M1&7$^{-}$$\rightarrow$ 7$^{-}$&\small{f,g}\\
\hline
3293.7 (2)&111.5 (1)&430 (30)&0.77 (4)&&M1/E2&8$^{-}$$\rightarrow$ 8$^{-}$&\small{a,f,g}\\
&311.6 (1)&903 (45)&0.63 (6)&-0.36 (13)&M1&8$^{-}$$\rightarrow$ 7$^{-}$&\small{f,g}\\
\hline
3313.0 (2)&655.1 (1)&135 (14)&1.28 (19)&&M1&5$^{-}$$\rightarrow$ 5$^{-}$&\small{b,f,g}\\
\hline
\hline
\end{tabular}
\end{center}
\label{tab:tabla_gammas_1}
\end{table}
\end{landscape}

\begin{landscape}
\begin{table}[thbp!]
\begin{center}
\begin{tabular}{|c|c|c|c|c|c|c|c|}
\hline
\hline
Level&Transition&Rel. $\gamma$-ray&(40$^{o}$/90$^{o}$)&&&&\\
Energy(keV)&Energy (keV)&Intensity&Anisotropy&\small{Polarization}&\small{Multipolarity}&I${_i}^{\pi}$$\rightarrow$I${_f}^{\pi}$&Comments\\
\hline
\hline
3411.8 (2)&380.9 (3)&28 (5)&0.65 (16)&-0.07 (24)&M1&4$^{+}$$\rightarrow$ 3$^{+}$&\small{g}\\
&415.3 (1)&33 (9)&1.24 (48)&-0.61 (48)&E1&4$^{+}$$\rightarrow$ 4$^{-}$&\small{b,f,g}\\
&800.2 (1)&14 (2)&1.23 (18)&&M1&4$^{+}$$\rightarrow$ 4$^{+}$&\small{f,g}\\
\hline
3416.5 (2)&804.9 (2)&6 (2)&1.02 (48)&0.5 (9)&M1&4$^{+}$$\rightarrow$ 4$^{+}$&\small{d,g}\\
&1444.6 (2)&10 (3)&1.24 (53)&0.3 (1.2)&E2&4$^{+}$$\rightarrow$ 2$^{+}$&\small{d,g}\\
&1837.2 (2)&74 (6)&0.63 (7)&&E1&4$^{+}$$\rightarrow$ 3$^{-}$&\small{d,f,g}\\
\hline
\hline
\end{tabular}
\end{center}
\label{tab:tabla_gammas_1}
\end{table}
\end{landscape}

\begin{landscape}
\begin{table}[thbp!]
\begin{center}
\begin{tabular}{|c|c|c|c|c|c|c|c|}
\hline
\hline
Level&Transition&Rel. $\gamma$-ray&(40$^{o}$/90$^{o}$)&&&&\\
Energy(keV)&Energy (keV)&Intensity&Anisotropy&\small{Polarization}&\small{Multipolarity}&I${_i}^{\pi}$$\rightarrow$I${_f}^{\pi}$&Comments\\
\hline
\hline
3547.5 (8)&3547.5 (8)&25 (5)&1.06 (30)&&E2&2$^{+}$$\rightarrow$ 0$^{+}$&\small{e}\\
\hline
3562.8 (2)&951.6 (2)&4 (1)&1.22 (43)&&E1 or E2&4$^{+}$,2$^{+}$$\rightarrow$ 4$^{+}$&\small{}\\
&1591.1 (3)&10 (3)&1.38 (65)&&E1 or E2&4$^{+}$,2$^{+}$$\rightarrow$ 2$^{+}$&\small{g}\\
&1983.1 (3)&37 (11)&0.77 (32)&&E1&4$^{+}$,2$^{+}$$\rightarrow$ 3$^{-}$&\small{g}\\
\hline
3585 (1)&2006 (1)&75 (12)&0.59 (13)&&&4,2 $\rightarrow$ 3$^{-}$&\small{$\divideontimes$,b,g}\\
\hline
3640 (1)&654.6 (6)&40 (20)&&&E2&0$^{+}$$\rightarrow$ 2$^{+}$&\small{e}\\
\hline
\hline
\end{tabular}
\end{center}
\label{tab:tabla_gammas_1}
\end{table}
\end{landscape}

\begin{landscape}
\begin{table}[thbp!]
\begin{center}
\begin{tabular}{|c|c|c|c|c|c|c|c|}
\hline
\hline
Level&Transition&Rel. $\gamma$-ray&(40$^{o}$/90$^{o}$)&&&&\\
Energy(keV)&Energy (keV)&Intensity&Anisotropy&\small{Polarization}&\small{Multipolarity}&I${_i}^{\pi}$$\rightarrow$I${_f}^{\pi}$&Comments\\
\hline
\hline
3854.0 (2)&671.7 (1)&37 (7)&0.79 (21)&&M1&7$^{-}$$\rightarrow$ 8$^{-}$&\small{f,g}\\
&755.2 (2)&18 (3)&0.78 (13)&&M1&7$^{-}$$\rightarrow$ 6$^{-}$&\small{f,g}\\
&871.8 (3)&52 (12)&1.05(34)&0.52 (75)&M1&7$^{-}$$\rightarrow$ 7$^{-}$&\small{f,g}\\
\hline
3864.8 (3)&436.3 (1)&438 (22)&0.63 (5)&0.39 (14)&E1&10$^{+}$$\rightarrow$ 9$^{-}$&\small{a,f,g}\\
\hline
3866.5 (2)&381.7 (3)&15 (6)&0.87 (49)&0.18 (90)&(E1)&(5$^{-}$) $\rightarrow$ 6$^{+}$&\small{}\\
&1255.2 (2)&3 (1)&0.52 (25)&&(E1)&(5$^{-}$) $\rightarrow$ 4$^{+}$&\small{}\\
\hline
\hline
\end{tabular}
\end{center}
\label{tab:tabla_gammas_1}
\end{table}
\end{landscape}

\begin{landscape}
\begin{table}[thbp!]
\begin{center}
\begin{tabular}{|c|c|c|c|c|c|c|c|}
\hline
\hline
Level&Transition&Rel. $\gamma$-ray&(40$^{o}$/90$^{o}$)&&&&\\
Energy(keV)&Energy (keV)&Intensity&Anisotropy&\small{Polarization}&\small{Multipolarity}&I${_i}^{\pi}$$\rightarrow$I${_f}^{\pi}$&Comments\\
\hline
\hline
4122 (2)&1511 (1)&8 (2)&0.77 (27)&&&5$^{-}$,3$^{-}$ $\rightarrow$ 4$^{+}$&\small{$\divideontimes$,e}\\
\hline
4131 (1)&1100 (1)&17 (4)&1.17 (28)&0.21 (60)&&3$^{+}$,5$^{+}$$\rightarrow$ 3$^{+}$&\small{}\\
\hline
4152 (1)&2573 (1)&20 (5)&0.45 (16)&&&2,4 $\rightarrow$ 3$^{-}$&\small{$\divideontimes$,g}\\
\hline
4166.4 (2)&1508.5 (3)&22 (5)&0.29 (10)&&&4,6 $\rightarrow$ 5$^{-}$&\small{g}\\
\hline
4179.4 (2)&1197.3 (2)&30 (5)&0.97 (23)&1.03 (76)&&(6$^{-}$) $\rightarrow$ 7$^{-}$&\small{g}\\
&1521.6 (4)&28 (5)&0.61 (15)&&&(6$^{-}$) $\rightarrow$ 5$^{-}$&\small{}\\
\hline
\hline
\end{tabular}
\end{center}
\label{tab:tabla_gammas_1}
\end{table}
\end{landscape}

\begin{landscape}
\begin{table}[thbp!]
\begin{center}
\begin{tabular}{|c|c|c|c|c|c|c|c|}
\hline
\hline
Level&Transition&Rel. $\gamma$-ray&(40$^{o}$/90$^{o}$)&&&&\\
Energy(keV)&Energy (keV)&Intensity&Anisotropy&\small{Polarization}&\small{Multipolarity}&I${_i}^{\pi}$$\rightarrow$I${_f}^{\pi}$&Comments\\
\hline
\hline
4389.5 (3)&1290.6 (6)&5 (3)&0.69 (49)&&&5,7 $\rightarrow$ 6$^{-}$&\small{}\\
\hline
4399.4 (3)&1300.5 (3)&66 (12)&0.71 (18)&&&5$^{-}$,7$^{-}$$\rightarrow$ 6$^{-}$&\small{$\divideontimes$,e}\\
&1741 (1)&8 (3)&&&&5$^{-}$,7$^{-}$$\rightarrow$ 5$^{-}$&\small{e}\\
\hline
4416.8 (3)&1123.2 (3)&12 (2)&1.31 (31)&0.2 (6)&&10$^{+}$,8$^{-}$$\rightarrow$ 8$^{-}$&\small{}\\
\hline
4459.0 (2)&1030.7 (5)&3 (2)&&&&7,9 $\rightarrow$ 9$^{-}$&\small{}\\
&1165.4 (5)&8 (3)&&&&7,9 $\rightarrow$ 8$^{-}$&\small{}\\
\hline
\hline
\end{tabular}
\end{center}
\label{tab:tabla_gammas_1}
\end{table}
\end{landscape}

\begin{landscape}
\begin{table}[thbp!]
\begin{center}
\begin{tabular}{|c|c|c|c|c|c|c|c|}
\hline
\hline
Level&Transition&Rel. $\gamma$-ray&(40$^{o}$/90$^{o}$)&&&&\\
Energy(keV)&Energy (keV)&Intensity&Anisotropy&\small{Polarization}&\small{Multipolarity}&I${_i}^{\pi}$$\rightarrow$I${_f}^{\pi}$&Comments\\
\hline
\hline
4729.5 (5)&1435.9 (2)&12 (2)&0.87 (21)&0.3 (5)&E1&(9$^{+}$,7$^{+}$) $\rightarrow$ 8$^{-}$&\small{}\\
&1547 (1)&5 (1)&&&E1&(9$^{+}$,7$^{+}$) $\rightarrow$ 8$^{-}$&\small{}\\
\hline
4780.5 (2)&1391.8 (2)&20 (4)&0.42 (12)&&&$\rightarrow$ 3 (1)&\small{}\\
\hline
4782 (1)&1800 (1)&6 (3)&0.37 (26)&&M1&8$^{-}$,6$^{-}$$\rightarrow$ 7$^{-}$&\small{}\\
\hline
4802 (1)&1703 (1)&9 (3)&&&&$\rightarrow$ 6$^{-}$&\small{}\\
\hline
4848 (2)&1554 (2)&11 (5)&0.27 (17)&&M1&9$^{-}$,7$^{-}$$\rightarrow$ 8$^{-}$&\small{}\\
\hline
\hline
\end{tabular}
\end{center}
\label{tab:tabla_gammas_19}
\end{table}
\end{landscape}

\begin{landscape}
\begin{table}[thbp!]
\begin{center}
\begin{tabular}{|c|c|c|c|c|c|c|c|}
\hline
\hline
Transition&Level&Rel. $\gamma$-ray&(40$^{o}$/90$^{o}$)&&&&\\
Energy (keV)&Energy(keV)&Intensity&Anisotropy&\small{Polarization}&\small{Multipolarity}&I${_i}^{\pi}$$\rightarrow$I${_f}^{\pi}$&Comments\\
\hline
\hline
111.5 (1)&3293.7 (2)&430 (30)&0.77 (4)&&M1/E2&8$^{-}$$\rightarrow$ 8$^{-}$&\small{a,f,g}\\
\hline
116.7 (2)&3098.9 (2)&22 (6)&&&M1&6$^{-}$$\rightarrow$ 7$^{-}$&\small{b}\\
\hline
125.9 (2)&4666.8 (2)&9 (3)&0.72 (34)&&(E2)&(12$^{+}$) $\rightarrow$ 10$^{+}$&\small{}\\
\hline
134.8 (1)&3428.5 (2)&536 (45)&0.77 (9)&&M1&9$^{-}$$\rightarrow$ 8$^{-}$&\small{a,f,g}\\
\hline
192.7 (1)&2164.7 (1)&4 (1)&0.69 (24)&&E2&0$^{+}$$\rightarrow$ 2$^{+}$&\small{c,d,e,f,g}\\
\hline
200.4 (1)&3182.4 (2)&1306 (60)&1.10 (7)&-0.27 (10)&M1/E2&8$^{-}$$\rightarrow$ 7$^{-}$&\small{a,d,f,g}\\
\hline
\hline
\end{tabular}
\end{center}
\caption{$\gamma$-rays observed in $^{146}$Gd in the $^{144}$Sm($\alpha$,2n) reaction at {\it E}$_{\alpha}$=26.3 MeV.}
\label{tab:tabla_gammas_21}
\end{table}
\end{landscape}

\begin{landscape}
\begin{table}[thbp!]
\begin{center}
\begin{tabular}{|c|c|c|c|c|c|c|c|}
\hline
\hline
Transition&Level&Rel. $\gamma$-ray&(40$^{o}$/90$^{o}$)&&&&\\
Energy (keV)&Energy(keV)&Intensity&Anisotropy&\small{Polarization}&\small{Multipolarity}&I${_i}^{\pi}$$\rightarrow$I${_f}^{\pi}$&Comments\\
\hline
\hline
639.6 (1)&2611.5 (1)&20 (2)&1.65 (23)&&E2&4$^{+}$$\rightarrow$ 2$^{+}$&\small{d,e,f,g}\\
\hline
654.6 (6)&3640 (1)&40 (20)&&&E2&0$^{+}$$\rightarrow$ 2$^{+}$&\small{e}\\
\hline
655.1 (1)&3313.0 (2)&135 (14)&1.28 (19)&&M1&5$^{-}$$\rightarrow$ 5$^{-}$&\small{b,f,g}\\
\hline
671.7 (1)&3854.0 (2)&37 (7)&0.79 (21)&&M1&7$^{-}$$\rightarrow$ 8$^{-}$&\small{f,g}\\
\hline
675.7 (1)&3287.2 (1)&46 (5)&0.89 (10)&0.26 (23)&M1&3$^{+}$$\rightarrow$ 4$^{+}$&\small{b,f,g}\\
\hline
676.3 (2)&4541.2 (3)&10 (5)&2.57 (73)&&M1&10$^{+}$$\rightarrow$ 10$^{+}$&\small{}\\
\hline
\hline
\end{tabular}
\end{center}
\label{tab:tabla_gammas_2}
\end{table}
\end{landscape}

\begin{landscape}
\begin{table}[thbp!]
\begin{center}
\begin{tabular}{|c|c|c|c|c|c|c|c|}
\hline
\hline
Transition&Level&Rel. $\gamma$-ray&(40$^{o}$/90$^{o}$)&&&&\\
Energy (keV)&Energy(keV)&Intensity&Anisotropy&\small{Polarization}&\small{Multipolarity}&I${_i}^{\pi}$$\rightarrow$I${_f}^{\pi}$&Comments\\
\hline
\hline
871.8 (3)&3854.0 (2)&52 (12)&1.05(34)&0.52 (75)&M1&7$^{-}$$\rightarrow$ 7$^{-}$&\small{f,g}\\
\hline
876.7 (3)&3907.9 (6)&8 (2)&1.59 (53)&&&(3$^{-}$) $\rightarrow$ 3$^{+}$&\small{e}\\
\hline
924.9 (1)&4107.6 (3)&58 (10)&1.20 (29)&-0.22 (35)&E1&8$^{+}$$\rightarrow$ 8$^{-}$&\small{f,g}\\
\hline
951.6 (2)&3562.8 (2)&4 (1)&1.22 (43)&&E1 or E2&4$^{+}$,2$^{+}$$\rightarrow$ 4$^{+}$&\small{}\\
\hline
977.8 (5)&4076.7 (3)&19 (4)&0.94 (32)&-1.5 (8)&& $\rightarrow$ 6$^{-}$&\small{}\\
\hline
1002.0 (1)&3659.9 (2)&129 (10)&0.83 (9)&&E1&6$^{+}$$\rightarrow$ 5$^{-}$&\small{f,g}\\
\hline
\hline
\end{tabular}
\end{center}
\label{tab:tabla_gammas_2}
\end{table}
\end{landscape}

\begin{landscape}
\begin{table}[thbp!]
\begin{center}
\begin{tabular}{|c|c|c|c|c|c|c|c|}
\hline
\hline
Transition&Level&Rel. $\gamma$-ray&(40$^{o}$/90$^{o}$)&&&&\\
Energy (keV)&Energy(keV)&Intensity&Anisotropy&\small{Polarization}&\small{Multipolarity}&I${_i}^{\pi}$$\rightarrow$I${_f}^{\pi}$&Comments\\
\hline
\hline
1125.5 (3)&4107.6 (3)&68 (12)&0.79 (20)&-0.49 (57)&&8$^{+}$$\rightarrow$ 7$^{-}$&\small{f,g}\\
\hline
1165.4 (5)&4459.0 (2)&8 (3)&&&&7,9 $\rightarrow$ 8$^{-}$&\small{}\\
\hline
1172.2 (1)&3783.6 (2)&28 (5)&0.63 (19)&0.15 (61)&M1&3$^{+}$,5$^{+}$ $\rightarrow$ 4$^{+}$&\small{f,g}\\
\hline
1185.2 (5)&4216.3 (3)&10 (3)&0.48 (20)&&&2$^{+}$,4$^{+}$ $\rightarrow$ 3$^{+}$&\small{e}\\
\hline
1191.5 (4)&5056.3 (3)&4 (1)&&0.6 (9)&&$\rightarrow$ 10$^{+}$&\small{}\\
\hline
1197.3 (2)&4179.4 (2)&30 (5)&0.97 (23)&1.03 (76)&&(6$^{-}$) $\rightarrow$ 7$^{-}$&\small{g}\\
\hline
\hline
\end{tabular}
\end{center}
\label{tab:tabla_gammas_2}
\end{table}
\end{landscape}

\begin{landscape}
\begin{table}[thbp!]
\begin{center}
\begin{tabular}{|c|c|c|c|c|c|c|c|}
\hline
\hline
Transition&Level&Rel. $\gamma$-ray&(40$^{o}$/90$^{o}$)&&&&\\
Energy (keV)&Energy(keV)&Intensity&Anisotropy&\small{Polarization}&\small{Multipolarity}&I${_i}^{\pi}$$\rightarrow$I${_f}^{\pi}$&Comments\\
\hline
\hline
1372.8 (6)&4354.9 (2)&12 (3)&0.85 (30)&&&5$^{-}$,6$^{+}$$\rightarrow$ 7$^{-}$&\small{g}\\
\hline
1388.0 (1)&2967.4 (1)&236 (30)&0.88 (16)&0.11 (29)&E1&4$^{+}$$\rightarrow$ 3$^{-}$&\small{b,c,f,g}\\
\hline
1391.8 (2)&4780.5 (2)&20 (4)&0.42 (12)&&&$\rightarrow$ 3 (1)&\small{}\\
\hline
1399 (1)&4580 (1)&20 (6)&0.80 (34)&&&7 $\rightarrow$ 8$^{-}$&\small{$\divideontimes$}\\
\hline
1408.8 (2)&3380.7 (5)&8 (1)&1.12 (20)&0.48 (40)&M1&2$^{+}$$\rightarrow$ 2$^{+}$&\small{d,e,g}\\
\hline
1416.7 (1)&3388.7 (1)&30 (5)&0.90 (21)&&&3 (1) $\rightarrow$ 2$^{+}$&\small{f,g}\\
\hline
\hline
\end{tabular}
\end{center}
\label{tab:tabla_gammas_2}
\end{table}
\end{landscape}

\begin{landscape}
\begin{table}[thbp!]
\begin{center}
\begin{tabular}{|c|c|c|c|c|c|c|c|}
\hline
\hline
Transition&Level&Rel. $\gamma$-ray&(40$^{o}$/90$^{o}$)&&&&\\
Energy (keV)&Energy(keV)&Intensity&Anisotropy&\small{Polarization}&\small{Multipolarity}&I${_i}^{\pi}$$\rightarrow$I${_f}^{\pi}$&Comments\\
\hline
\hline
1579.4 (1)&1579.4 (1)&10000&1.63 (10)&0.55 (20)&E3&3$^{-}$$\rightarrow$ 0$^{+}$&\small{a,b,c,d,e,f,g}\\
\hline
1591.1 (3)&3562.8 (2)&10 (3)&1.38 (65)&&E1 or E2&4$^{+}$,2$^{+}$$\rightarrow$ 2$^{+}$&\small{g}\\
\hline
1604.7 (6)&4898.3 (3)&9 (3)&0.59 (28)&&M1&9$^{-}$,7$^{-}$$\rightarrow$ 8$^{-}$&\small{}\\
\hline
1606.1 (4)&3185.8 (2)&55 (7)&0.84 (15)&-0.98 (61)&E1&2$^{+}$$\rightarrow$ 3$^{-}$&\small{$\divideontimes$,c,e,g}\\
\hline
1653.0 (4)&3232.2 (2)&43 (4)&0.80 (11)&&E1&2$^{+}$$\rightarrow$ 3$^{-}$&\small{c,d,e,g}\\
\hline
1684.3 (1)&3656.2 (2)&12 (2)&0.75 (18)&&&3 $^{}$$\rightarrow$ 2$^{+}$&\small{g}\\
\hline
\hline
\end{tabular}
\end{center}
\label{tab:tabla_gammas_2}
\end{table}
\end{landscape}

\begin{landscape}
\begin{table}[thbp!]
\begin{center}
\begin{tabular}{|c|c|c|c|c|c|c|c|}
\hline
\hline
Transition&Level&Rel. $\gamma$-ray&(40$^{o}$/90$^{o}$)&&&&\\
Energy (keV)&Energy(keV)&Intensity&Anisotropy&\small{Polarization}&\small{Multipolarity}&I${_i}^{\pi}$$\rightarrow$I${_f}^{\pi}$&Comments\\
\hline
\hline
1826 (1)&4484 (2)&24 (6)&0.64 (23)&&(E1)&(4$^{+}$) $\rightarrow$ 5$^{-}$&\small{$\divideontimes$,e}\\
\hline
1837.2 (2)&3416.5 (2)&74 (6)&0.63 (7)&&E1&4$^{+}$$\rightarrow$ 3$^{-}$&\small{d,f,g}\\
\hline
1843.8 (2)&3423.2 (2)&62 (6)&0.75 (10)&&M1/E2&3$^{-}$$\rightarrow$ 3$^{-}$&\small{b,e,f,g}\\
\hline
1857.0 (3)&3436.2 (2)&20 (4)&0.32 (9)&&E1&4$^{+}$$\rightarrow$ 3$^{-}$&\small{$\divideontimes$,g}\\
\hline
1877.0 (2)&3456.4 (2)&43 (13)&1.80 (77)&&&(3,5$^{-}$)$\rightarrow$ 3$^{-}$&\small{$\divideontimes$,f,g}\\
\hline
1881.4 (2)&3853.5 (2)&14 (3)&1.30 (39)&&&(3$^{-}$) $\rightarrow$ 2$^{+}$&\small{$\divideontimes$,e,g}\\
\hline
\hline
\end{tabular}
\end{center}
\label{tab:tabla_gammas_2}
\end{table}
\end{landscape}

\begin{landscape}
\begin{table}[thbp!]
\begin{center}
\begin{tabular}{|c|c|c|c|c|c|c|c|}
\hline
\hline
Transition&Level&Rel. $\gamma$-ray&(40$^{o}$/90$^{o}$)&&&&\\
Energy (keV)&Energy(keV)&Intensity&Anisotropy&\small{Polarization}&\small{Multipolarity}&I${_i}^{\pi}$$\rightarrow$I${_f}^{\pi}$&Comments\\
\hline
\hline
2274 (1)&3853.5 (2)&24 (5)&1.55 (46)&&&(3$^{-}$) $\rightarrow$ 3$^{-}$&\small{e,g}\\
\hline
2329 (1)&3907.9 (6)&8 (2)&&&&(3$^{-}$) $\rightarrow$ 3$^{-}$&\small{b,e,g}\\
\hline
2394 (1)&3973 (1)&16 (5)&0.99 (44)&&&(3$^{-}$) $\rightarrow$ 3$^{-}$&\small{$\divideontimes$,e}\\
\hline
2408 (1)&3987 (1)&30 (7)&1.06 (35)&&&$\rightarrow$ 3$^{-}$&\small{$\divideontimes$,g}\\
\hline
2427 (1)&4006.6 (5)&2 (1)&&&&(4$^{+}$) $\rightarrow$ 3$^{-}$&\small{e}\\
\hline
2534 (1)&4113 (1)&12 (4)&0.94 (44)&&& $\rightarrow$ 3$^{-}$&\small{}\\
\hline
\hline
\end{tabular}
\end{center}
\label{tab:tabla_gammas_29}
\end{table}
\end{landscape}

\begin{figure}[!htbp]
\begin{center}
\includegraphics[width=13.8cm,angle=0.]{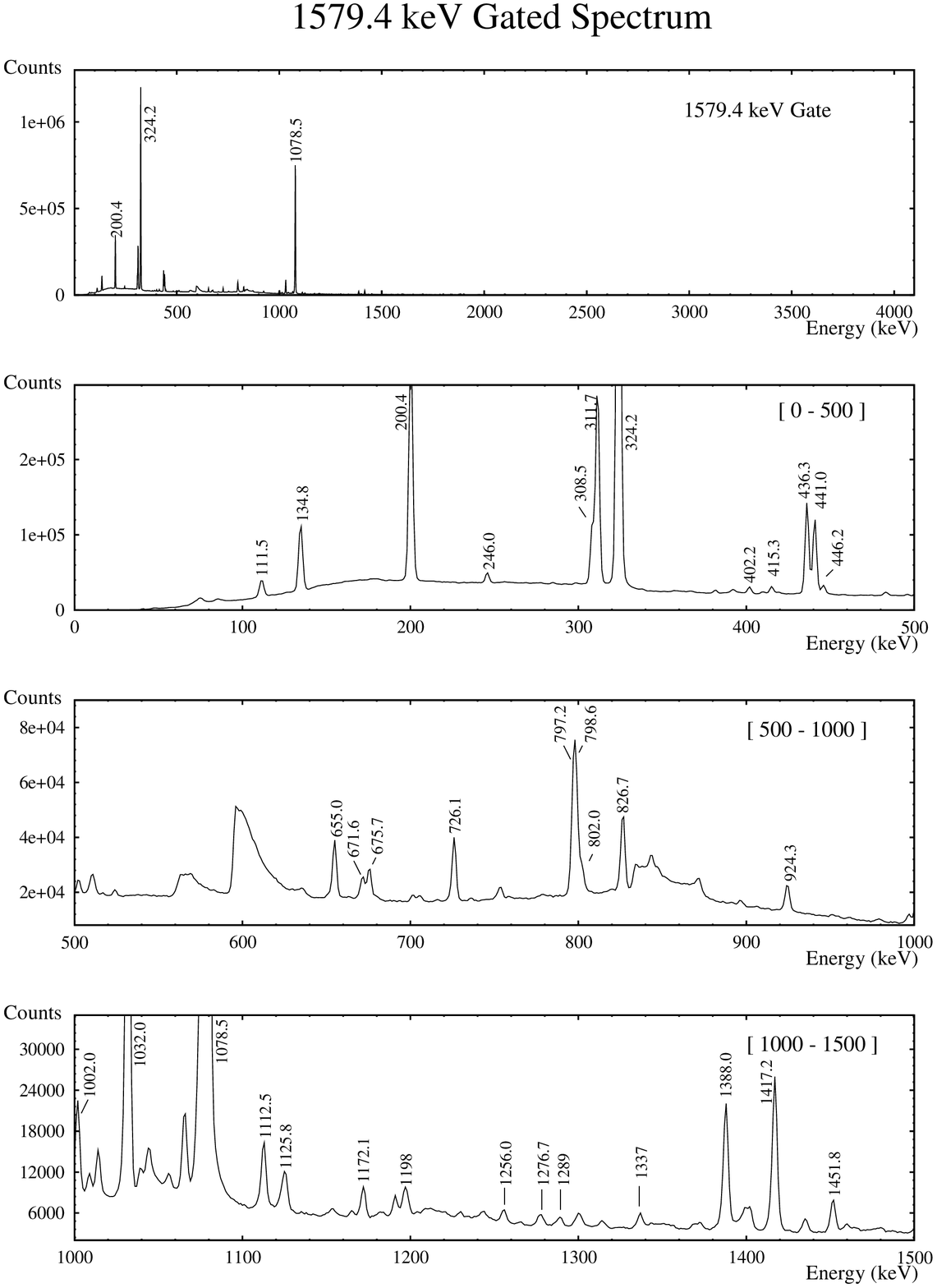}
\end{center}
\caption{The panels show the 1579.4 keV gated spectrum in ranges of 500 keV. Only the most intense peaks are labeled.}
\label{fig:gate1972_b}
\end{figure}

\begin{figure}[!htbp]
\begin{center}
\includegraphics[width=13.8cm,angle=0.]{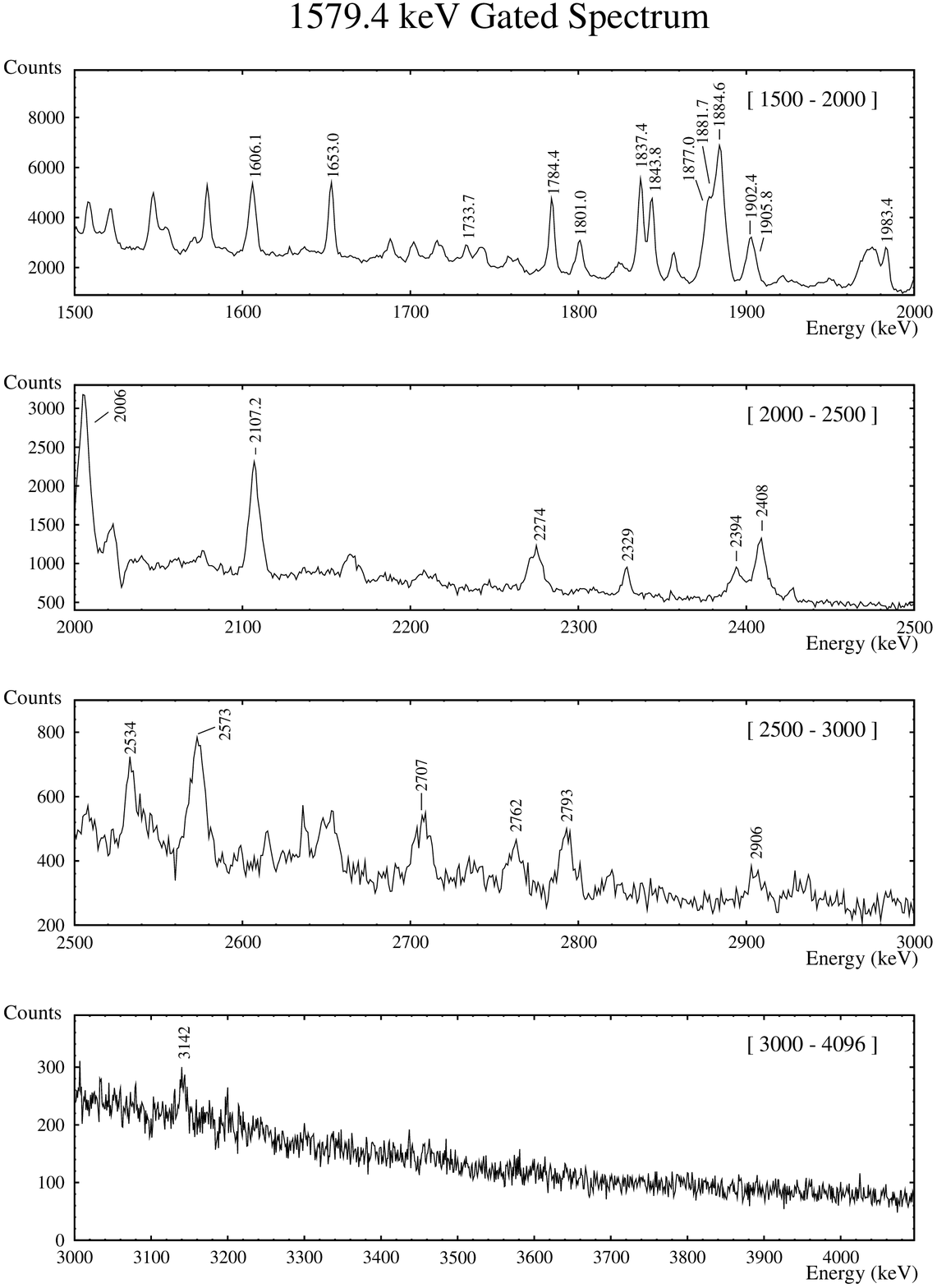}
\end{center}
\caption{The panels show the 1579.4 keV gated spectrum in ranges of 500 keV (except the last panel). Only the most intense peaks are labeled.}
\label{fig:gate1972_b}
\end{figure}

\begin{figure}[!htbp]
\begin{center}
\includegraphics[width=13.8cm,angle=0.]{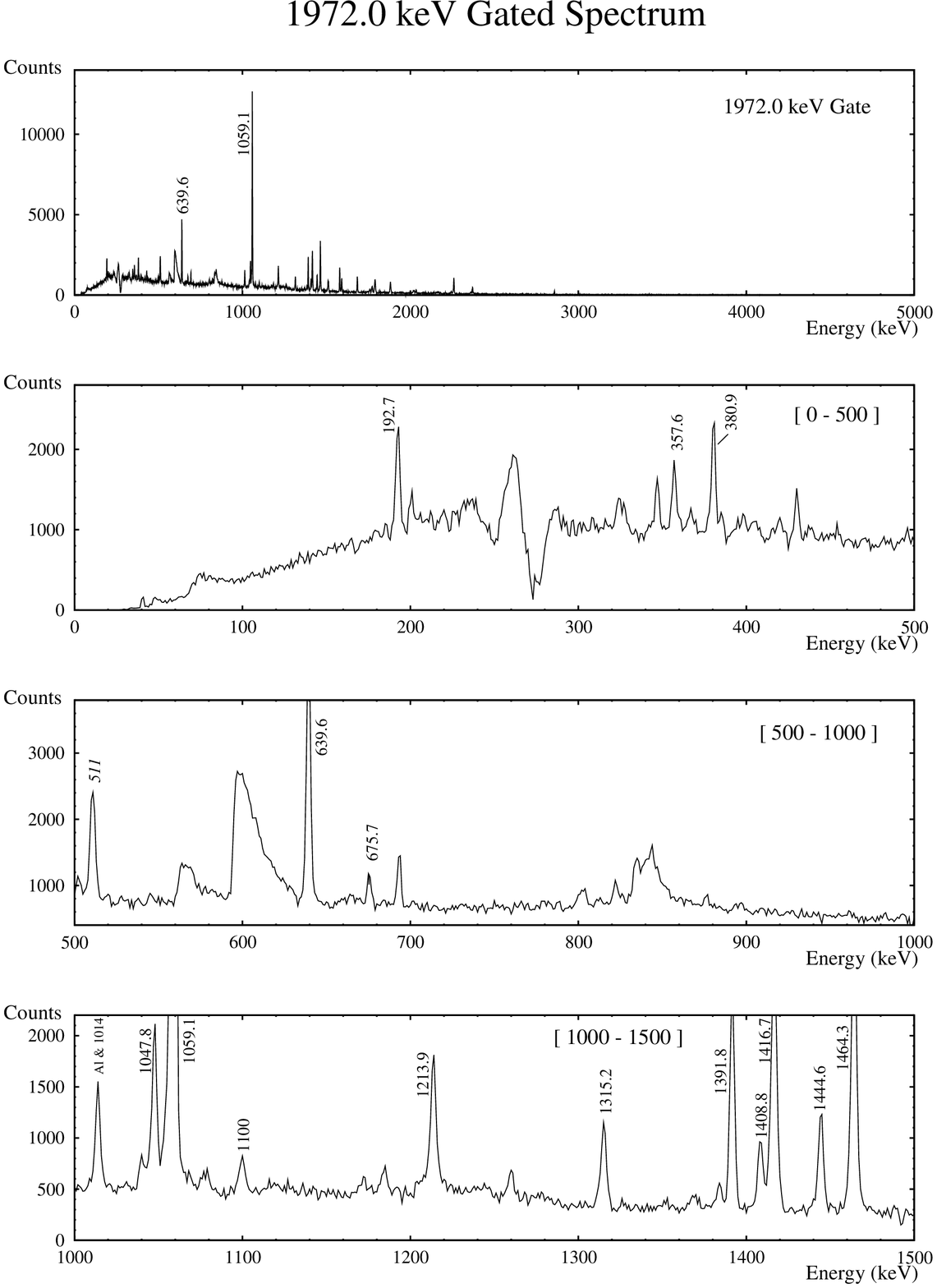}
\end{center}
\caption{The panels show the 1972.0 keV gated spectrum in ranges of 500 keV. Only the most intense peaks are labeled.}
\label{fig:gate1972_b}
\end{figure}

\begin{figure}[!htbp]
\begin{center}
\includegraphics[width=13.8cm,angle=0.]{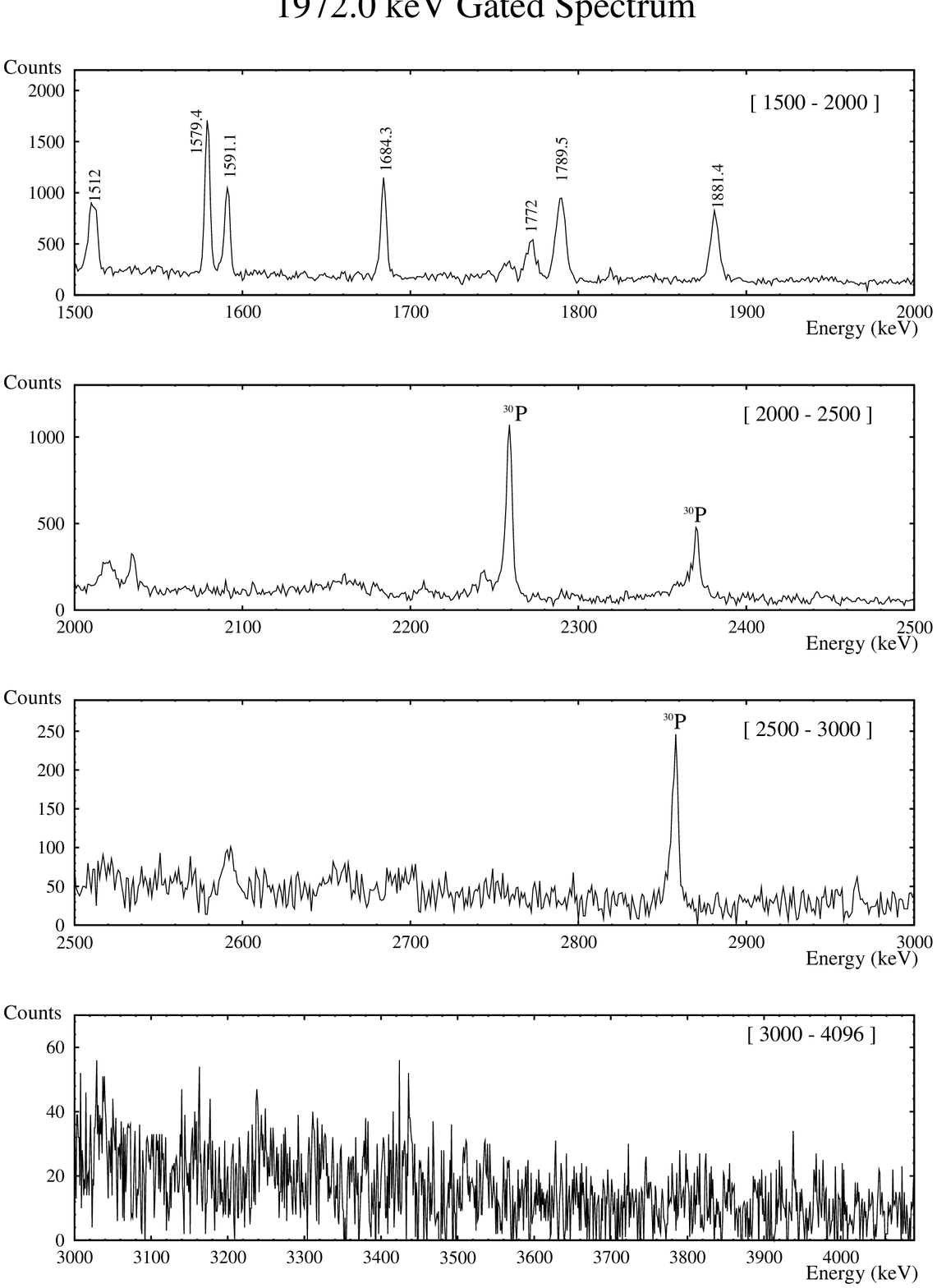}
\end{center}
\caption{The panels show the 1972.0 keV gated spectrum in ranges of 500 keV (except the last panel). Only the most intense peaks are labeled.}
\label{fig:gate1972_b}
\end{figure}

\begin{landscape}
\begin{figure}[thbp!]
\begin{center}
\vskip -3.0cm
\hskip -3.0cm
\includegraphics[width=17.0 cm,angle=90.]{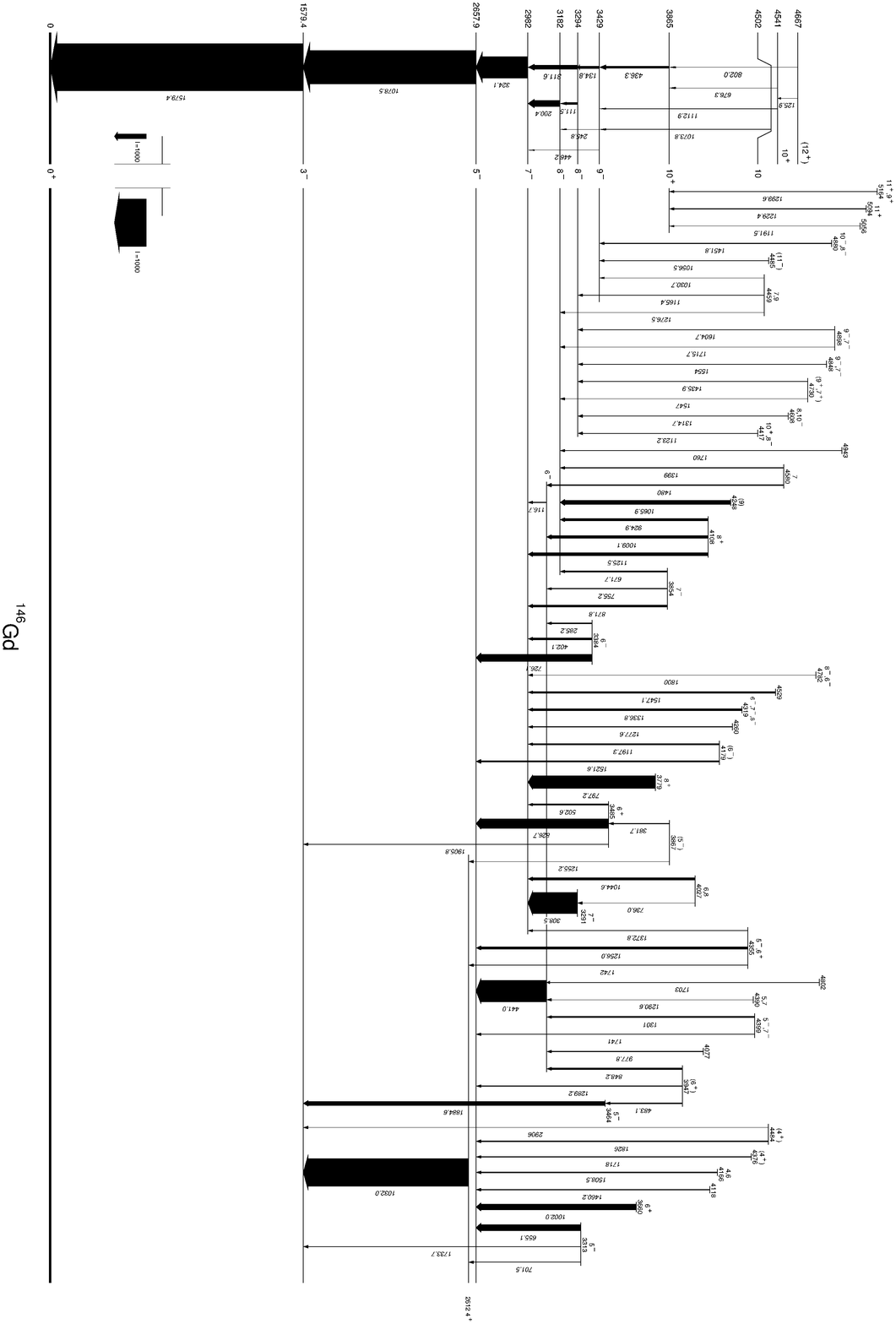}
\end{center}
\caption{The level scheme of $^{146}$Gd deduced from this work. Yrast and near
yrast levels populated in fussion-evaporation reactions with heavy ions are shown to the left. The thicknesses 
of the transitions represents their $\gamma$-ray intensities. Note the change in the scale from the left to the right 
part of the scheme, part 1.}
\label{fig:level1}
\end{figure}

\begin{figure}[thbp!]
\begin{center}
\vskip -3.0cm
\hskip -3.0cm
\includegraphics[width=17.0 cm,angle=90.]{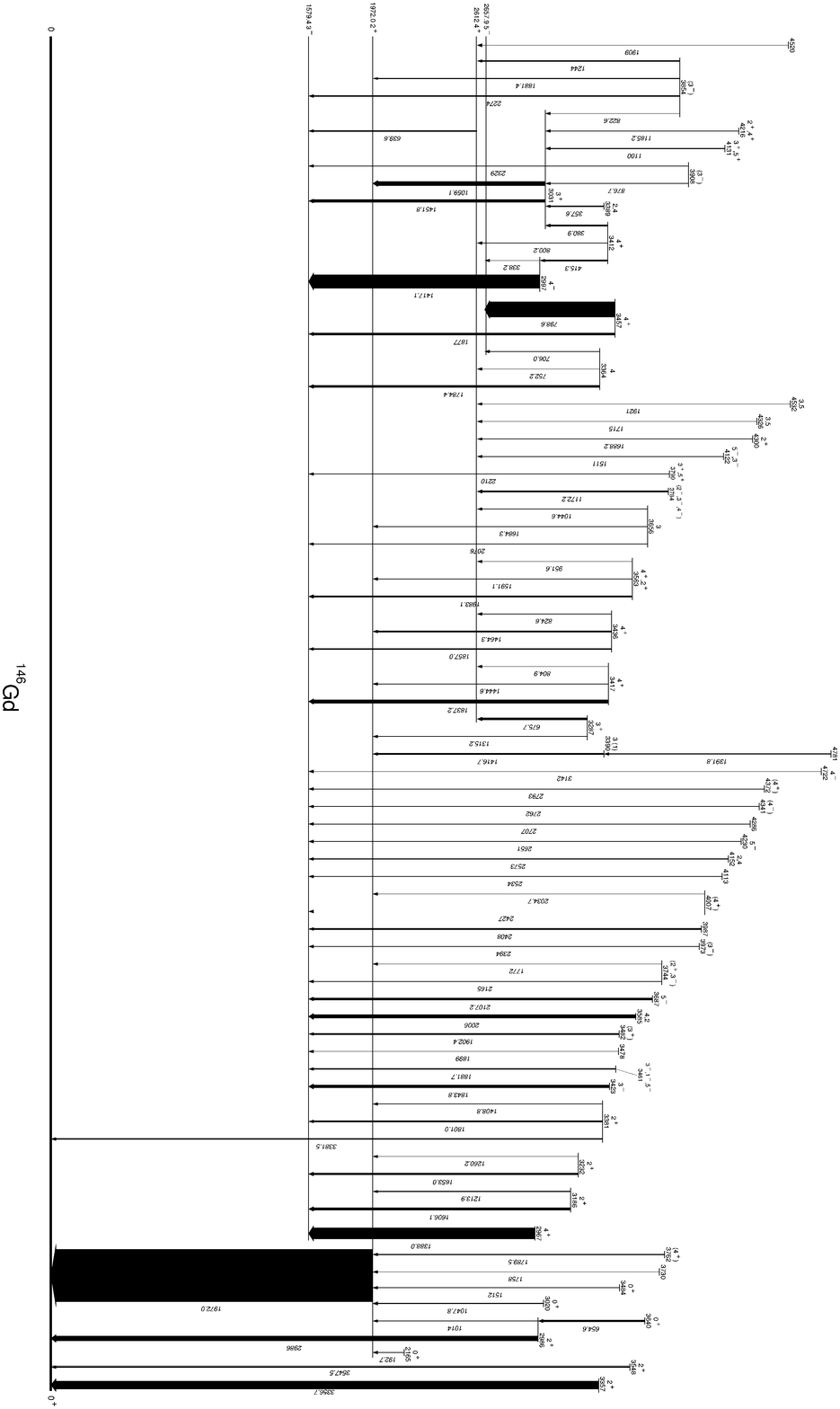}
\end{center}
\caption{The level scheme of $^{146}$Gd deduced from this work. Yrast and near
yrast levels populated in fussion-evaporation reactions with heavy ions are shown to the left. The thicknesses of the 
transitions represents their $\gamma$-ray intensities. Note the change in the scale from the left to the right part of the 
scheme, part 2.}
\label{fig:level1}
\end{figure}
\end{landscape}
\newpage
\chapter{Discussion of the results}

{\it  In this chapter we will study the particle-hole multiplets of $^{146}$Gd, first calculating the energies where the multiplets members should lie and later assigning configurations to the observed states. The two-phonon octupole states in $^{146}$Gd will be discussed as well.}

\section{Nucleon-nucleon multiplets}

As it mentioned in Chapter 1, one of the main goals of this work is to extract the nucleon-nucleon residual interaction in the $^{146}$Gd region. For this purpose we have to calculate the particle-hole (p-h) multiplet unperturbed energies and compare them with our experimental values. Since there are many levels with the same J$^{\pi}$ assignment, we need to estimate the expected energy of the multiplets members in order to identify them. There are two ways to estimate them: 
\begin{itemize}
\item[-] From the experimental knowledge in neighbouring nuclei. The procedure is first to calculate the energy (unperturbed energy) at which the multiplet is expected in the neighbouring nucleus with a ``clean'' multiplet nucleon-nucleon configuration, and compare it with the experimental data for this nucleus where the multiplet members have been clearly identified. The difference between the calculated and the experimental value in each multiplet member is the so-called {\it residual nucleon-nucleon interaction energy} in this nucleus.

\begin{equation}
E^{Res} = E^{Exp} - E^{Unp}
\end{equation}\

Then, if we assume that the residual interaction is the same in neighbouring nuclei (the only difference between their nuclear structure is the presence or absence of at most two nucleons), we can add it to the calculated unperturbed energy of the nucleus in consideration and estimate the energies of the multiplet members.

\item[-] In cases where we cannot be helped by experimental data, we can calculate the residual interaction by taking, for instance, a Surface Delta Interaction (SDI) that, together with a Coulomb interaction estimation of $\sim$300 keV (this interaction is only present when we calculate proton multiplets), will give us the energy correction of the calculated unperturbed multiplet energies.
\end{itemize}

As mentioned before, the best way to extract this information is to look at experimental levels with clean nucleon-nucleon configurations. Following this reasoning, the best nucleus to extract the $\pi$$\pi$ residual interaction energy would be $^{148}$Dy, with two protons outside the $^{146}$Gd core. Analogously, the best way to extract information about $\pi^{-1}$$\pi^{-1}$ configurations would be $^{144}$Sm. With this philosophy, the ideal nucleus to identify $\pi$$\pi^{-1}$ excitations and $\nu$$\nu^{-1}$ excitations is $^{146}$Gd itself. In the next sections we will calculate the two-particle state multiplet energies.\

\subsection{The $\pi$h$_{11/2}^{2}$ multiplet}



As we mentioned above, the best nucleus to extract the $\pi$$\pi$ interaction would be $^{148}$Dy. Then, we first calculate the $\pi$h$_{11/2}^{2}$ multiplet unperturbed energy in $^{148}$Dy. This method of calculating unperturbed energies is well described in~\cite{BLOMQ83}.

\begin{figure}[h]
\begin{center}
\includegraphics[width=13.cm,height=3.4cm,angle=0.]{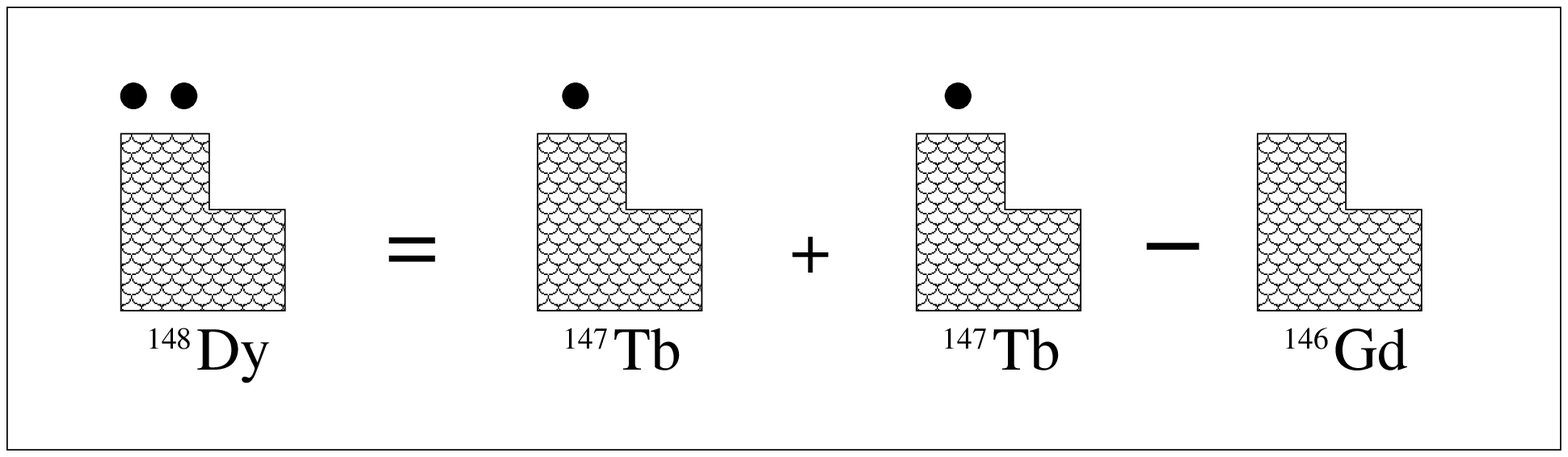}
\end{center}
\label{fig:calc}
\end{figure}

\begin{equation}
[E^{Unp}(\pi h_{11/2}^{2},^{148}Dy)+(\textrm{\it Mass of }^{148}Dy)] = [E'(\pi h_{11/2},^{147}Tb)+(\textrm{\it Mass of }^{147}Tb)]\nonumber
\end{equation}
\begin{equation}
+[E'(\pi h_{11/2},^{147}Tb)+(\textrm{\it Mass of }^{147}Tb)]-[E'(^{146}Gd)+(\textrm{\it Mass of }^{146}Gd)]\\
\end{equation}\\

where E$^{\prime}$($\pi$h$_{11/2}$,$^{147}$Tb) is the experimental energy of the h$_{11/2}$ state in $^{147}$Tb and E$^{\prime}$($^{146}$Gd) is the $^{146}$Gd ground state. Then, the $\pi$h$_{11/2}^{2}$ multiplet unperturbed energy in $^{148}$Dy is

\begin{equation}
E^{Unp}(\pi h_{11/2}^{2},^{148}Dy) = 2\times E'(\pi h_{11/2},^{147}Tb) + \;\;
\begin{tabular}[tb]{|c|}
\hline
-1 \\
\hline
+2 \\
\hline
\multicolumn{1}{||c||}{-1} \\
\hline
\end{tabular} \;= \\ \nonumber
\end{equation}\
\begin{equation}
= (2 \times 51)\; +\; 2497 = 2599 \; keV
\end{equation}\

where

\begin{equation}
\begin{tabular}[tb]{|c|}
\hline
-1 \\
\hline
+2 \\
\hline
\multicolumn{1}{||c||}{-1} \\
\hline
\end{tabular} \;= -(\textrm{\it Mass of }^{148}Dy) + 2\times(\textrm{\it Mass of }^{147}Tb) - (\textrm{\it Mass of }^{146}Gd)
\end{equation}\\

The mass values have been taken from~\cite{AUDI}. We have to repeat the same type of calculation in order to calculate the $\pi$h$_{11/2}^{2}$ multiplet unperturbed energy in $^{146}$Gd.

\begin{equation}
E^{Unp}(\pi h_{11/2}^{2},^{146}Gd) = 2\times E'(\pi h_{11/2},^{145}Eu) + \;\;
\begin{tabular}[tb]{|c|}
\hline
\multicolumn{1}{||c||}{-1} \\
\hline
+2 \\
\hline
-1 \\
\hline
\end{tabular} \;= \\ \nonumber
\end{equation}\
\begin{equation}
= (2 \times 716)\; +\; 2070 = 3502 \; keV
\end{equation}\

where

\begin{equation}
\begin{tabular}[tb]{|c|}
\hline
\multicolumn{1}{||c||}{-1} \\
\hline
+2 \\
\hline
-1 \\
\hline
\end{tabular} \;= -(\textrm{\it Mass of }^{146}Gd) + 2\times(\textrm{\it Mass of }^{145}Eu) - (\textrm{\it Mass of }^{144}Sm)
\end{equation}\

We observe that the expected energy for a two-proton multiplet is lower in the ``two-proton'' nucleus $^{148}$Dy than in  $^{146}$Gd where we have to excite the core. Now, assuming that the residual interactions in $^{148}$Dy and in $^{146}$Gd for the $\pi$h$_{11/2}^{2}$ multiplet are the same, we can calculate the estimated energies (E$^{\star}$=E$^{Unp}$+E$^{Res}$) of the multiplet members in $^{146}$Gd. The results are shown in the next table and in Figure~[\ref{fig:posmult}].

\begin{center}
\begin{tabular}{|c|c|c|c|c|c|c|}
\hline
\hline
\multicolumn{7}{|c|}{E$^{Unp}$($\pi$h$_{11/2}^{2}$,$^{148}$Dy) = 2599 keV} \\
\hline
J$^{\pi}$&0$^{+}$&2$^{+}$&4$^{+}$&6$^{+}$&8$^{+}$&10$^{+}$\\
\hline
E$^{Exp}$($\pi$h$_{11/2}^{2}$,$^{148}$Dy)(keV)&0&1678&2428&2732&2833&2919\\
\hline
E$^{Res}$(keV)&-2599&-921&-171&133&234&320\\
\hline
\hline
\multicolumn{7}{|c|}{E$^{Unp}$($\pi$h$_{11/2}^{2}$,$^{146}$Gd) = 3502 keV} \\
\hline
E$^{\star}$($\pi$h$_{11/2}^{2}$,$^{146}$Gd)(keV)&903&2581&3331&3635&3736&3822\\
\hline
\hline
\end{tabular}\\ 
\end{center}\

The $^{148}$Dy experimental data were taken from~\cite{TAIN148DY}.\\


\subsection{The $\pi$s$_{1/2}^{2}$ multiplet}

Similarly, the best nucleus to extract the $\pi$$\pi$ interaction is $^{148}$Dy. Unfortunately, there is no experimental information on the $\pi$s$_{1/2}^{2}$ multiplet in $^{148}$Dy and, consequently, we have to calculate the residual interaction in $^{146}$Gd from a SDI. First, we calculate the unperturbed multiplet energy in $^{146}$Gd:

\begin{equation}
E^{Unp}(\pi s_{1/2}^{2},^{146}Gd) = 2\times E'(\pi s_{1/2},^{145}Eu) + \;\;
\begin{tabular}[tb]{|c|}
\hline
\multicolumn{1}{||c||}{-1} \\
\hline
+2 \\
\hline
-1 \\
\hline
\end{tabular} \;= \\ \nonumber
\end{equation}\
\begin{equation}
= (2 \times 809)\; +\; 2070 = 3688 \; keV
\end{equation}\

Once we have this value, we can estimate the multiplet energy in $^{146}$Gd by adding the residual interaction and taking into account the Coulomb interaction of the two protons. Results are shown in the next table and in Figure~[\ref{fig:posmult}]. 

\begin{center}
\begin{tabular}{|c|c|}
\hline
\hline
J$^{\pi}$&0$^{+}$\\
\hline
E$^{Res}$\small{({\it SDI} Calculation)}\normalsize{(keV)}&-171\\
\hline
Coulomb Interaction (keV)&+300\\
\hline
E$^{\star}$($\pi$s$_{1/2}^{2}$,$^{146}$Gd)(keV)&3817\\
\hline
\hline
\end{tabular} 
\end{center}


\subsection{The $\pi$d$_{3/2}^{2}$ multiplet}

Again, this is a $\pi$$\pi$ interaction and the best nucleus to extract the $\pi$$\pi$ interaction is $^{148}$Dy. In this case, there is no experimental information on the $\pi$d$_{3/2}^{2}$ multiplet energy in $^{148}$Dy, and we calculate the residual interaction in $^{146}$Gd from a SDI. First, we calculate the unperturbed multiplet energy in $^{146}$Gd:

\begin{equation}
E^{Unp}(\pi d_{3/2}^{2},^{146}Gd) = 2\times E'(\pi d_{3/2},^{145}Eu) + \;\;
\begin{tabular}[tb]{|c|}
\hline
\multicolumn{1}{||c||}{-1} \\
\hline
+2 \\
\hline
-1 \\
\hline
\end{tabular} \;= \\ \nonumber
\end{equation}\
\begin{equation}
= (2 \times 1042)\; +\; 2070 = 4156 \; keV
\end{equation}\

Once we have this value we can estimate the multiplet energy in $^{146}$Gd by adding the residual interaction and taking into account the Coulomb interaction of the two protons. Results are shown in the next table and in Figure~[\ref{fig:posmult}].

\begin{center}
\begin{tabular}{|c|c|c|}
\hline
\hline
J$^{\pi}$&0$^{+}$&2$^{+}$\\
\hline
E$^{Res}$\small{({\it SDI} Calculation)}\normalsize{(keV)}&-343&-69\\
\hline
Coulomb Interaction (keV)&+300&+300\\
\hline
E$^{\star}$($\pi$d$_{3/2}^{2}$,$^{146}$Gd)(keV)&4113&4387\\
\hline
\hline
\end{tabular} 
\end{center}



\subsection{The $\pi$h$_{11/2}$$\pi$s$_{1/2}$ multiplet}

As mentioned earlier, the best nucleus to extract the $\pi$$\pi$ interaction is $^{148}$Dy. Then, we first calculate the $\pi$h$_{11/2}$$\pi$s$_{1/2}$ multiplet unperturbed energy in $^{148}$Dy.


\begin{equation}
E^{Unp}(\pi h_{11/2},\pi s_{1/2},^{148}Dy) = E'(\pi h_{11/2},^{147}Tb) + E'(\pi s_{1/2},^{147}Tb) + \;\;
\begin{tabular}[tb]{|c|}
\hline
-1 \\
\hline
+2 \\
\hline
\multicolumn{1}{||c||}{-1} \\
\hline
\end{tabular} \;= \\ \nonumber
\end{equation}\
\begin{equation}
= 0\; + \;51\; +\; 2497 = 2548 \; keV
\end{equation}\


We have to repeat the same type of calculation in order to calculate the $\pi$h$_{11/2}$$\pi$s$_{1/2}$ multiplet unperturbed energy in $^{146}$Gd.

\begin{equation}
E^{Unp}(\pi h_{11/2},\pi s_{1/2},^{146}Gd) = E'(\pi h_{11/2},^{145}Eu) + E'(\pi s_{1/2},^{145}Eu) + \;\;
\begin{tabular}[tb]{|c|}
\hline
\multicolumn{1}{||c||}{-1} \\
\hline
+2 \\
\hline
-1 \\
\hline
\end{tabular} \;= \\ \nonumber
\end{equation}\
\begin{equation}
= 803\; + \;716\; +\; 2070 = 3589 \; keV
\end{equation}\

Now, assuming that the residual interactions in $^{148}$Dy and in $^{146}$Gd for the $\pi$h$_{11/2}$$\pi$s$_{1/2}$ multiplet are the same, we can calculate the estimated energies (E$^{\star}$=E$^{Unp}$+E$^{Res}$) of the multiplet members in $^{146}$Gd. Results are shown in the next table and in Figure~[\ref{fig:negmult}].

\begin{center}
\begin{tabular}{|c|c|c|}
\hline
\hline
\multicolumn{3}{|c|}{E$^{Unp}$($\pi$h$_{11/2}$,$\pi$s$_{1/2}$,$^{148}$Dy) = 2548 keV} \\
\hline
J$^{\pi}$&5$^{-}$&6$^{-}$\\
\hline
E$^{Exp}$($\pi$h$_{11/2}$,$\pi$s$_{1/2}$,$^{148}$Dy)(keV)&2350&2854\\
\hline
E$^{Res}$(keV)&-198&306\\
\hline
\hline
\multicolumn{3}{|c|}{E$^{Unp}$($\pi$h$_{11/2}$,$\pi$s$_{1/2}$,$^{146}$Gd) = 3589 keV} \\
\hline
E$^{\star}$($\pi$h$_{11/2}$,$\pi$s$_{1/2}$,$^{146}$Gd)(keV)&3391&3895\\
\hline
\hline
\end{tabular}\\ 
\end{center} 

The $^{148}$Dy experimental data were taken from~\cite{TAIN148DY}.\


\subsection{The $\pi$h$_{11/2}$$\pi$d$_{3/2}$ multiplet}

In this case we have experimental data of the $\pi$$\pi$ interaction is $^{148}$Dy. Then, we first calculate the $\pi$h$_{11/2}$$\pi$d$_{3/2}$ multiplet unperturbed energy in $^{148}$Dy.

\begin{equation}
E^{Unp}(\pi h_{11/2},\pi d_{3/2},^{148}Dy) = E'(\pi h_{11/2},^{147}Tb) + E'(\pi d_{3/2},^{147}Tb) + \;\;
\begin{tabular}[tb]{|c|}
\hline
-1 \\
\hline
+2 \\
\hline
\multicolumn{1}{||c||}{-1} \\
\hline
\end{tabular} \;= \\ \nonumber
\end{equation}\
\begin{equation}
= 253\; + \;51\; +\; 2497 = 2801 \; keV
\end{equation}\

We have to repeat the same type of calculation in order to calculate the $\pi$h$_{11/2}$$\pi$d$_{3/2}$ multiplet unperturbed energy in $^{146}$Gd.

\begin{equation}
E^{Unp}(\pi h_{11/2},\pi d_{3/2},^{146}Gd) = E'(\pi h_{11/2},^{145}Eu) + E'(\pi d_{3/2},^{145}Eu) + \;\;
\begin{tabular}[tb]{|c|}
\hline
\multicolumn{1}{||c||}{-1} \\
\hline
+2 \\
\hline
-1 \\
\hline
\end{tabular} \;= \\ \nonumber
\end{equation}\
\begin{equation}
= 1042\; + \;716\; +\; 2070 = 3828 \; keV
\end{equation}\

If we assume that the residual interactions in $^{148}$Dy and in $^{146}$Gd for the $\pi$h$_{11/2}$$\pi$d$_{3/2}$ multiplet are the same, we can calculate the estimated energies (E$^{\star}$=E$^{Unp}$+E$^{Res}$) of the multiplet members in $^{146}$Gd. Results are shown in the next table and in Figure~[\ref{fig:negmult}].

\begin{center}
\begin{tabular}{|c|c|c|c|c|}
\hline
\hline
\multicolumn{5}{|c|}{E$^{Unp}$($\pi$h$_{11/2}$,$\pi$d$_{3/2}$,$^{148}$Dy) = 2801 keV}\\
\hline
J$^{\pi}$&4$^{-}$&5$^{-}$&6$^{-}$&7$^{-}$\\
\hline
E$^{Exp}$($\pi$h$_{11/2}$,$\pi$d$_{3/2}$,$^{148}$Dy)(keV)&2995&3172&3323&2739\\
\hline
E$^{Res}$(keV)&194&371&522&-62\\
\hline
\hline
\multicolumn{5}{|c|}{E$^{Unp}$($\pi$h$_{11/2}$,$\pi$d$_{3/2}$,$^{146}$Gd) = 3828 keV}\\
\hline
E$^{\star}$($\pi$h$_{11/2}$,$\pi$d$_{3/2}$,$^{146}$Gd)(keV)&4002&4199&4350&3766\\
\hline
\hline
\end{tabular}\\ 
\end{center}

The $^{148}$Dy experimental data were taken from~\cite{TAIN148DY}.\


\subsection{The $\pi$s$_{1/2}$$\pi$d$_{3/2}$ multiplet}

There is no experimental information of the $\pi$s$_{1/2}$$\pi$d$_{3/2}$ multiplet in $^{148}$Dy. Therefore, we have to calculate the residual interaction in $^{146}$Gd from a SDI. First, we calculate the unperturbed multiplet energy in $^{146}$Gd:

\begin{equation}
E^{Unp}(\pi s_{1/2},\pi d_{3/2},^{146}Gd) = E'(\pi s_{1/2},^{145}Eu) + E'(\pi d_{3/2},^{145}Eu) + \;\;
\begin{tabular}[tb]{|c|}
\hline
\multicolumn{1}{||c||}{-1} \\
\hline
+2 \\
\hline
-1 \\
\hline
\end{tabular} \;= \\ \nonumber
\end{equation}\
\begin{equation}
= 803\; + \;1042\; +\; 2070 = 3915 \; keV
\end{equation}\

Once we have this value, we can estimate the multiplet energy in $^{146}$Gd by adding the residual interaction and taking into account the Coulomb interaction of the protons. Results are shown in the next table and in Figure~[\ref{fig:posmult}]. 

\begin{center}
\begin{tabular}{|c|c|c|}
\hline
\hline
J$^{\pi}$&1$^{+}$&2$^{+}$\\
\hline
E$^{Res}$\small{({\it SDI} Calculation)}\normalsize{(keV)}&0&-137\\
\hline
Coulomb Interaction (keV)&+300&+300\\
\hline
E$^{\star}$($\pi$s$_{1/2}$,$\pi$d$_{3/2}$,$^{146}$Gd)(keV)&4215&4078\\
\hline
\hline
\end{tabular} 
\end{center}



\subsection{The $\pi$d$_{5/2}^{-2}$ multiplet}

Now we are considering a two proton-hole multiplet and, therefore, the best nucleus to extract the $\pi^{-1}$$\pi^{-1}$ interaction is $^{144}$Sm. Then, we first calculate the $\pi$d$_{5/2}^{-2}$ multiplet unperturbed energy in $^{144}$Sm. This multiplet has been identified by Rico~\cite{RICO} and the experimental data are taken from this work.

\begin{equation}
E^{Unp}(\pi d_{5/2}^{-2},^{144}Sm) = 2\times E'(\pi d_{5/2}^{-1},^{145}Eu) + \;\;
\begin{tabular}[tb]{|c|}
\hline
\multicolumn{1}{||c||}{-1} \\
\hline
+2 \\
\hline
-1 \\
\hline
\end{tabular} \;= \\ \nonumber
\end{equation}\
\begin{equation}
= (2 \times 0)\; +\; 2070 = 2070 \; keV
\end{equation}\

We have to repeat the same type of calculation in order to determine the $\pi$d$_{5/2}^{-2}$ multiplet unperturbed energy in $^{146}$Gd.

\begin{equation}
E^{Unp}(\pi d_{5/2}^{-2},^{146}Gd) = 2\times E'(\pi d_{5/2}^{-1},^{147}Tb) + \;\;
\begin{tabular}[tb]{|c|}
\hline
-1 \\
\hline
+2 \\
\hline
\multicolumn{1}{||c||}{-1} \\
\hline
\end{tabular} \;= \\ \nonumber
\end{equation}\
\begin{equation}
= (2 \times 354)\; +\; 2497 = 3205 \; keV
\end{equation}\

Now, assuming that the residual interactions in $^{144}$Sm and in $^{146}$Gd for the $\pi$d$_{5/2}^{-2}$ multiplet are the same, we can calculate the estimated energies (E$^{\star}$=E$^{Unp}$+E$^{Res}$) of the multiplet members in $^{146}$Gd. Results are shown in the next table and in Figure~[\ref{fig:posmult}].\\

\begin{center}
\begin{tabular}{|c|c|c|c|}
\hline
\hline
\multicolumn{4}{|c|}{E$^{Unp}$($\pi$d$_{5/2}^{-2}$,$^{144}$Sm) = 2070 keV}\\
\hline
J$^{\pi}$&0$^{+}$&2$^{+}$&4$^{+}$\\
\hline
E$^{Exp}$($\pi$d$_{5/2}^{-2}$,$^{144}$Sm)(keV)&0&1660&2190\\
\hline
E$^{Res}$(keV)&-2070&-410&120\\
\hline
\hline
\multicolumn{4}{|c|}{E$^{Unp}$($\pi$d$_{5/2}^{-2}$,$^{146}$Gd) = 3205 keV}\\
\hline
E$^{\star}$($\pi$d$_{5/2}^{-2}$,$^{146}$Gd)(keV)&1127&2787&3317\\
\hline
\hline
\end{tabular}
\end{center} 

~\\


\subsection{The $\pi$d$_{5/2}^{-1}$$\pi$g$_{7/2}^{-1}$ multiplet}

In a similar way, the best nucleus to extract the $\pi^{-1}$$\pi^{-1}$ interaction is $^{144}$Sm. Then, we first calculate the $\pi$d$_{5/2}^{-1}$$\pi$g$_{7/2}^{-1}$ multiplet unperturbed energy in $^{144}$Sm. This multiplet has been identified also by Rico~\cite{RICO} and the experimental data are taken from this work.

\begin{equation}
E^{Unp}(\pi d_{5/2}^{-1},\pi g_{7/2}^{-1},^{144}Sm) = E'(\pi d_{5/2}^{-1},^{145}Eu) + E'(\pi g_{7/2}^{-1},^{145}Eu) + \;\;
\begin{tabular}[tb]{|c|}
\hline
\multicolumn{1}{||c||}{-1} \\
\hline
+2 \\
\hline
-1 \\
\hline
\end{tabular} \;= \\ \nonumber
\end{equation}\
\begin{equation}
= 0\; + \;330\; +\; 2070 = 2400 \; keV
\end{equation}\

We have to repeat the same type of calculation in order to determine the $\pi$d$_{5/2}^{-1}$$\pi$g$_{7/2}^{-1}$ multiplet unperturbed energy in $^{146}$Gd.

\begin{equation}
E^{Unp}(\pi d_{5/2}^{-1},\pi g_{7/2}^{-1},^{146}Gd) = E'(\pi d_{5/2}^{-1},^{147}Tb) + E'(\pi g_{7/2}^{-1},^{147}Tb) + \;\;
\begin{tabular}[tb]{|c|}
\hline
-1 \\
\hline
+2 \\
\hline
\multicolumn{1}{||c||}{-1} \\
\hline
\end{tabular} \;= \\ \nonumber
\end{equation}\
\begin{equation}
= 354\; + \;719\; +\; 2497 = 3570 \; keV
\end{equation}\

Now, assuming that the residual interactions in $^{144}$Sm and in $^{146}$Gd for the $\pi$d$_{5/2}^{-1}$$\pi$g$_{7/2}^{-1}$ multiplet are the same, we can calculate the estimated energies (E$^{\star}$) of the multiplet members in $^{146}$Gd. Results are shown in the next table and in Figure~[\ref{fig:posmult}].

\begin{center}
\begin{tabular}{|c|c|c|c|c|c|c|}
\hline
\hline
\multicolumn{7}{|c|}{E$^{Unp}$($\pi$d$_{5/2}^{-1}$,$\pi$g$_{7/2}^{-1}$,$^{144}$Sm) = 2400 keV}\\
\hline
J$^{\pi}$&1$^{+}$&2$^{+}$&3$^{+}$&4$^{+}$&5$^{+}$&6$^{+}$\\
\hline
E$^{Exp}$($\pi$d$_{5/2}^{-1}$,$\pi$g$_{7/2}^{-1}$,$^{144}$Sm)(keV)&2645&2661&2688&2588&2707&2323\\
\hline
E$^{Res}$(keV)&245&261&288&188&307&-77\\
\hline
\hline
\multicolumn{7}{|c|}{E$^{Unp}$($\pi$d$_{5/2}^{-1}$,$\pi$g$_{7/2}^{-1}$,$^{146}$Gd) = 3570 keV}\\
\hline
E$^{\star}$($\pi$d$_{5/2}^{-1}$,$\pi$g$_{7/2}^{-1}$,$^{146}$Gd)(keV)&3815&3831&3858&3758&3877&3493\\
\hline
\hline
\end{tabular}\\ 
\end{center}\


\subsection{The $\pi$g$_{7/2}^{-2}$ multiplet}

As mentioned previously, the best nucleus to extract the $\pi^{-1}$$\pi^{-1}$ interaction would be $^{144}$Sm. Then, we first calculate the $\pi$g$_{7/2}^{-2}$ multiplet unperturbed energy in $^{144}$Sm. This multiplet has been identified by Rico~\cite{RICO} and the experimental data are taken from this work.

\begin{equation}
E^{Unp}(\pi g_{7/2}^{-2},^{144}Sm) = 2\times E'(\pi g_{7/2}^{-1},^{145}Eu) + \;\;
\begin{tabular}[tb]{|c|}
\hline
\multicolumn{1}{||c||}{-1} \\
\hline
+2 \\
\hline
-1 \\
\hline
\end{tabular} \;= \\ \nonumber
\end{equation}\
\begin{equation}
= (2 \times 330)\; +\; 2070 = 2730 \; keV
\end{equation}\

We have to repeat the same type of calculation in order to determine the $\pi$g$_{7/2}^{-2}$ multiplet unperturbed energy in $^{146}$Gd.

\begin{equation}
E^{Unp}(\pi g_{7/2}^{-2},^{146}Gd) = 2\times E'(\pi g_{7/2}^{-1},^{147}Tb) + \;\;
\begin{tabular}[tb]{|c|}
\hline
-1 \\
\hline
+2 \\
\hline
\multicolumn{1}{||c||}{-1} \\
\hline
\end{tabular} \;= \\ \nonumber
\end{equation}\
\begin{equation}
= (2 \times 719)\; +\; 2497 = 3935 \; keV
\end{equation}\

Now, assuming that the residual interactions in $^{144}$Sm and in $^{146}$Gd for the $\pi$g$_{7/2}^{-2}$ multiplet are the same, we can calculate the estimated energies (E$^{\star}$=E$^{Unp}$+E$^{Res}$) of the multiplet members in $^{146}$Gd. Results are shown in the next table and in Figure~[\ref{fig:posmult}].\\

\begin{center}
\begin{tabular}{|c|c|c|c|c|}
\hline
\hline
\multicolumn{5}{|c|}{E$^{Unp}$($\pi$g$_{7/2}^{-2}$,$^{144}$Sm) = 2730 keV}\\
\hline
J$^{\pi}$&0$^{+}$&2$^{+}$&4$^{+}$&6$^{+}$\\
\hline
E$^{Exp}$($\pi$g$_{7/2}^{-2}$,$^{144}$Sm)(keV)&2478&2800&3019&3079\\
\hline
E$^{Res}$(keV)&-252&70&289&349\\
\hline
\hline
\multicolumn{5}{|c|}{E$^{Unp}$($\pi$g$_{7/2}^{-2}$,$^{146}$Gd) = 3935 keV}\\
\hline
E$^{\star}$($\pi$g$_{7/2}^{-2}$,$^{146}$Gd)(keV)&3683&4005&4224&4284\\
\hline
\hline
\end{tabular}\\ 
\end{center}\ 



\subsection{The $\pi$s$_{1/2}$$\pi$d$_{5/2}^{-1}$ multiplet}

There is no previous identification of this multiplet in $^{146}$Gd. In order to identify possible candidates, first, we calculate the unperturbed multiplet energy in $^{146}$Gd and later we will calculate the residual interaction in $^{146}$Gd from a SDI and estimate the multiplet energies.

\begin{equation}
E^{Unp}(\pi s_{1/2},\pi d_{5/2}^{-1},^{146}Gd) = E'(\pi s_{1/2},^{147}Tb) + E'(\pi d_{5/2}^{-1},^{145}Eu) + \;\;
\begin{tabular}[tb]{|c|}
\hline
+1 \\
\hline
\multicolumn{1}{||c||}{-2} \\
\hline
+1 \\
\hline
\end{tabular} \;= \\ \nonumber
\end{equation}\
\begin{equation}
= 0\; + \;0\; +\; 3438 = 3438 \; keV
\end{equation}\

Once we have this value, we can estimate the multiplet energy in $^{146}$Gd by adding the residual interaction and taking into account the Coulomb interaction of the protons. Results are shown in the next table and in Figure~[\ref{fig:posmult}].\\

\begin{center}
\begin{tabular}{|c|c|c|c|c|c|c|c|c|}
\hline
\hline
J$^{\pi}$&2$^{+}$&3$^{+}$\\
\hline
E$^{Res}$\small{({\it SDI} Calculation)}\normalsize{(keV)}&-34&171\\
\hline
Coulomb Interaction (keV)&-300&-300\\
\hline
E$^{\star}$($\pi$s$_{1/2}$,$\pi$d$_{5/2}^{-1}$,$^{146}$Gd)(keV)&3104&3309\\
\hline
\hline
\end{tabular}\\ 
\end{center}


\subsection{The $\pi$s$_{1/2}$$\pi$g$_{7/2}^{-1}$ multiplet}

There is no previous identification of this multiplet in $^{146}$Gd. In order to identify possible candidates, first, we calculate the unperturbed multiplet energy in $^{146}$Gd and later we will calculate the residual interaction in $^{146}$Gd from a SDI and estimate the multiplet energies.

\begin{equation}
E^{Unp}(\pi s_{1/2},\pi g_{7/2}^{-1},^{146}Gd) = E'(\pi s_{1/2},^{147}Tb) + E'(\pi g_{7/2}^{-1},^{145}Eu) + \;\;
\begin{tabular}[tb]{|c|}
\hline
+1 \\
\hline
\multicolumn{1}{||c||}{-2} \\
\hline
+1 \\
\hline
\end{tabular} \;= \\ \nonumber
\end{equation}\
\begin{equation}
= 0\; + \;330\; +\; 3438 = 3768 \; keV
\end{equation}\

Once we have this value we can estimate the multiplet energy in $^{146}$Gd by adding the residual interaction and taking into account the Coulomb interaction of the protons. Results are shown in the next table and in Figure~[\ref{fig:posmult}].\\

\begin{center}
\begin{tabular}{|c|c|c|c|c|c|c|c|c|}
\hline
\hline
J$^{\pi}$&3$^{+}$&4$^{+}$\\
\hline
E$^{Res}$\small{({\it SDI} Calculation)}\normalsize{(keV)}&171&19\\
\hline
Coulomb Interaction (keV)&-300&-300\\
\hline
E$^{\star}$($\pi$s$_{1/2}$,$\pi$g$_{7/2}^{-1}$,$^{146}$Gd)(keV)&3639&3487\\
\hline
\hline
\end{tabular}\\ 
\end{center}


\subsection{The $\pi$h$_{11/2}$$\pi$d$_{5/2}^{-1}$ multiplet}

The best nucleus to extract this $\pi$$\pi^{-1}$ interaction is $^{146}$Gd itself. However, this multiplet has been identified in $^{146}$Gd and in $^{144}$Sm, and it is interesting to compare both results and extract an idea of how good our assumption of a constant residual interaction is. Then, we first calculate the $\pi$h$_{11/2}$$\pi$d$_{5/2}^{-1}$ multiplet unperturbed energy in $^{144}$Sm.

\begin{equation}
E^{Unp}(\pi h_{11/2},\pi d_{5/2}^{-1},^{144}Sm) = E'(\pi h_{11/2},^{145}Eu) + E'(\pi d_{5/2}^{-1},^{143}Pm) + \;\;
\begin{tabular}[tb]{|c|}
\hline
\multicolumn{1}{||c||}{} \\
\hline
+1 \\
\hline
-2 \\
\hline
+1 \\
\hline
\end{tabular} \;= \\ \nonumber
\end{equation}\
\begin{equation}
= 716\; + \;0\; +\; 2979 = 3695 \; keV
\end{equation}\

We have to repeat the same type of calculation in order to determine the $\pi$h$_{11/2}$$\pi$d$_{5/2}^{-1}$ multiplet unperturbed energy in $^{146}$Gd.

\begin{equation}
E^{Unp}(\pi h_{11/2},\pi d_{5/2}^{-1},^{146}Gd) = E'(\pi h_{11/2},^{147}Tb) + E'(\pi d_{5/2}^{-1},^{145}Eu) + \;\;
\begin{tabular}[tb]{|c|}
\hline
+1 \\
\hline
\multicolumn{1}{||c||}{-2} \\
\hline
+1 \\
\hline
\end{tabular} \;= \\ \nonumber
\end{equation}\
\begin{equation}
= 51\; + \;0\; +\; 3438 = 3489 \; keV
\end{equation}\

Now, we can compare both residual interactions. This multiplet has been identified in $^{144}$Sm by Rico~\cite{RICO} and the experimental data are taken from this work. Results are shown in the next table and in Figure~[\ref{fig:negmult}].\\

\begin{center}
\begin{tabular}{|c|c|c|c|c|c|c|}
\hline
\hline
\multicolumn{7}{|c|}{E$^{Unp}$($\pi$h$_{11/2}$,$\pi$d$_{5/2}^{-1}$,$^{144}$Sm) = 3695 keV}\\
\hline
J$^{\pi}$&3$^{-}$&4$^{-}$&5$^{-}$&6$^{-}$&7$^{-}$&8$^{-}$\\
\hline
E$^{Exp}$($\pi$h$_{11/2}$,$\pi$d$_{5/2}^{-1}$,$^{144}$Sm)(keV)&1810&3118&2826&3266&3124&3519\\
\hline
E$^{Res}$($\pi$h$_{11/2}$,$\pi$d$_{5/2}^{-1}$,$^{144}$Sm)(keV)&-1885&-577&-869&-429&-571&-176\\
\hline
\hline
\multicolumn{7}{|c|}{E$^{Unp}$($\pi$h$_{11/2}$,$\pi$d$_{5/2}^{-1}$,$^{146}$Gd) = 3489 keV}\\
\hline
E$^{Exp}$($\pi$h$_{11/2}$,$\pi$d$_{5/2}^{-1}$,$^{146}$Gd)(keV)&1579&2997&2658&3099&2982&3183\\
\hline
E$^{Res}$($\pi$h$_{11/2}$,$\pi$d$_{5/2}^{-1}$,$^{146}$Gd)(keV)&-1910&-492&-831&-390&-507&-306\\
\hline
\hline
\end{tabular}
\end{center}

~\\


\subsection{The $\pi$d$_{3/2}$$\pi$d$_{5/2}^{-1}$ multiplet}
~\\

Only the 4$^{+}$ member of this multiplet has been previously identified in $^{146}$Gd. In order to identify the other members, first, we calculate the unperturbed multiplet energy in $^{146}$Gd and later we will calculate the residual interaction in $^{146}$Gd from a SDI and estimate the multiplet energies.

\begin{equation}
E^{Unp}(\pi d_{3/2},\pi d_{5/2}^{-1},^{146}Gd) = E'(\pi d_{3/2},^{147}Tb) + E'(\pi d_{5/2}^{-1},^{145}Eu) + \;\;
\begin{tabular}[tb]{|c|}
\hline
+1 \\
\hline
\multicolumn{1}{||c||}{-2} \\
\hline
+1 \\
\hline
\end{tabular} \;= \\ \nonumber
\end{equation}\
\begin{equation}
= 253\; + \;0\; +\; 3438 = 3681 \; keV
\end{equation}\

Once we have this value we can estimate the multiplet energy in $^{146}$Gd by adding the residual interaction and taking into account the Coulomb interaction of the protons. Results are shown in the next table and in Figure~[\ref{fig:posmult}].\\

\begin{center}
\begin{tabular}{|c|c|c|c|c|c|c|c|c|}
\hline
\hline
J$^{\pi}$&1$^{+}$&2$^{+}$&3$^{+}$&4$^{+}$\\
\hline
E$^{Res}$\small{({\it SDI} Calculation)}\normalsize{(keV)}&308&93&64&25\\
\hline
Coulomb Interaction (keV)&-300&-300&-300&-300\\
\hline
E$^{\star}$($\pi$d$_{3/2}$,$\pi$d$_{5/2}^{-1}$,$^{146}$Gd)(keV)&3689&3474&3445&3406\\
\hline
\hline
\end{tabular}\\ 
\end{center}

~\\
~\\


\subsection{The $\pi$h$_{11/2}$$\pi$g$_{7/2}^{-1}$ multiplet}

Similar to the $\pi$h$_{11/2}$$\pi$d$_{5/2}^{-1}$ case, some members of this multiplet have been identified in $^{144}$Sm. We will again compare the residual interactions in $^{144}$Sm and $^{146}$Gd as a consistency check.

\begin{equation}
E^{Unp}(\pi h_{11/2},\pi g_{7/2}^{-1},^{144}Sm) = E'(\pi h_{11/2},^{145}Eu) + E'(\pi g_{7/2}^{-1},^{143}Pm) + \;\;
\begin{tabular}[tb]{|c|}
\hline
\multicolumn{1}{||c||}{} \\
\hline
+1 \\
\hline
-2 \\
\hline
+1 \\
\hline
\end{tabular} \;= \\ \nonumber
\end{equation}\
\begin{equation}
= 716\; + \;272\; +\; 2979 = 3967 \; keV
\end{equation}\

We have to repeat the same type of calculation in order to determine the $\pi$h$_{11/2}$$\pi$g$_{7/2}^{-1}$ multiplet unperturbed energy in $^{146}$Gd.

\begin{equation}
E^{Unp}(\pi h_{11/2},\pi g_{7/2}^{-1},^{146}Gd) = E'(\pi h_{11/2},^{147}Tb) + E'(\pi g_{7/2}^{-1},^{145}Eu) + \;\;
\begin{tabular}[tb]{|c|}
\hline
+1 \\
\hline
\multicolumn{1}{||c||}{-2} \\
\hline
+1 \\
\hline
\end{tabular} \;= \\ \nonumber
\end{equation}\
\begin{equation}
= 51\; + \;330\; +\; 3438 = 3819 \; keV
\end{equation}\

Now, we can compare the residual interactions in $^{146}$Gd and $^{144}$Sm. This multiplet has been identified in $^{144}$Sm by Rico~\cite{RICO} and the experimental data are taken from this work. Results are shown in the next table and in Figure~[\ref{fig:negmult}].

\begin{center}
\begin{tabular}{|c|c|c|c|c|c|c|c|c|}
\hline
\hline
\multicolumn{9}{|c|}{E$^{Unp}$($\pi$h$_{11/2}$,$\pi$g$_{7/2}^{-1}$,$^{144}$Sm) = 3967 keV}\\
\hline
J$^{\pi}$&2$^{-}$&3$^{-}$&4$^{-}$&5$^{-}$&6$^{-}$&7$^{-}$&8$^{-}$&9$^{-}$\\
\hline
E$^{Exp}$($\pi$h$_{11/2}$,$\pi$g$_{7/2}^{-1}$,$^{144}$Sm)(keV)&-&3360&3307&3469&3534&3444&3376&3460\\
\hline
E$^{Res}$($\pi$h$_{11/2}$,$\pi$g$_{7/2}^{-1}$,$^{144}$Sm)(keV)&-&-607&-660&-498&-433&-523&-600&-507\\
\hline
\hline
\multicolumn{9}{|c|}{E$^{Unp}$($\pi$h$_{11/2}$,$\pi$g$_{7/2}^{-1}$,$^{146}$Gd) = 3819 keV}\\
\hline
E$^{Exp}$($\pi$h$_{11/2}$,$\pi$g$_{7/2}^{-1}$,$^{146}$Gd)(keV)&-&3389&3389&3313&3384&3290&3294&3428\\
\hline
E$^{Res}$($\pi$h$_{11/2}$,$\pi$g$_{7/2}^{-1}$,$^{146}$Gd)(keV)&-&-430&-430&-506&-435&-529&-525&-391\\
\hline
\hline
\end{tabular}\\ 
\end{center}


\subsection{The $\pi$d$_{3/2}$$\pi$g$_{7/2}^{-1}$ multiplet}

In this case, there is no previous identification of this multiplet in $^{146}$Gd. In order to identify possible candidates, first, we calculate the unperturbed multiplet energy in $^{146}$Gd and later we will calculate the residual interaction in $^{146}$Gd from a SDI and estimate the multiplet energies.

\begin{equation}
E^{Unp}(\pi d_{3/2},\pi g_{7/2}^{-1},^{146}Gd) = E'(\pi d_{3/2},^{147}Tb) + E'(\pi g_{7/2}^{-1},^{145}Eu) + \;\;
\begin{tabular}[tb]{|c|}
\hline
+1 \\
\hline
\multicolumn{1}{||c||}{-2} \\
\hline
+1 \\
\hline
\end{tabular}\; = \\ \nonumber
\end{equation}\
\begin{equation}
= 253\; + \;330\; +\; 3438 = 4021 \; keV
\end{equation}\

Once we have this value, we can estimate the multiplet energy in $^{146}$Gd by adding the residual interaction and taking into account the Coulomb interaction of the protons. Results are shown in the next table and in Figure~[\ref{fig:posmult}].

\begin{center}
\begin{tabular}{|c|c|c|c|c|c|}
\hline
\hline
J$^{\pi}$&2$^{+}$&3$^{+}$&4$^{+}$&5$^{+}$\\
\hline
E$^{Res}$\small{({\it SDI} Calculation)}\normalsize{(keV)}&-59&131&-44&228\\
\hline
Coulomb Interaction (keV)&-300&-300&-300&-300\\
\hline
E$^{\star}$($\pi$d$_{3/2}$,$\pi$g$_{7/2}^{-1}$,$^{146}$Gd)(keV)&3660&3852&3677&3949\\
\hline
\hline
\end{tabular}\\ 
\end{center}


\section{Discussion of the multiplets}

As discussed in the introduction, the appearance of the energy gap at Z=64 in the single-particle energy spectrum (see Figure~[\ref{fig:gaps}]) is large enough to give $^{146}$Gd many of the features of a doubly closed shell nucleus. However, we have to keep in mind that the proton gap is not as large as the neutron gap at N=82. Therefore, most of the states below 3.5 MeV will correspond to two-proton excitations. Until now, we have either estimated or identified all the $\pi$$\pi$, $\pi$$\pi^{-1}$ and $\pi^{-1}$$\pi^{-1}$ multiplet energies considering that the valence protons can be in any of the available orbitals between Z=50 and Z=82. However, in order to have a clear overview of all expected states below 3.5 MeV, we have to consider:\

\begin{itemize}
\item[{\it a})] The (3$^{-}$$\times$2$^{+}$) and (3$^{-}$$\times$3$^{-}$) two-phonon multiplets. In Figure~[\ref{fig:negmult}] and~[\ref{fig:posmult}] these multiplets are shown at their unperturbed energy

\begin{equation}
E^{Unp}((3^{-}\times2^{+}),^{146}Gd) = E^{Exp}(3^{-},^{146}Gd) +  E^{Exp}(2^{+},^{146}Gd) = 3551\; keV
\end{equation}

\begin{equation}
E^{Unp}((3^{-}\times3^{-}),^{146}Gd) = 2\times E^{Exp}(3^{-},^{146}Gd) = 3159 \; keV
\end{equation}\

\item[{\it b})] We should have an idea about the lowest neutron excitations in $^{146}$Gd. The most clear cut identification of a $\nu$$\nu^{-1}$ excitation in $^{146}$Gd, comes from the $\beta$-decay of the 5$^{-}$ isomer in $^{146}$Tb. This decay proceeds in two ways

\begin{equation}
^{146}Tb\;  [\pi h_{11/2},\nu d_{3/2}^{-1}]_{5^{-}}\rightarrow \; ^{146}Gd\; [\pi h_{11/2},\pi d_{5/2}^{-1}]_{4^{-},\;  5^{-},\; 6^{-}}
\nonumber
\end{equation}\

and

\begin{equation}
^{146}Tb\;  [\pi h_{11/2},\nu d_{3/2}^{-1}]_{5^{-}}\rightarrow \; ^{146}Gd\; [\nu h_{9/2},\nu d_{3/2}^{-1}]_{4^{-},\; 5^{-},\; 6^{-}}
\nonumber
\end{equation}\\

These two Gamow-Teller transitions were observed in the $\beta$-decay of the 5$^{-}$ isomer in $^{146}$Tb by Styczen et al.~\cite{STYCZEN}. In this paper, they identified the states at 4720 and 4829 keV as the multiplet members of the ($\nu$h$_{9/2}$,$\nu$$d_{3/2}^{-1}$)$_{4^{-},5^{-}}$ configuration. In the same work, they have observed a 1296 keV $\gamma$-ray transition from the ($\nu$h$_{9/2}$,$\nu$$d_{3/2}^{-1}$)$_{4^{-}}$ state to the state at 3423 keV and assigned it as ($\nu$f$_{7/2}$,$\nu$$d_{3/2}^{-1}$)$_{3^{-}}$. The reason is that this transition is very close to the 1397.0 keV 9/2$^{-}$ to 7/2$^{-}$ transition in $^{147}$Gd identified as $\nu$h$_{9/2}$$\rightarrow$$\nu$f$_{7/2}$. Although this assignment looks very reasonable, when calculating the expected energy for this level a discrepancy arises; the experimental value suggested in~\cite{STYCZEN} is much lower (about 550 keV) than the expected level energy (E$^{\star}$).

\begin{equation}
E^{Unp}(\nu f_{7/2},\nu d_{3/2}^{-1},^{146}Gd) = E'(\nu f_{7/2},^{147}Gd) + E'(\nu d_{3/2}^{-1},^{145}Gd) + \;\;
\begin{tabular}[tb]{|c|}
\hline
+1 \\
\hline
\multicolumn{1}{||c||}{-2} \\
\hline
+1 \\
\hline
\end{tabular} \;= \\ \nonumber
\end{equation}\
\begin{equation}
= 0\; + \;27\; +\; 3878 = 3905 \; keV
\end{equation}\

\begin{equation}
E^{Res}(\nu f_{7/2},\nu d_{3/2}^{-1},^{146}Gd)\small{({\it SDI} Calculation)}= 65 \; keV
\end{equation}

\begin{equation}
E^{\star}(\nu f_{7/2},\nu d_{3/2}^{-1},^{146}Gd) = 3905 + 65 =3970 \;keV
\end{equation}\

\begin{center}
\begin{tabular}{|c|c|c|c|c|c|}
\hline
\hline
E$^{\star}$($\nu$f$_{7/2}$,$\nu$d$_{3/2}^{-1}$,$^{146}$Gd)&3970 keV\\
\hline
E$^{Exp}$($\nu$f$_{7/2}$,$\nu$d$_{3/2}^{-1}$,$^{146}$Gd)&3423 keV\\
\hline
\hline
\end{tabular}\\ 
\end{center}

It sounds unreasonable that the discrepancies in the calculated SDI residual interactions are so large. As we have noted, the level was tentatively assigned to the $\nu$f$_{7/2}$$\nu$d$_{3/2}^{-1}$ configuration in~\cite{STYCZEN} and as was stated in that work, ``the excitation is somewhat too low to be compatible with the N=82 energy gap''. As we will see later in the negative-parity multiplet discussion section, we prefer to assign the level at 3423.2 keV to the (3$^{-}$$\times$2$^{+}$) multiplet. Thus, we will assume in the present work that there are no neutron excitations below 3.5 MeV.
\end{itemize}

An important remark, before we go into the multiplets discussion and possible assignments, is that the number of expected levels below 3.55 MeV shown in Figure~[\ref{fig:negmult}] and~[\ref{fig:posmult}] is 38 and the number of experimentally observed levels is 38. Thus, the concordance is excellent.

\subsection{Negative-parity multiplets}

\begin{figure}[h]
\begin{center}
\includegraphics[width=13.cm,height=15.cm,angle=0.]{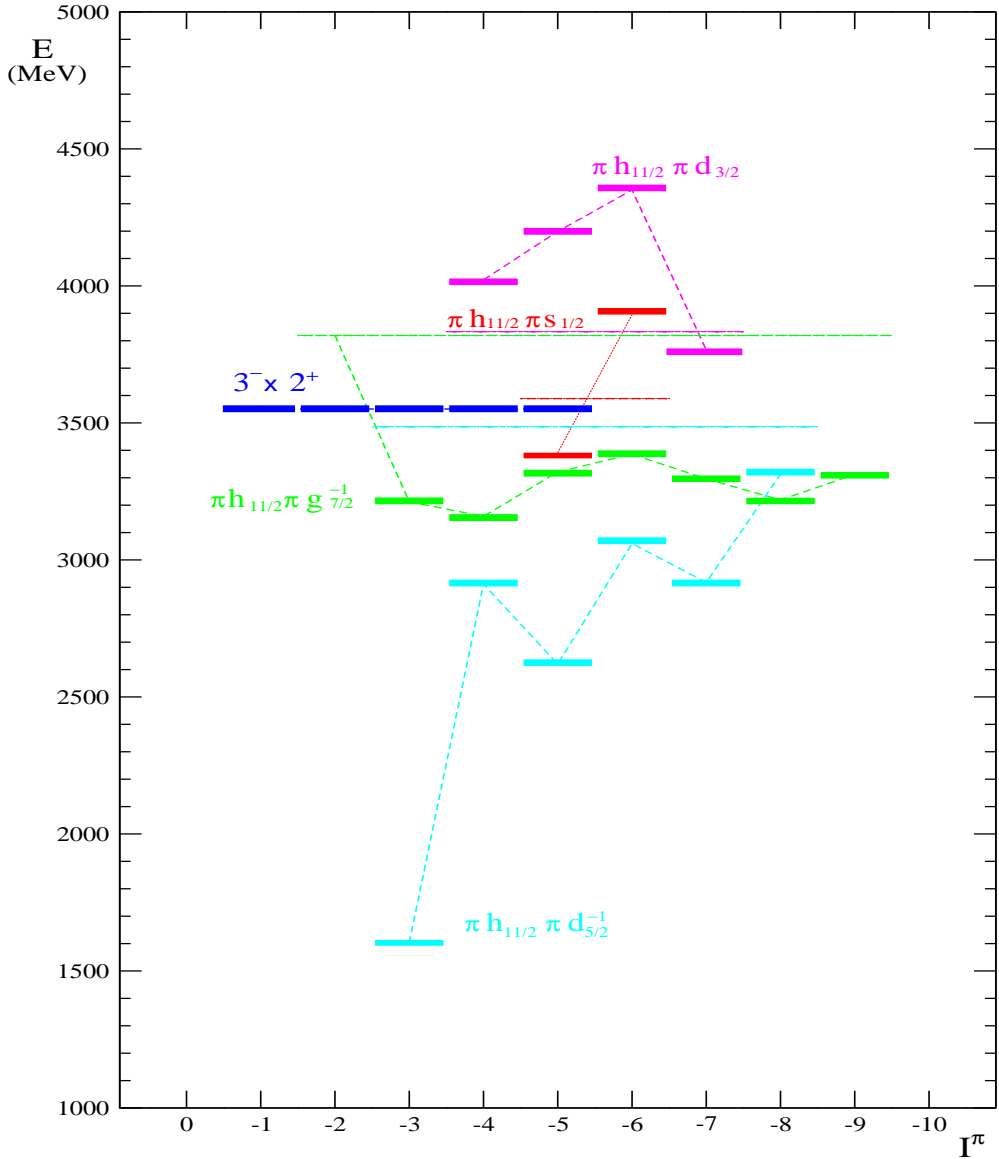}
\end{center}
\caption{Estimated energies of negative-parity two-proton multiplets in $^{146}$Gd. Multiplet members are shown with the same colour (dashed lines). Unperturbed energies are also shown with their corresponding colour. }
\label{fig:negmult}
\end{figure}

There are four multiplets expected with negative parity involving the $\pi$h$_{11/2}$ proton orbital and a fifth multiplet from the (3$^{-}$$\times$2$^{+}$) coupling. We will discuss the possible configuration assignments for the observed states in $^{146}$Gd (see Figure~[\ref{fig:negmult}]).\\

\begin{itemize}
\item[-] {\it 1$^{-}$ levels}. Only the 1$^{-}$ state at 3551 keV from the (3$^{-}$$\times$2$^{+}$) multiplet is expected.  There is no firm 1$^{-}$ level assignment in our results and the only possible candidate for this J$^{\pi}$ is the level at 3461.1 keV. The energy agreement is quite good and since no strong anharmonicities are expected in this multiplet member, we point to this level as the (3$^{-}$$\times$2$^{+}$) 1$^{-}$ member.
\item[-] {\it 2$^{-}$ levels}. Two 2$^{-}$ states are expected. One is the (3$^{-}$$\times$2$^{+}$) member expected at 3551 keV, while we do not know at what energy to expect the $\pi$h$_{11/2}$$\pi$g$_{7/2}^{-1}$ state, since we do not have experimental data of this level in $^{144}$Sm. But from SDI calculations we expect it at roughly the same energy as the other multiplet members. However, there are no firm 2$^{-}$  assignments from our data. There are possible 2$^{-}$ states at 3388.8 keV and at 3790 keV. In case they were the expected 2$^{-}$ levels, the most probable assignment is that the lower in energy corresponds to the $\pi$h$_{11/2}$$\pi$g$_{7/2}^{-1}$ multiplet, since the rest of the members lie about the same energy. Thus, the upper candidate would correspond to the (3$^{-}$$\times$2$^{+}$) multiplet.
\item[-] {\it 3$^{-}$ levels}. Three 3$^{-}$ states are expected and three candidates were found in the present work. The lowest in energy is the $\pi$h$_{11/2}$$\pi$d$_{5/2}^{-1}$ 3$^{-}$ state with a strong octupole component~\cite{KLEIN783-}. In addition to this level, we have identified a possible 3$^{-}$ state at 3388.7 keV and the firm 3$^{-}$ at 3423.2 keV, both known from Yates et al.~\cite{YATES}. The second was tentatively assigned to the $\nu$f$_{7/2}$$\nu$d$_{3/2}^{-1}$ configuration in~\cite{STYCZEN} but as was noted in that work, ``the excitation is somewhat too low to be compatible with the N=82 energy gap''. Thus, we prefer to assign the $\pi$h$_{11/2}$$\pi$g$_{7/2}^{-1}$ configuration to the lowest level at 3388.7 keV and the (3$^{-}$$\times$2$^{+}$) multiplet to the level at 3423.2 keV. However, it should be noted that we have made this assignment considering only energy arguments.
\item[-] {\it 4$^{-}$ levels}. Four 4$^{-}$ states are expected. Only one firm 4$^{-}$ level assignment was possible from our data: the well-known $\pi$h$_{11/2}$$\pi$d$_{5/2}^{-1}$ 4$^{-}$ state at 2996.6 keV which is directly populated in the $^{146}$Tb (5$^{-}$) decay~\cite{STYCZEN}. Levels assigned as 4 at 3363.8 and 3585 keV appear in our data. We think that the most probable assignment is that the lowest corresponds to the $\pi$h$_{11/2}$$\pi$g$_{7/2}^{-1}$ multiplet and the higher to the (3$^{-}$$\times$2$^{+}$) multiplet. No firm candidate was found for the $\pi$h$_{11/2}$$\pi$d$_{3/2}$ multiplet.
\item[-] {\it 5$^{-}$ levels}. Five 5$^{-}$ states are expected and five firm 5$^{-}$ states are assigned in the present work. The lowest expected 5$^{-}$ excited state is the well known $\pi$h$_{11/2}$$\pi$d$_{5/2}^{-1}$ 5$^{-}$ state at 2657.9 keV from~\cite{KLEIN783-}, which is directly populated in the $^{146}$Tb (5$^{-}$) decay~\cite{STYCZEN}. The next observed 5$^{-}$ state is at 3313.0 keV, which is very close in energy to the 3321 keV of the 5$^{-}$ member of the $\pi$h$_{11/2}$$\pi$g$_{7/2}^{-1}$ multiplet predicted in our calculations. This configuration was already assigned in~\cite{YATES}. Then, we have levels at 3464.0 and 3686.6 keV which can correspond to the $\pi$h$_{11/2}$$\pi$s$_{1/2}$ and (3$^{-}$$\times$2$^{+}$) multiplets, respectively. The last 5$^{-}$ state found is at 4230 keV, which is very close to the 4199 keV energy expected of the 5$^{-}$ member of the $\pi$h$_{11/2}$$\pi$d$_{3/2}$ multiplet. It should be noted that the agreement between the experimental and expected 5$^{-}$ levels is remarkable.
\item[-] {\it 6$^{-}$ levels}. Four 6$^{-}$ states are expected. The first two firm 6$^{-}$ states we find were assigned to the $\pi$h$_{11/2}$$\pi$d$_{5/2}^{-1}$ and $\pi$h$_{11/2}$$\pi$g$_{7/2}^{-1}$ multiplets. The first, at 3098.9 keV was assigned to that configuration in~\cite{STYCZEN} where this state was strongly populated. The second was assigned in~\cite{YATES} and we agree with this proposition since the energy of the level (at 3384.0 keV) fits extremely well with the 3386 keV predicted energy. Additionally, two possible 6$^{-}$ states were found at 4026.6 and 4179.4 keV. Each of them would fit quite well with the expected $\pi$h$_{11/2}$$\pi$s$_{1/2}$ member, but we cannot conclude anything. For the last 6$^{-}$ level expected at 4350 keV  and corresponding to the $\pi$h$_{11/2}$$\pi$d$_{3/2}$ multiplet, we make a speculative assignment to this multiplet for the possible 6$^{-}$ level at 4318.8 keV.   
\item[-] {\it 7$^{-}$ levels}. Three 7$^{-}$ levels are predicted and three firm 7$^{-}$ states are found in excellent agreement with the expected energies. The lowest in energy is the level at 2982.0 keV, identified in~\cite{KLEIN783-} as the $\pi$h$_{11/2}$$\pi$d$_{5/2}^{-1}$ multiplet member. The second 7$^{-}$ state is at 3290.5 keV and was identified in~\cite{YATES} as the $\pi$h$_{11/2}$$\pi$g$_{7/2}^{-1}$ multiplet member. Finally, the 7$^{-}$ state at 3854.0 keV was assigned in~\cite{YATES} to the $\pi$h$_{11/2}$$\pi$d$_{3/2}$ multiplet expected at 3766 keV. We agree with all the above mentioned configuration assignments. 
\item[-] {\it 8$^{-}$ levels}. Two 8$^{-}$ states were expected. The $\pi$h$_{11/2}$$\pi$d$_{5/2}^{-1}$ multiplet member predicted to be at 3313 keV was assigned in~\cite{KLEINAR} to the level seen at 3182 keV. The other 8$^{-}$ state at 3293.7 keV corresponds to the $\pi$h$_{11/2}$$\pi$g$_{7/2}^{-1}$ multiplet member (first assignment in~\cite{KLEINAR}) predicted at 3219 keV. One might think that this assignment is inverted, but the estimated energies in $^{146}$Gd are calculated from the residual interaction extracted from the experimental data in $^{144}$Sm. Initially, in $^{144}$Sm the $\pi$h$_{11/2}$$\pi$d$_{5/2}^{-1}$ 8$^{-}$ state was expected at lower energy than the $\pi$h$_{11/2}$$\pi$g$_{7/2}^{-1}$ state, but the experimental data inverted the order, and this inversion is transmitted to the $^{146}$Gd multiplet energy prediction. In the case of $^{146}$Gd the 8$^{-}$ member of the $\pi$h$_{11/2}$$\pi$g$_{7/2}^{-1}$ multiplet is clearly identified through the strong 134.8 keV transition from the $\pi$h$_{11/2}$$\pi$g$_{7/2}^{-1}$ 9$^{-}$ level at 3428.5 keV to the 3293.7 keV level, which is 81 times stronger than the de-excitation to the lower 8$^{-}$ level at 3182.0 keV. To the contrary, in $^{144}$Sm, the lowest 8$^{-}$ state at 3376 keV has been identified as the $\pi$h$_{11/2}$$\pi$g$_{7/2}^{-1}$ member because a pure {\it M}1 transition has been observed from the 9$^{-}$ level at 3460 keV to this state.
\item[-] {\it 9$^{-}$ levels}. The $\pi$h$_{11/2}$$\pi$g$_{7/2}^{-1}$ 9$^{-}$ member is expected at 3312 keV and a firm 9$^{-}$ level is observed at 3428.5 keV. This assignment was made in~\cite{KLEINAR} and we are in agreement with it.
\end{itemize}

\subsection{Positive-parity multiplets}

\begin{figure}[h]
\begin{center}
\includegraphics[width=13.cm,height=15.cm,angle=0.]{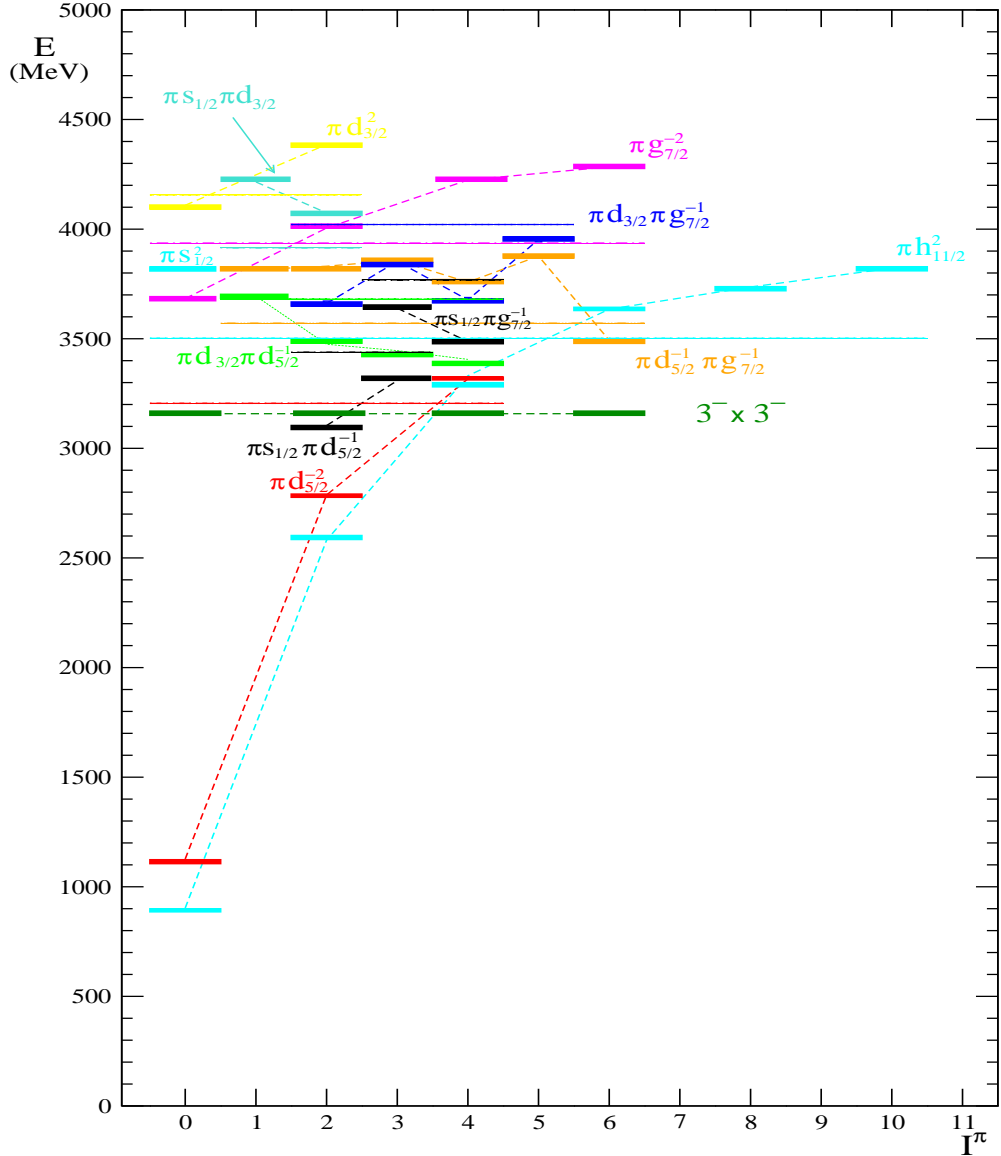}
\end{center}
\caption{Estimated energies of positive-parity two-proton multiplets in $^{146}$Gd. Multiplet members are shown with the same colour (dashed lines). Unperturbed energies are also shown with their corresponding colour.}
\label{fig:posmult}
\end{figure}

The configuration assignments are harder in the case of the positive-parity multiplets than for the negative-parity states. Twelve multiplets can be formed with the five available orbitals, including the 3$^{-}$$\times$3$^{-}$ two-phonon octupole multiplet. As can be seen in Figure~[\ref{fig:posmult}] the density of expected levels above 3.4 MeV complicates the level identification.

\begin{itemize}
\item[-] {\it 0$^{+}$ levels}. In addition to the ground state, 0$^{+}$ levels have been identified in (p,t) reaction experiments at 2165 and 3020 keV(~\cite{FLYNN} and~\cite{MANN}). The large cross section observed in the (p,t) reaction~\cite{FLYNN} for the state at 3020 keV, clearly identifies it as the neutron pairing vibrational state. Based on energy arguments, the lower 0$^{+}$ state at 2165 keV was identified as the proton pairing vibrational state~\cite{JULIN}. These two levels and the 0$^{+}$ states at 3484 and 3640 keV were clearly identified as 0$^{+}$ states in the conversion-electron measurements from~\cite{JULIN} and~\cite{YATES87}. They were speculatively assigned in~\cite{YATES} as $\pi$g$_{7/2}^{-2}$ and $\pi$s$_{1/2}^{-2}$, respectively. The lower in energy was discarded as the 3$^{-}$$\times$3$^{-}$ 0$^{+}$ state because it should present small anharmonicities and is expected to lie at 3159 keV. An alternative assignment is that the level at 3484 keV corresponds to the two-phonon octupole multiplet and the level at 3640 keV to the $\pi$g$_{7/2}^{-2}$ multiplet. In this case, the 0$^{+}$ and 6$^{+}$ members of the 3$^{-}$$\times$3$^{-}$ multiplet would lie at almost the same energy. However, this assignment is also speculative and based on level order arguments. No additional 0$^{+}$ states were found in the present work, in agreement with the conversion-electron experiments.
\item[-] {\it 1$^{+}$ levels}. Three 1$^{+}$ states are expected at relatively high energy, where the J$^{\pi}$ assignments are difficult. We do not have any firm 1$^{+}$ assignment. There is a possible 1$^{+}$ level at 3730 keV that could belong either to the $\pi$d$_{3/2}$$\pi$d$_{5/2}^{-1}$ or to the $\pi$d$_{5/2}^{-1}$$\pi$g$_{7/2}^{-1}$ multiplet.
\item[-] {\it 2$^{+}$ levels}. The 2$^{+}$ state at 1972.0 keV was identified as the $\pi$s$_{1/2}$$\pi$d$_{5/2}^{-1}$ multiplet member. In addition, there are seven firm 2$^{+}$ levels. It is worth noting the two different decay patterns that we observe in these states. The levels at 2986, 3356.7 and 3547.5 keV decay either only or mainly to the ground state while the levels at 3185.8, 3232.2 and 3380.7 keV decay mainly to the 3$^{-}$ state by {\it E}1 transitions. Based on these decay patterns, we assign the 2986 keV state to the $\pi$h$_{11/2}^{-2}$ multiplet and the level at 3356.7 keV to the $\pi$d$_{5/2}^{-2}$ multiplet. The level at 3547.5 keV would correspond to the $\pi$d$_{3/2}$$\pi$g$_{7/2}^{-1}$ multiplet. We consider both the level at 3185.8 keV and the one at 3232.2 keV as good candidates for the (3$^{-}$$\times$3$^{-}$) 2$^{+}$ state. Finally, there is a firm 2$^{+}$ state at 4299.6 keV, which could be the $\pi$d$_{3/2}^{-2}$ multiplet member expected at 4386 keV.
\item[-] {\it 3$^{+}$ levels}. Five 3$^{+}$ levels are expected, but we only have firm 3$^{+}$ assignments for three. The lowest in energy is the level at 3031.2 keV, identified in~\cite{YATES} as the $\pi$s$_{1/2}$$\pi$d$_{5/2}^{-1}$. Then, there is a firm 3$^{+}$ level at 3287.2 keV that can be assigned either to the $\pi$d$_{3/2}$$\pi$d$_{5/2}^{-1}$ or to the $\pi$s$_{1/2}$$\pi$g$_{7/2}^{-1}$ multiplet. Finally, we have found a probable 3$^{+}$ level at 3783.6 keV, which again can correspond to one of the two levels expected at about 3850 keV, namely the $\pi$d$_{3/2}$$\pi$g$_{7/2}^{-1}$ and $\pi$d$_{5/2}^{-1}$$\pi$g$_{7/2}^{-1}$ multiplets members. 
\item[-] {\it 4$^{+}$ levels}. There are five firm 4$^{+}$ states seen in the present work. The level at 2611.5 keV was assigned to the $\pi$d$_{3/2}$$\pi$d$_{5/2}^{-1}$ multiplet in~\cite{OGAWA}. The next level is at 2967.4 keV, and in~\cite{YATES} it was suggested to be the (3$^{-}$$\times$3$^{-}$) 4$^{+}$ state based on energy reasons since it is the closest to the predicted energy. Then, four levels are predicted at energies below 3.7 MeV and four firm 4$^{+}$ levels were found in this work. The level at 3411.8 keV is assigned to the $\pi$h$_{11/2}^{-2}$ multiplet, the level at 3416.5 keV assigned to the $\pi$d$_{5/2}^{-2}$ multiplet, a level at 3436.2 keV corresponding to $\pi$s$_{1/2}$$\pi$g$_{7/2}^{-1}$, and the level at 3456.5 keV is assigned to the $\pi$d$_{3/2}$$\pi$g$_{7/2}^{-1}$ multiplet. It should be noted that these assignments are somewhat arbitrary and based on the energy ordering of the levels. However, the fact that we see the same number of 4$^{+}$ levels as expected theoretically tells us that we have seen the two-phonon octupole 4$^{+}$ member.
\item[-] {\it 5$^{+}$ levels}. The 5$^{+}$ levels are expected at relatively high energies. There are 5$^{+}$ states expected at 3877 and 3953 keV. We see possible 5$^{+}$ states at 4131, 4326 and 4389.5 keV. The first could correspond to the $\pi$d$_{3/2}$$\pi$g$_{7/2}^{-1}$ multiplet. The others seem to be too far away from the predicted energy of the $\pi$d$_{5/2}^{-1}$$\pi$g$_{7/2}^{-1}$ multiplet member.
\item[-] {\it 6$^{+}$ levels}. There are four 6$^{+}$ levels predicted and the same number is found in the present work. The first is the (3$^{-}$$\times$3$^{-}$) 6$^{+}$ state found at 3484.7 keV, which decays by a cascade of two {\it E}3 transitions to the ground state and will be extensively discussed in the next chapter. The 6$^{+}$ state at 3457.6 keV is assigned to the $\pi$d$_{5/2}^{-1}$$\pi$g$_{7/2}^{-1}$ multiplet. The 6$^{+}$ state at 3659.6 keV is assigned to the $\pi$h$_{11/2}^{-2}$ configuration. The last level found with probable 6$^{+}$ assignment is at 4354.9 keV, which is very close to the predicted for the $\pi$g$_{7/2}^{-2}$ multiplet. This assignment differs from that by Yates et al.~\cite{YATES}.
\item[-] {\it 7$^{+}$ levels}. There are no 7$^{+}$ levels expected below 4 MeV and none has been found in the present work, which is an indication of the quality of our data and the correctness of our analysis.
\item[-] {\it 8$^{+}$ levels}. The $\pi$h$_{11/2}^{-2}$ 8$^{+}$ member is expected at 3736 keV and a firm 8$^{+}$ state was found at 3779.2 keV and assigned to this configuration in~\cite{YATES}.
\item[-] {\it 9$^{+}$ levels}. There are no 9$^{+}$ levels expected below 4 MeV and none has been found in the present work.
\item[-] {\it 10$^{+}$ levels}. The 10$^{+}$ level found at 3864.8 keV was assigned to the $\pi$h$_{11/2}^{-2}$ 10$^{+}$ member in ~\cite{KLEINAR}. This assignment is in accord with the expected energy of 3822 keV. We have found an additional firm 10$^{+}$ level at 4541.2 keV, which probably belongs to a four-particle configuration since it is at high energy an no other two-particle 10$^{+}$ level is predicted there.
\end{itemize}


\chapter{Two-Phonon Excitations in $^{146}$Gd}

{\it  In this chapter the two-phonon multiplet landscape and candidates of its members will be discussed. The results of the first observation of a 6$^{+}$$\rightarrow$3$^{-}$$\rightarrow$0$^{+}$ double E3 cascade in the decay of a two-phonon octupole state are presented.}
  
\vspace*{0.6cm}
\section{ Two-phonon multiplet previous knowledge}

Since the 3$^{-}$ first excited state in $^{208}$Pb was first identified and interpreted as a collective octupole phonon state, many attempts have been made to find the members of the anticipated even-parity 0,2,4,6 two-phonon octupole quartet. This quest has not ended; in particular, it has not been possible to find the aligned (3$^{-}$$\times$3$^{-}$)6$^{+}$ member of the quartet. The experimental situation is, however, somewhat more favourable in the case of $^{146}$Gd, the only other even-even nucleus known to have a 3$^{-}$ first excited state. Since the angular momentum transferred in the $^{144}$Sm($\alpha$,2n) fusion-evaporation reaction is low, it leads to the population of a range of low- to intermediate-spin states lying above the yrast line in $^{146}$Gd. As we have often mentioned in this work, twenty years ago this reaction was used to study non-yrast states and in particular to search for the double-octupole excitations. They are expected at twice the energy of the one-phonon 3$^{-}$ state for the 0$^{+}$, 2$^{+}$ and 4$^{+}$ multiplet members, while the 6$^{+}$ member is expected at 3568 keV~\cite{LUNARDI}. This slightly higher energy is due to the Pauli principle for the mayor component of the 3$^{-}$ configuration, namely $\pi$h$_{11/2}$$\pi$d$^{-1}$$_{5/2}$. In these experiments two Ge(Li) detectors were used to record $\gamma$-$\gamma$ coincidences and $\gamma$-ray angular distributions. In a separate measurement with a broad-range electron spectrometer, conversion coefficients were obtained. These experiments resulted in a substantial extension of the level scheme. Amongst the newly observed states two possible 6$^{+}$ states were identified (3457 and 3484 keV) and interpreted as the two-phonon octupole quartet member and the nearby expected ( $\pi${\it d}$^{-1}$$_{5/2}$,$\pi${\it g}$^{-1}$$_{7/2}$) two-proton hole state, but the data could not distinguish conclusively between the two assignments. As we have mentioned in the first chapter, great strides have been made in methods of $\gamma$-ray detection, so we repeated this experiment with a modern array of large volume Ge detectors.\\

\section{ 0$^{+}$ Two-phonon multiplet member}  

As mentioned in Chapter 4, the lowest 0$^{+}$ state after the two pairing vibrational states is at 3484 keV, which is far away from the expected 3159 keV of the two-phonon octupole member. This level was discarded as the (3$^{-}$$\times$3$^{-}$) member~\cite{YATES} because small anharmonicities where expected in this state together with the fact that no {\it E}3 transition to the 3$^{-}$ state was observed. However, it seems that all the 0$^{+}$ levels below that level are pushed up in energy. In addition, as we will discuss later, we have identified the two-phonon octupole 6$^{+}$ state at 3484.7 keV, i.e., at the same energy. In that case we see a weak transition to the 3$^{-}$ state, with a big uncertainty in the intensity, so it cannot be discarded that the 0$^{+}$$\rightarrow$3$^{-}$ transition (of the same energy) could be contained in that peak. Of course, this assignment of the 3484 keV level as the two-phonon octupole member is speculative.

\section{ 2$^{+}$ Two-phonon multiplet member}  

We have found equally good candidates for the 2$^{+}$ two-phonon multiplet member at 3185.8 and 3232.2 keV that decay with similar patterns: a weak transition to the 2$^{+}$ state and a stronger {\it E}1 to the 3$^{-}$ state.

\section{ 4$^{+}$ Two-phonon multiplet member}  

In the previous $^{144}$Sm($\alpha$,2n) fusion-evaporation reaction work by Yates et al.~\cite{YATES} the 4$^{+}$ level at 2967.4 keV was assigned to the (3$^{-}$$\times$3$^{-}$) multiplet. This level de-excites by a strong {\it E}1 transition to the 3$^{-}$ state. We agree with this level assignment since we see the complete set of expected 4$^{+}$ levels (see Chapter 4).

\section{ 6$^{+}$ Two-phonon multiplet member}

\begin{figure}[ht]
\begin{center}
\includegraphics[width=11.cm,height=8.cm,angle=0.]{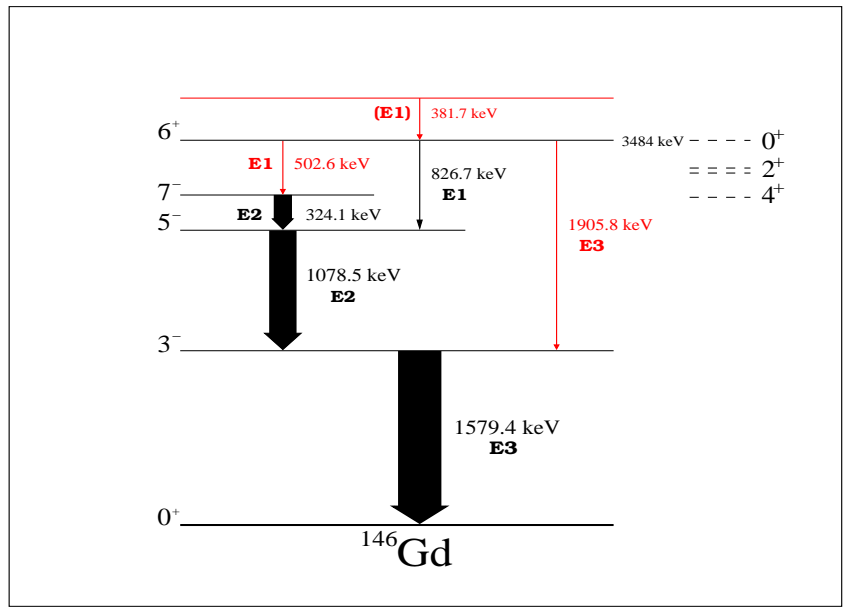}
\end{center}
\caption{A partial level scheme for $^{146}$Gd showing transitions and levels
related to the 3484 keV 6$^{+}$ state, which from the present experiment is identified as the 6$^{+}$ member of the double octupole quartet. The crucial new data for this assignment are shown in red. Previous knowledge is shown in black~\cite{YATES}. All transition multipolarities are measured (see also Table~[\ref{tab:tabla_gammas_11}]). 0$^{+}$, 2$^{+}$ and 4$^{+}$ double octupole candidates are shown in dashed lines.}
\label{fig:level}
\end{figure}

\begin{figure}[!h]
\begin{center}
\includegraphics[width=8.6cm,height=4.8cm,angle=0.]{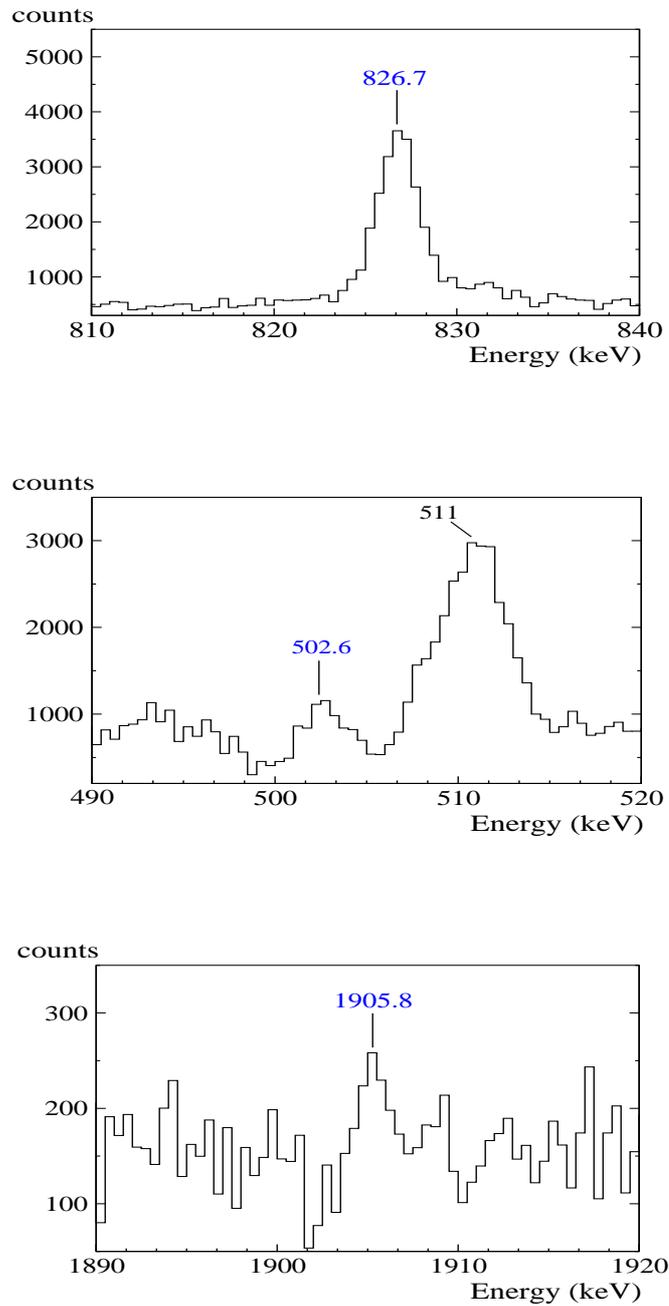}
\caption{Three sections of the gamma-ray spectrum in coincidence with the weak 381.7 keV transition feeding the 3484 keV 6$^{+}$ state. The marked in blue peaks correspond to the three gamma rays de-exciting the 3484 keV level.}
\label{fig:1905}
\end{center}
\end{figure}

A 6$^{+}$ level at 3484 keV was observed in the previous work~\cite{YATES}. It was known to decay by a stretched {\it E}1 transition of 826.7 keV to the yrast 5$^{-}$ state and thus was assigned as a possible 6$^{+}$ state (see Figure~[\ref{fig:level}]). In our work we have observed a new 502.6 keV transition from this level to the yrast 7$^{-}$ state. This $\gamma$-ray has a negative anisotropy (a$_{2}$$<$0) and positive polarization (see Table ~[\ref{tab:tabla_gammas_11}]), which classifies it unequivocally as a stretched {\it E}1 transition. This additional {\it E}1 transition confirms the 6$^{+}$ assignment. Furthermore, we also observed a third $\gamma$-ray of 1905.8 keV (with positive anisotropy and polarization) feeding the 3$^{-}$ octupole state. If these three $\gamma$-rays de-excite the same level, we have identified a firmly assigned 6$^{+}$ state decaying by a cascade of two {\it E}3 transitions to the ground state; it would be clearly assigned as the 6$^{+}$ member of the double octupole quartet. Conclusive evidence came from the spectrum of $\gamma$-rays in coincidence with the weak 381.7 keV feeding transition (see Figure~[\ref{fig:1905}]), which clearly shows all three de-exciting transitions, thus demonstrating that they all arise from the same state at 3484 keV.\\\\

\

It should be noted that, for $\gamma$-rays  above 1 MeV de-exciting levels with half lives of less than one ps, the Doppler shifts of the emitted $\gamma$-rays were clearly seen in our experiments. No such shift was observed for the 1905.8 keV $\gamma$-ray. Taking for this transition the theoretical strength of 57 W.u. \cite{LUNARDI}, we calculated with the measured 5\% {\it E}3 intensity branching a level half life of 10 ps, in agreement with the non-observation of a Doppler shift. For the {\it E}1 transitions this result gives a retardation factor of the order of 5 $\times$ 10$^{-5}$, in accord with the general experimental systematics for {\it E}1 transitions \cite{ENDT}.\\\\
 
\newpage

\begin{figure}[!h]
\begin{center}
\includegraphics[width=8.6cm,height=4.8cm,angle=0.]{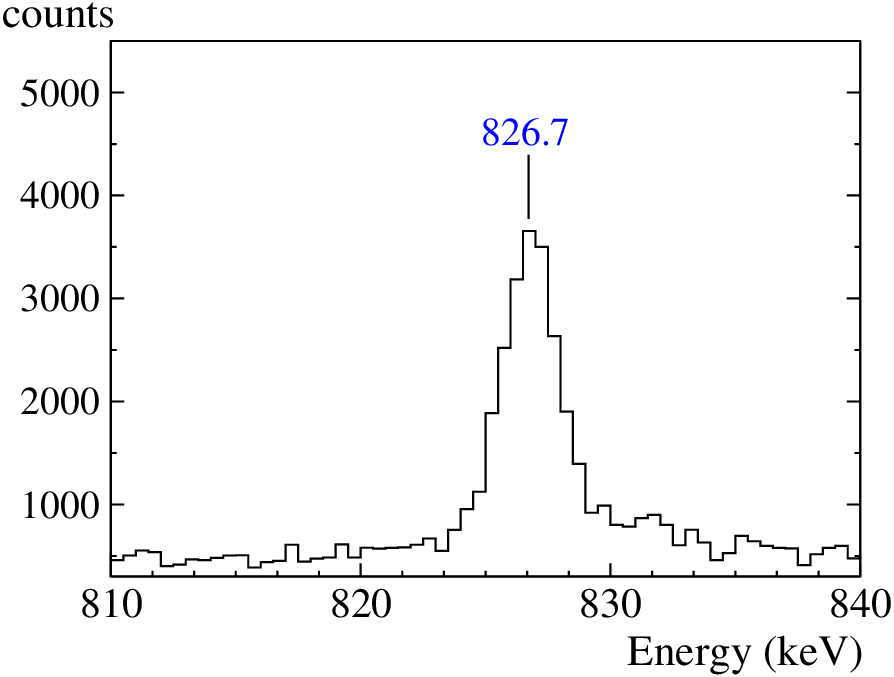}
\label{fig:827}
\end{center}
\end{figure}

\begin{figure}[!h]
\begin{center}
\includegraphics[width=8.6cm,height=4.8cm,angle=0.]{./figs/Deb_Proc_Fig_2.eps}
\label{fig:503}
\end{center}
\end{figure}

\

Finally it should be noted that in the previous work~\cite{YATES} a level at 3457 keV with possible J$^{\pi}$=6$^{+}$ was identified by its {\it E}1 decay to the 5$^{-}$ yrast state. Further, it was also suggested that there was a possible 1877 keV, {\it E}3 transition to the 3$^{-}$ state, thus making it a reasonable candidate for the double octupole state. However, the present results rule out {\it E}3 multipolarity since we observe a Doppler shift for the 1877 keV transition. If this does indeed de-excite the 3457 keV level, then it represents the decay of a 4$^{+}$ state, not 6$^{+}$. If the two transitions de-excite different levels, then the 6$^{+}$ assignment for the 3457 keV level is probably correct, but there is no {\it E}3 de-excitation to the 3$^{-}$ state.\

In summary, we have identified the 6$^{+}$ member of the two-phonon octupole quartet in $^{146}$Gd at 3484 keV by making a firm J$^{\pi}$ assignment for the level and then, more importantly, by identifying the {\it E}3 branch to the one phonon 3$^{-}$ state. This results present the first conclusive observation of a 6$^{+}$$\rightarrow$3$^{-}$$\rightarrow$0$^{+}$ double {\it E}3 cascade in the decay of a two-phonon octupole state. The reader should note that, although we have commented some candidates for the other members of the double octupole multiplet, only in the case of the 6$^{+}$ member we have a clear evidence of its (3$^{-}$$\times$3$^{-}$) nature.\\

\chapter{Conclusions}

{\it  The conclusions from this work will be exposed in this chapter as a summary.}
  
\vspace*{0.6cm}
\section{Conclusions}
In the present work, we have studied the $^{144}$Sm($\alpha$,2n) fusion-evaporation reaction using 26.3 MeV $\alpha$-particles impinging on 3.0 mg/cm$^{2}$ thick samarium target enriched to 97.6\% in $^{144}$Sm supported by a 0.5 mg/cm$^{2}$ thick Au backing. The target was surrounded by a compact array of nine individual Ge detectors set at 90, $\pm$ 45 and $\pm$ 35 degrees to the beam direction. Five of them had anti-Compton shields. In addition, a EUROBALL CLUSTER detector was placed at 90$^{o}$ to act as a non-orthogonal $\gamma$-ray Compton polarimeter.\\
In a previous preparatory $^{144}$Sm($\alpha$,2n) experiment~\cite{TESINA}, a total of 21 new $\gamma$-rays from 16 new levels were identified, as well as 19 new $\gamma$-rays corresponding to 13 previously known levels. Also, 7 $\gamma$-rays were seen for the first time in an in-beam experiment.\\ 
In the present work, a total of 35 new  $\gamma$-rays have been identified 
for the first time, corresponding to 28 new states
(44 if we include previous experiment)
as well as 31 new $\gamma$-rays corresponding to 26 previously known levels. Also, 3 $\gamma$-rays were seen for the first time in an in-beam experiment.\\
If we put together all the new information from both experiments, we can estimate the improvement of $^{146}$Gd knowledge that the two experiments have provided. In total, 56 new $\gamma$-rays from 44 new levels were identified as well as 50 new $\gamma$-rays corresponding to 39 previously known levels. Also, 10 $\gamma$-rays were seen for the first time in an in-beam experiment.\\ 
We have, in general, confirmed previous results and in only few cases modified them. The angular anisotropy and polarization data contributed to the identification of states and to the level spin and parity assignments. From our data, new candidates for the two-particle configurations have been found as well as for the (3$^{-}$$\times$2$^{+}$) and (3$^{-}$$\times$3$^{-}$) two-phonon multiplets, although clear evidence of its identification is reduced to the 6$^{+}$ case (see later). It is particularly remarkable that we find excellent agreement between the experiment and expectations of the number of levels below 3.55 MeV; 38 states were expected in the present experiment and we have identified 38 states below that energy.\\
A very important result is the unequivocal identification of the 6$^{+}$ member of the two-phonon octupole multiplet in $^{146}$Gd by identifying the {\it E}3 branching to the one-phonon 3$^{-}$ state. This results present the first conclusive observation of a 6$^{+}$$\rightarrow$3$^{-}$$\rightarrow$0$^{+}$ double {\it E}3 cascade in the decay of a two-phonon octupole state.\
 
The other double octupole member that could be identified by its characteristic decay pattern is the 0$^{+}$ state. This characteristic pattern will be an {\it E}3 transition to the 3$^{-}$ state as in the 6$^{+}$ case. The possible 0$^{+}$ member at 3484 keV is very close to the energy of the 6$^{+}$ member at 3484.7 keV. The level at 3484 keV is de-excited by an {\it E}2 transition of 1512 keV and 12 units of intensity (referred to the 10000 units of the 1579.4 keV $\gamma$-ray) to the first 2$^{+}$ state. If we place a gate on the 3$^{-}$ state, the possible {\it E}3 transition to the 3$^{-}$ will be a $\gamma$-ray of 1905 keV and will be obscured by the 1905.8 keV $\gamma$-ray that de-excites the 6$^{+}$ state. Consequently, the only possibility to observe this transition is to find a $\gamma$-ray feeding the 0$^{+}$ level and use it as the gating condition. Assuming 50 W.u. for the {\it E}3 transition and assuming a feeding transition of similar characteristics to the one feeding the 6$^{+}$ level, we estimate that we need about a factor 500 more in statistics than in the present work to be able to observe this effect. The only experimental future device that can achieve such improvement in efficiency is the planned 4$\pi$ detector array AGATA.\ 

Concerning the 2$^{+}$ and 4$^{+}$ multiplet members, we have proposed candidates but they cannot be conclusively assigned since more states of these spin and parity are expected in that region and, in addition, the double octupole members will not present a characteristic decay pattern for unequivocally identify them.

\appendix


\chapter{Level J$^{\pi}$ assignments.}\label{Appendix1}

\vspace*{0.6cm}

In this appendix we will discuss all the levels placed in the $^{146}$Gd level scheme shown in Chapter 3, which summarizes the analysis and results of the present experiment. We will give the arguments used to assign spins and parities to these levels. We will make use of the anisotropy and polarization data in order to determine the transition character and multipolarity and, thus, to assign spins and parities. We will also take into account the results from previous experiments. An example would be the (p,t) experiment by Mann et al.~\cite{MANN}, where states of two-neutron hole character in $^{146}$Gd are strongly populated, and there is a strong selection rule against populating states of unnatural parity. However, it should be noted that at high excitation energy and, due to the large uncertainty in the level energy extracted from that experiment, the correspondence with our data was not always unequivocal. \


\begin{itemize}
\item[-] {\it Level at 1579.4 keV}. First excited state in $^{146}$Gd and the octupole phonon. Its spin and parity (J$^{\pi}$=3$^{-}$) are well determined from previous work~\cite{KLEIN783-}. It decays to the ground state by an {\it E}3 transition. Our measured positive angular distribution and polarization data confirm the transition multipolarity and level assignment.
\item[-] {\it Level at 1972.0 keV}. 2$^{+}$ quadrupole phonon state in $^{146}$Gd. It decays to the ground state and shows a Doppler shift (its half life is less than 0.32 {\it ps}~\cite{TESINA}) as expected for an {\it E}2 transition of that energy. Its measured angular distribution is positive, which confirms the {\it E}2 transition multipolarity. 
\item[-] {\it Level at 2164.7 keV}. Proton pairing vibration identified in ~\cite{JULIN} and ~\cite{OGAWA}. It decays by a $\gamma$-ray of 192.7 keV to the 1972.0 keV 2$^{+}$ state. Since the transition is very weak, it was not possible to extract information about its angular distribution and polarization.  
\item[-] {\it Level at 2611.5 keV}. Well-known as 4$^{+}$. It decays by a 639.6 keV {\it E}2 $\gamma$-ray with positive angular distribution to the 1972.0 keV 2$^{+}$ state and by a 1032.0 keV {\it E}1 transition ({\it$\alpha_{k}$} from~\cite{YATES}) with negative angular distribution and positive polarization to the 1579.4 keV 3$^{-}$ state.    
\item[-] {\it Level at 2657.9 keV}. 5$^{-}$ yrast state. It is de-excited by a 1078.5 keV transition to the 1579.4 keV 3$^{-}$ state with positive angular distribution and positive polarization, as expected for an {\it E}2 ({\it$\alpha_{k}$} from~\cite{YATES}) transition.  
\item[-] {\it Level at 2967.4 keV}. The level is de-excited by a 1388.0 keV $\gamma$-ray to the 1579.4 keV 3$^{-}$ state. The negative angular distribution and positive polarization indicates that it is an {\it E}1 transition. Its two possible assignments are 4$^{+}$ and 2$^{+}$, but we prefer the first option because we did not observe the usual transition to the ground state of a 2$^{+}$ state. Thus, the level assignment is 4$^{+}$. 
\item[-] {\it Level at 2982.0 keV}. 7$^{-}$ yrast state. It decays by a 324.1 keV $\gamma$-ray to the 5$^{-}$ yrast state. The transition has positive angular distribution and polarization as expected for an {\it E}2 ({\it$\alpha_{k}$} from~\cite{YATES}) multipolarity.
\item[-] {\it Level at 2986 keV}. This 2$^{+}$ level was first observed in an in-beam conversion  electron study~\cite{YATES87}. In the Tb(1$^{+}$) {\it $\beta$}-decay work~\cite{KLEINAR}, the authors mentioned a 1407 keV transition from this level to the 1579.4 keV 3$^{-}$ state that was wrongly placed (they settled the transition there only because the energies fit well). This transition does not feed that level but the 2$^{+}$ state at 1972.0 keV, as we deduced from our coincidence matrices. What we observe is a 2986 keV $\gamma$-ray to the ground state, which has a positive angular distribution as expected for an {\it E}2 transition. We also see a 1014 keV $\gamma$-ray to the 1972.0 keV 2$^{+}$ state but we could not extract any information about its intensity because the transition is contaminated by the 1014 keV aluminum line (the target frame and beam pipes are made of aluminum). We can argue that the transition exists since it presents an anisotropy which is in opposition to the isotropic radiation distribution of the aluminum line. This anisotropy comes from the transition to the 1972.0 keV 2$^{+}$ state and is positive as expected for this {\it M}1 transition. 
\item[-] {\it Level at 2996.6 keV}. Assigned previously as 4$^{-}$ in ~\cite{YATES} and ~\cite{STYCZEN}. It is de-excited by a 1417.1 transition to the 1579.4 keV 3$^{-}$ state that presents negative angular distribution and polarization as expected. In addition, we see a new transition of 338.2 keV to the 2657.9 keV 5$^{-}$ yrast state. Its intensity is too weak to extract further information about its character and multipolarity. 
\item[-] {\it Level at 3019.8 keV}. The strong population of this level in the (p,t) reaction experiment~\cite{FLYNN} led to the identification of the level as the neutron paring vibration 0$^{+}$ state. We have observed a de-excitation of the level to the 2$^{+}$ state at 1972.0 keV by a weak $\gamma$-ray of 1047.8 keV.
\item[-] {\it Level at 3031.2 keV}. 3$^{+}$ state known from previous work~\cite{YATES}. Our data confirm this assignment. The level de-excites by a 1059.1 keV $\gamma$-ray with negative angular distribution and positive polarization to the 1972.0 keV 2$^{+}$ state and by a 1451.8 keV $\gamma$-ray with negative angular distribution to the 1579.4 keV 3$^{-}$ state. The 1059.1 keV $\gamma$-ray {\it $\alpha$$_{k}$} value was measured in~\cite{YATES}. 
\item[-] {\it Level at 3098.9 keV}. Firm 6$^{-}$ state known from previous work~\cite{STYCZEN}. The positive angular distribution and negative polarization of the 441.0 keV transition to the 2657.9 keV 5$^{-}$ yrast state confirm the level assignment. We also observe the 116.7 keV transition seen in~\cite{STYCZEN} to the 7$^{-}$ yrast state at 2982.0 keV.
\item[-] {\it Level at 3182.4 keV}. 8$^{-}$ level known from previous work~\cite{KLEIN79}. It is de-excited by an {\it M}1/{\it E}2 transition ({\it $\alpha$$_{k}$} value was measured in~\cite{YATES}) to the 2982.0 keV 7$^{-}$ yrast state by a 200.4 keV $\gamma$-ray with measured positive angular distribution and negative polarization that confirms its multipolarity.
\item[-] {\it Level at 3185.8 keV}. Level known from~\cite{KLEINAR} with 2$^{+}$ assignment. We observe the two transitions seen in that work but not seen in previous in-beam experiments. The slightly negative angular distribution of the 1213.9 keV $\gamma$-ray de-excitation to the 1972.0 keV 2$^{+}$ state indicates mixed multipolarity of the transition. For the other de-exciting transition of the level of 1606.1 keV to the 1579.4 keV 3$^{-}$ state, the negative angular distribution agrees with its {\it E}1 multipolarity but the polarization does not. This could be due to the difficulty in extracting its polarization since this $\gamma$-ray is in a doublet.
\item[-] {\it Level at 3232.2 keV}. Level seen in~\cite{FLYNN} and identified in~\cite{MANN} as 2$^{+}$. The level was also seen in Tb(1$^{+}$) {\it $\beta$}-decay~\cite{KLEINAR}. We see a weak de-excitation of the level of 1260.2 keV to the 1972.0 keV 2$^{+}$ state and a de-excitation of 1653.0 keV $\gamma$-ray with negative angular distribution to the 1579.4 keV 3$^{-}$ state. This is in agreement with the 2$^{+}$ assignment of the level. We do not see the transition to the ground state observed in Tb(1$^{+}$) {\it $\beta$}-decay.
\item[-] {\it Level at 3287.2 keV}. 3$^{+}$ level previously seen in Tb(5$^{-}$) {\it $\beta$}-decay~\cite{STYCZEN} and in an ({\it  $\alpha$},{\it 2n}) in-beam experiment~\cite{YATES}. We have seen the same 675.7 keV $\gamma$-ray de-excitation  with negative angular distribution and positive polarization to the 2611.5 keV 4$^{+}$ state. The polarization is not in agreement with its {\it M}1 multipolarity ({\it $\alpha$$_{k}$} measured in~\cite{YATES}) probably because the $\gamma$-ray is in a doublet. Our data shows an additional transition of 1315.2 keV to the 1972.0 keV 2$^{+}$ state, which presents a negative angular distribution in agreement with the 3$^{+}$ assignment of the level. 
\item[-] {\it Level at 3290.5 keV}. Firm 7$^{-}$ level seen in an ({\it  $\alpha$},{\it 2n}) in-beam experiment~\cite{YATES}. Our positive angular distribution and positive polarization of the 308.5 keV $\gamma$-ray to the 2982.0 keV 7$^{-}$ yrast state confirm the {\it M}1/{\it E}2 multipolarity of the transition and the level assignment. 
\item[-] {\it Level at 3293.7 keV}. Level seen in previous work with an 8$^{-}$ level assignment from~\cite{KLEIN79}. We see the two well-known $\gamma$-ray de-excitations of 111.5 keV and 311.6 keV to the 8$^{-}$ and 7$^{-}$ states, respectively. The negative angular distributions of both transitions and the negative polarization measured in the case of the 311.6 keV $\gamma$-ray confirm the 8$^{-}$ level assignment. 
\item[-] {\it Level at 3313.0 keV}. Level seen in Tb(5$^{-}$) {\it $\beta$}-decay~\cite{STYCZEN} and in an ({\it  $\alpha$},{\it 2n}) in-beam experiment~\cite{YATES}. We observe the 655.1 and 701.5 keV known $\gamma$-rays to the 2657.9 keV 5$^{-}$ yrast state and to the 2611.5 keV 4$^{+}$ state, respectively, and an additional 1733.7 keV $\gamma$-ray to the 1579.4 keV 3$^{-}$ state. The measured angular distribution and polarization values confirm the level assignment. An exception is the angular distribution of the 1733.7 keV transition were we obtain a negative value.
\item[-] {\it Level at 3356.7 keV}. 2$^{+}$ level known from the (p,t) reaction experiment~\cite{MANN}. We see the first evidence of the level in a $\gamma$-ray measurement from a transition to the ground state. This $\gamma$-ray shows a Doppler shift, as expected for an {\it E}2 transition of that energy.  
\item[-] {\it Level at 3363.8 keV}. Level seen for the first time. It has branches of 706.0, 752.2 and 1784.4 keV. Their angular distributions lead to 4 as the level spin assignment. We cannot conclude anything about the level parity.
\item[-] {\it Level at 3380.7 keV}. Level seen in (p,t) reaction experiment by~\cite{FLYNN} (they measured L=2). We see level branches of 1408.8 , 1801.0 and 3381.5 keV. The angular distributions and polarizations measured for the first two mentioned $\gamma$-rays confirm the 2$^{+}$ level assignment. The later $\gamma$-ray de-excites the level to the ground state.   
\item[-] {\it Level at 3384.0 keV}. 6$^{-}$ level known previously in Tb(5$^{-}$) {\it $\beta$}-decay~\cite{STYCZEN} and in an ({\it  $\alpha$},{\it 2n}) in-beam experiment~\cite{YATES}. We see all known level de-excitations and our angular distribution and polarization data confirm the 6$^{-}$ assignment.
\item[-] {\it Level at 3388.7 keV}. The energies of this and the next level fit very closely, but the angular distribution data indicate that they cannot be the same. In addition, we see a $\gamma$-ray feeding this level not seen for the other. This level was previously known in an ({\it  $\alpha$},{\it 2n}) in-beam experiment~\cite{YATES}. We see the same 1416.7 keV $\gamma$-ray with negative angular distribution to the 1972.0 keV 2$^{+}$ state. Then, the possible spin assignments could be 3 and 1. We prefer the spin-3 assignment because of the absence of a transition to the ground state, as expected in the case of being a spin-1 state.
\item[-] {\it Level at 3388.8 keV}. Level seen for the first time. From the negative angular distribution of the 357.6 keV $\gamma$-ray de-exciting the level to the 3031.2 keV 3$^{+}$ state, the spin assignment could be 2 or 4.
\item[-] {\it Level at 3411.8 keV}. Level known from previous works (~\cite{YATES87} and~\cite{STYCZEN}). We see a 380.9 keV branch in addition to the known 415.3 and 800.2 keV branches. The angular distributions and polarizations give to the level a firm 4$^{+}$ assignment.
\item[-] {\it Level at 3416.5 keV}. This level was known from a previous ({\it  $\alpha$},{\it 2n}) in-beam experiment~\cite{YATES}. We see an 804.9 keV transition with a flat angular distribution and positive polarization to the 2611.5 keV 4$^{+}$ state and a 1444.6 keV transition with positive angular distribution and positive polarization to the 1972.0 keV 2$^{+}$ state, in addition to the previously known de-excitation of 1837.2 keV with negative angular distribution to the 1579.4 keV 3$^{-}$ state. Thus, the level assignment is 4$^{+}$.
\item[-] {\it Level at 3423.2 keV}. This level was previously known from Tb(5$^{-}$) {\it $\beta$}-decay~\cite{STYCZEN} and in an ({\it  $\alpha$},{\it 2n}) in-beam experiment~\cite{YATES} and assigned as 3$^{-}$. We see the known 1843.8 keV $\gamma$-ray to the 1579.4 keV 3$^{-}$ state with a negative angular distribution, suggestive of {\it M}1/{\it E}2 multipolarity.   
\item[-] {\it Level at 3428.5 keV}. 9$^{-}$ level known from~\cite{KLEIN79}. We see the 134.8, 245.8 and 446.2 keV known $\gamma$-rays. Their angular distributions and polarizations confirm the assignment. 
\item[-] {\it Level at 3436.2 keV}. Level known from Tb(5$^{-}$) {\it $\beta$}-decay~\cite{STYCZEN}. It was assigned as J=3. In addition to the known 1464.3 keV $\gamma$-ray to the 1579.4 keV 3$^{-}$ state that exhibits a positive angular distribution and positive polarization, we see two new de-excitations: a 824.6 keV $\gamma$-ray to the 2611.5 keV 4$^{+}$ state with positive angular distribution and a 1857.0 keV $\gamma$-ray with negative angular distribution that shows a Doppler shift to the 1579.4 keV 3$^{-}$ state. Thus, the level assignment could be 4$^{+}$ or 2$^{+}$ but we prefer the former since there is no de-excitation to the ground state, as should occur in the later case.
\item[-] {\it Level at 3456.5 keV}. This level was observed in a previous ({\it  $\alpha$},{\it 2n}) in-beam experiment~\cite{YATES} but was split into two levels because the side feeding was smaller than expected for a 4$^{+}$ state, as can be seen in Figure~[\ref{fig:side_feeding}]. But the fact that we observe a Doppler shift in the 1877.0 keV $\gamma$-ray to the 1579.4 keV 3$^{-}$ state indicates its {\it E}1, {\it M}1 or {\it E}2 nature. Then, taking also into account the angular distribution and polarization of the 798.6 keV $\gamma$-ray to the 2657.9 keV 5$^{-}$ yrast state, the level assignment is 4$^{+}$.
\item[-] {\it Level at 3461.1 keV}. This level and/or the next in energy was seen in the ({\it  p},{\it t}) reaction experiment~\cite{MANN} and assigned as (2). We now see two different levels. A 1881.7 keV $\gamma$-ray de-excitation to the 1579.4 keV 3$^{-}$ state with positive angular distribution and Doppler shift, together with the fact that the ({\it  p},{\it t}) reaction strongly populates natural parity states, lead to a 3$^{-}$, 1$^{-}$ or 5$^{-}$ assignment.
\item[-] {\it Level at 3464.0 keV}. A 1884.6 keV de-excitation to the 1579.4 keV 3$^{-}$ state with positive angular distribution and Doppler shift, together with the fact that the ({\it  p},{\it t}) reaction strongly populates natural parity states, lead to a 3$^{-}$, 1$^{-}$ or 5$^{-}$ assignment. If we take also into account that the level is fed from the 3947.2 keV 6$^{+}$ state we can discard the 1$^{-}$ and 3$^{-}$ assignments. So finally, our level assignment is 5$^{-}$.  
\item[-] {\it Level at 3478 keV}. Level seen for the first time. It de-excites by a 1899 keV $\gamma$-ray to the 1579.4 keV 3$^{-}$ state. No information about its angular distribution and polarization was extracted, and we cannot make a level assignment.
\item[-] {\it Level at 3481.8 keV}. Level seen for the first time. It de-excites by a 1902.4 keV $\gamma$-ray to the 1579.4 keV 3$^{-}$ state with negative angular distribution and negative polarization (both with large uncertainties) and has a Doppler shift that suggest its {\it E}1 nature. Thus, the level assignment is (3$^{+}$).
\item[-] {\it Level at 3484 keV}. Previously known 0$^{+}$ state~\cite{YATES87}. We now see a de-excitation to the 1972.0 keV 2$^{+}$ state. 
\item[-] {\it Level at 3484.7 keV}. This is the 6$^{+}$ two-phonon octupole state discussed in Chapter 5.
\item[-] {\it Level at 3547.5 keV}. 2$^{+}$ level known from the (p,t) reaction experiment~\cite{MANN}. We see the first evidence of the level in a $\gamma$-ray measurement by a transition to the ground state. This transition shows a Doppler shift as expected for an {\it E}2 transition of that energy.
\item[-] {\it Level at 3562.8 keV}. Level seen for the first time. The level has three de-excitations: a 951.6 keV $\gamma$-ray with positive angular distribution to the 2611.5 keV 4$^{+}$ state, a 1591.1 keV $\gamma$-ray with positive angular distribution to the 1972.0 keV 2$^{+}$ state and 1983.1 keV $\gamma$-ray with negative angular distribution to the 1579.4 keV 3$^{-}$ state. Then, the possible level assignments are 4$^{+}$ or 2$^{+}$.
\item[-] {\it Level at 3585 keV}. Level seen for the first time. A 2006 keV $\gamma$-ray with negative angular distribution and Doppler shift de-excites the level to the 1579.4 keV 3$^{-}$ state. The possible level spin assignments are 4 and 2. We cannot determine the level parity.
\item[-] {\it Level at 3640 keV}. 0$^{+}$ level identified in the {\it E}0 decay study~\cite{YATES87}. We see a 654.6 keV $\gamma$-ray de-exciting this level to the 2986  keV 2$^{+}$ state. 
\item[-] {\it Level at 3656.2 keV}. Level seen for the first time. The level de-excites by a 1044.6 keV $\gamma$-ray to the 2611.5 keV 4$^{+}$ state, by a 1684.3 keV $\gamma$-ray with negative angular distribution to the 1972.0 keV 2$^{+}$ state and by a 2076 keV $\gamma$-ray to the 1579.4 keV 3$^{-}$ state that exhibits a Doppler shift. We can conclude that the level spin is 3, but we cannot determine the parity.
\item[-] {\it Level at 3659.9 keV}. Level seen in the previous ({\it  $\alpha$},{\it 2n}) in-beam experiment~\cite{YATES} and assigned as 6$^{+}$. We see the 1002.0 keV $\gamma$-ray with negative angular distribution to the 2657.9 keV 5$^{-}$ yrast state in agreement with the previous level assignment.
\item[-] {\it Level at 3686.6 keV}. Level known from the ({\it  p},{\it t}) reaction experiment~\cite{MANN}. In that work the proposed level spin is (5). We see a level de-excitation $\gamma$-ray of 2107.2 keV with positive angular distribution and Doppler shift feeding the 1579.4 keV 3$^{-}$ state. It should be recalled that the ({\it  p},{\it t}) reaction strongly populates natural parity states. Then, the level assignment is 5$^{-}$.
\item[-] {\it Level at 3730 keV}. Level seen for the first time. The 1758 keV de-excitation to the 1972.0 keV 2$^{+}$ state is too weak to extract information about the angular distribution and polarization, so we cannot say anything about the level spin and parity.
\item[-] {\it Level at 3744 keV}. Level known from the (p,t) reaction experiment by~\cite{MANN} but no conclusive level assignment was extracted. The assignment in that work was (2,3). We see a 1772 keV $\gamma$-ray with positive angular distribution and a Doppler shift, which de-excites the level to the 1972.0 keV 2$^{+}$ state. We also see a 2165 keV $\gamma$-ray to the 1579.4 keV 3$^{-}$ state, which does not exhibit a Doppler shift and is suggestive of {\it M}1 multipolarity. Then, taking into account that in the ({\it  p},{\it t}) reaction natural parity states are populated the level assignment will be (2$^{+}$,3$^{-}$). 
\item[-] {\it Level at 3761.5 keV}. Level known from (p,t) reaction experiment~\cite{MANN} and its assignment was (5$^{-}$). An inspection on the angular cross section distribution in that work shows that it could also be a L=4. The level de-excites by a 1789.5 keV $\gamma$-ray with positive angular distribution and a Doppler shift to the 1972.0 keV 2$^{+}$ state. Then, our level assignment is (4$^{+}$).  
\item[-] {\it Level at 3779.2 keV}. 8$^{+}$ level known from the previous ({\it  $\alpha$},{\it 2n}) in-beam experiment~\cite{YATES}. We see the level de-excitation of 797.2 keV with negative angular distribution and positive polarization to the 2982.0 keV 7$^{-}$ yrast state thus confirming the previous level assignment.
\item[-] {\it Level at 3783.6 keV}. Level known from the previous ({\it  $\alpha$},{\it 2n}) in-beam experiment by~\cite{YATES} but no conclusive level assignment was extracted (only the {\it $\alpha$$_{k}$} value of the $\gamma$-ray to the 2611.5 keV 4$^{+}$ state that indicated its {\it M}1 or {\it E}2 multipolarity). We have seen the same 1172.2 keV $\gamma$-ray and have measured its angular distribution (negative) and polarization (positive). Our polarization is opposed to the {\it  $\alpha$$_{k}$} value but, since the polarization is slightly positive with a big uncertainty, we believe that the transition has {\it M}1 or {\it E}2 multipolarity. Then, the level assignment is 3$^{+}$ or 5$^{+}$.
\item[-] {\it Level at 3790 keV}. Level seen for the first time. It is de-excited by a 2210 keV $\gamma$-ray to the 1579.4 keV 3$^{-}$ state. It does not exhibit a Doppler shift, which suggests {\it M}1 multipolarity. Thus, the level assignment can be (2$^{-}$,3$^{-}$,4$^{-}$).
\item[-] {\it Level at 3853.5 keV}. Level known from the (p,t) reaction experiment by~\cite{MANN} and assigned as (5). We see four transitions not seen before: a 822.6 keV $\gamma$-ray to the 3031.2 keV 3$^{+}$ state, a 1244 keV $\gamma$-ray with negative angular distribution to the 2611.5 keV 4$^{+}$ state, a 1881.4 keV $\gamma$-ray with positive angular distribution to the 1972.0 keV 2$^{+}$ state and a 2274 keV $\gamma$-ray with positive angular distribution to the 1579.4 keV 3$^{-}$ state. We cannot firmly assign spin and parity to the level, but we consider that the best option to be (3$^{-}$), although this does not agree with the 1881.4 keV $\gamma$-ray positive angular distribution.
\item[-] {\it Level at 3854.0 keV}. 7$^{-}$ level known from the previous ({\it  $\alpha$},{\it 2n}) in-beam experiment by~\cite{YATES}. We see all three {\it M}1 de-excitations seen in that work and confirm the level assignment.
\item[-] {\it Level at 3864.8 keV}. 10$^{+}$ level known from  previous work~\cite{KLEIN79}. We see the 436.3 keV $\gamma$-ray with negative angular distribution and positive polarization to the 3428.5  keV 9$^{-}$ state that confirms the level assignment.
\item[-] {\it Level at 3866.5 keV}. Level seen for the first time. It is de-excited by a 381.7 keV $\gamma$-ray with negative angular distribution and positive polarization to the 6$^{+}$ two-phonon octupole state. It is also de-excited by a 1255.2 keV $\gamma$-ray with negative angular distribution to the 2611.5 keV 4$^{+}$ state. Then, our level assignment is (5$^{-}$). 
\item[-] {\it Level at 3907.9 keV}. Level known from the (p,t) reaction experiment~\cite{MANN} but no level assignment could be extracted. We see a de-excitation of 876.7 keV with positive angular distribution to the 3031.2 keV 3$^{+}$ state, and a 2329 keV (without Doppler shift) to the 1579.4 keV 3$^{-}$ state. Then, taking into account that in the ({\it  p},{\it t}) reaction natural parity states are strongly populated, the level assignment will be (3$^{-}$).   
\item[-] {\it Level at 3947.2 keV}. Level seen for the first time. We see three level de-excitations: a 483.1 keV $\gamma$-ray with negative angular distribution and positive polarization to the proposed 5$^{-}$ level at 3464.0 keV, a 848.1 keV $\gamma$-ray with positive angular distribution to the 3098.9 keV 6$^{-}$ state, and a 1289.2 keV $\gamma$-ray with ``flat'' angular distribution to the 2657.9 keV 5$^{-}$ yrast state. Then, the level assignment is 6$^{+}$. 
\item[-] {\it Level at 3973 keV}. Level known from the (p,t) reaction experiment~\cite{MANN}. We see a 2394 keV $\gamma$-ray with ``flat'' angular distribution and Doppler shift to the 1579.4 keV 3$^{-}$ state. Taking also into account that in the ({\it  p},{\it t}) reaction natural parity states are strongly populated, the level assignment will be (3$^{-}$).  
\item[-] {\it Level at 3987 keV}. Level seen for the first time. It is de-excited by a 2408 keV $\gamma$-ray with ``flat'' angular distribution and a Doppler shift to the 1579.4 keV 3$^{-}$ state. We cannot determine level spin and parity assignment.  
\item[-] {\it Level at 4006.6 keV}. Level known from (p,t) reaction experiment by~\cite{MANN} where the possible spin assignments were (4,5). We see two level de-excitations: a 2034.7 keV $\gamma$-ray to the 1972.0 keV 2$^{+}$ state and a 2427 keV $\gamma$-ray to the 1579.4 keV 3$^{-}$ state. Taking also into account that in the ({\it  p},{\it t}) reaction natural parity states are strongly populated, the level assignment will be (4$^{+}$). 
\item[-] {\it Level at 4026.6 keV}. Level seen for the first time. We see two level de-excitations: a 736.0 keV $\gamma$-ray with negative angular distribution to the 3290.5 keV 7$^{-}$ state and a 1044.6 keV with negative angular distribution to the 2982.0 keV 7$^{-}$ yrast state. Then, the two possible spin assignments are 6 and 8.
\item[-] {\it Level at 4076.7 keV}. Level seen for the first time. It de-excites by a 977.8 keV $\gamma$-ray to the 3098.9 keV 6$^{-}$ state. The big uncertainties in the angular distribution and polarization data make a level assignment difficult.
\item[-] {\it Level at 4076.7 keV}. 8$^{+}$ level known from the previous ({\it  $\alpha$},{\it 2n}) in-beam experiment~\cite{YATES}. We see the known 924.9 keV $\gamma$-ray with positive angular distribution and negative polarization to the 3182.4 keV 8$^{-}$ yrast state and the 1125.5 keV $\gamma$-ray with negative angular distribution and negative polarization to the 2982.0 keV 7$^{-}$ yrast state. In addition, we see a new transition of 1009.1 keV with negative angular distribution to the 3098.9 keV 6$^{-}$. The negative angular distribution of the new transition is in conflict with the firm level assignment. It is also strange that such an intense $\gamma$-ray was not seen in~\cite{YATES}.
\item[-] {\it Level at 4113 keV}. Level seen for the first time. It de-excites by a 2534 keV $\gamma$-ray to the 1579.4 keV 3$^{-}$ state. Its angular distribution is almost ``flat'' so we cannot make a level assignment.
\item[-] {\it Level at 4118.1 keV}. Level seen for the first time. It de-excites by a 1460.2 keV $\gamma$-ray to the 2657.9 keV 5$^{-}$ yrast state. No angular distribution or polarization information could be extracted and we cannot provide a level assignment.
\item[-] {\it Level at 4122 keV}. Level known from the (p,t) reaction experiment~\cite{MANN}, where the possible spin assignments were (4,5). We see a level de-excitation of 1511 keV with negative angular distribution to the 2611.5 keV 4$^{+}$ state. Since in the ({\it  p},{\it t}) reaction natural parity states are strongly populated, our level assignment is 5$^{-}$ or 3$^{-}$.
\item[-] {\it Level at 4131 keV}. Level seen for the first time. The level is de-excited by a 1100 keV $\gamma$-ray with positive angular distribution and positive polarization to the 3031.2 keV 3$^{+}$ state and, thus, the level assignment can be 3$^{+}$ or 5$^{+}$.
\item[-] {\it Level at 4152 keV}.  Level seen for the first time. The level is de-excited by a 2573 keV $\gamma$-ray with negative angular distribution and a Doppler shift to the 1579.4 keV 3$^{-}$ state. Then, the possible level assignments are 2 and 4.
\item[-] {\it Level at 4166.4 keV}. Level seen for the first time. The level is de-excited by a 1508.5 keV $\gamma$-ray with negative angular distribution to the 2657.9 keV 5$^{-}$ yrast state. Then, the possible level assignments are 4 and 6. 
\item[-] {\it Level at 4179.4 keV}. Level seen for the first time. Two $\gamma$-rays de-excite the level: a 1197.3 keV $\gamma$-ray with ``flat'' angular distribution and positive polarization to the 2982.0 keV 7$^{-}$ yrast state, and a 1521.6 keV $\gamma$-ray with negative angular distribution to the 2657.9 keV 5$^{-}$ yrast state. Then, our level assignment is (6$^{-}$).
\item[-] {\it Level at 4216.3 keV}. Level known from the (p,t) reaction experiment~\cite{MANN} where a spin assignment could not be extracted. We see a level de-excitation of 1185.2 keV with negative angular distribution to the 3031.2 keV 3$^{+}$ state. Since in the ({\it  p},{\it t}) reaction natural parity states are strongly populated, our level assignment is 2$^{+}$ or 4$^{+}$.
\item[-] {\it Level at 4230 keV}. Level known from the (p,t) reaction experiment~\cite{MANN} where the L-transfer measured was (5). The level de-excites by a 2651 keV $\gamma$-ray with a Doppler shift to the 1579.4 keV 3$^{-}$ state. Since in the ({\it  p},{\it t}) reaction natural parity states are strongly populated, our level assignment is 5$^{-}$. The Doppler shift observed is in agreement with the {\it E}2 multipolarity of the transition.
\item[-] {\it Level at 4248.3 keV}. Level known from the previous ({\it  $\alpha$},{\it 2n}) in-beam experiment~\cite{YATES} and assigned as (9). We see the same 1065.9 keV $\gamma$-ray with negative angular distribution to the 3182.4 keV 8$^{-}$ state. We cannot add new information about the level spin and parity.
\item[-] {\it Level at 4259.6 keV}. Level seen for the first time. There is a de-excitation of the level by a 1277.6 keV $\gamma$-ray with a ``flat'' angular distribution to the 2982.0 keV 7$^{-}$ yrast state. We cannot conclude anything about the level assignment.  
\item[-] {\it Level at 4286 keV}. Level seen for the first time. It is de-excited by a 2707 keV $\gamma$-ray to the 1579.4 keV 3$^{-}$ state. We cannot conclude anything about the level assignment.  
\item[-] {\it Level at 4299.6 keV}. Level known from the (p,t) reaction experiment~\cite{MANN} where the L-transfer measured was (2). We see a 1688.2 keV $\gamma$-ray with positive angular distribution that de-excites the level to the 2611.5 keV 4$^{+}$ state. Since in the ({\it  p},{\it t}) reaction natural parity states are strongly populated, the level assignment is 2$^{+}$. Our angular distribution data agrees with this assignment.
\item[-] {\it Level at 4318.8 keV}. Level seen for the first time. We see a 1336.8 keV $\gamma$-ray with a ``flat'' angular distribution and negative polarization that de-excites the level to the 2982.0 keV 7$^{-}$ yrast state. Then, the possible level assignments are 6$^{-}$, 7$^{-}$ and 8$^{-}$.
\item[-] {\it Level at 4326 keV}. Level seen for the first time. It is de-excited by a 1715 keV $\gamma$-ray with negative angular distribution to the 2611.5 keV 4$^{+}$ state. Thus, the possible level spins are 3 or 5.
\item[-] {\it Level at 4341 keV}. In the (p,t) reaction experiment~\cite{MANN}, a level at 4336 keV assigned as (4$^{+}$) was seen, but we are not sure it is the same level we see. Our level is de-excited by a 2762 keV $\gamma$-ray without a Doppler shift, which in our experiment is suggestive of {\it M}1 multipolarity. Then, from our data the level is 4$^{-}$.
\item[-] {\it Level at 4354.9 keV}. Level seen for the first time. Three $\gamma$-rays de-excite the level: a 1256.0 keV $\gamma$-ray with negative angular distribution to the 3098.9 keV 6$^{-}$ state, a 1372.8 keV $\gamma$-ray with negative angular distribution to the 2982.0 keV 7$^{-}$ yrast state, and a 1742 keV $\gamma$-ray with positive angular distribution to the 2611.5 keV 4$^{+}$ state. The big uncertainties in the data of the latter two mentioned $\gamma$-rays make the spin and parity assignment difficult. The possible level assignments are then 5$^{-}$ and 6$^{+}$.
\item[-] {\it Level at 4372 keV}. Level known from the (p,t) reaction experiment~\cite{MANN} and its assignment was (4). The level is de-excited by a 2793 keV $\gamma$-ray to the 1579.4 keV 3$^{-}$ state. We cannot add anything new to the (4$^{+}$) level assignment. We give positive parity because in the ({\it  p},{\it t}) reaction populates natural parity states.
\item[-] {\it Level at 4376 keV}. This level is very close in energy to the previous level so we cannot be sure which of them is the level known from the (p,t) reaction experiment~\cite{MANN}. We see a de-excitation of the level by a 1718 keV $\gamma$-ray to the 2657.9 keV 5$^{-}$ yrast state. Since we have no angular distribution or polarization data, our assignment cannot differ from the previous level: (4$^{+}$).
\item[-] {\it Level at 4389.5 keV}. Level seen for the first time. The level is de-excited by a 1290.6 keV $\gamma$-ray with negative angular distribution to the 3098.9 keV 6$^{-}$ state. The possible level assignments are 5 and 7.
\item[-] {\it Level at 4399.4 keV}. Level known from the (p,t) reaction experiment~\cite{MANN} but no conclusive level assignment was extracted. We see two $\gamma$-rays de-exciting the level: a 1300.5 keV $\gamma$-ray with negative angular distribution to the 3098.9 keV 6$^{-}$ state, and a 1741 keV $\gamma$-ray to the 2657.9 keV 5$^{-}$ yrast state. Then, taking into account that in the ({\it  p},{\it t}) reaction natural parity states are strongly populated, our assignments are 5$^{-}$ and 7$^{-}$. 
\item[-] {\it Level at 4416.8 keV}. Level seen for the first time. The level is de-excited by a 1123.2 keV $\gamma$-ray with positive angular distribution and positive polarization. Then, the possible level assignments are 10$^{+}$ and 8$^{-}$.
\item[-] {\it Level at 4459.0 keV}. Level seen for the first time. Three $\gamma$-rays de-excite the level: a 1030.7 keV $\gamma$-ray to the 3428.5 keV 9$^{-}$ state, a 1165.4 keV $\gamma$-ray to the 3293.7 keV 8$^{-}$ state and a 1276.5 keV $\gamma$-ray with negative angular distribution to the 3182.4 keV 8$^{-}$ state. Then, the level assignment is 7 or 9.
\item[-] {\it Level at 4484 keV}. Level known from the (p,t) reaction experiment by~\cite{MANN}. We see two new de-excitations: a 1826 keV $\gamma$-ray with negative angular distribution to the 2657.9 keV 5$^{-}$ yrast state and a 2906 keV $\gamma$-ray to the 1579.4 keV 3$^{-}$ state. Since in the ({\it  p},{\it t}) reaction natural parity states are strongly populated, the level assignment is (4$^{+}$).
\item[-] {\it Level at 4484.9 keV}. Level seen for the first time and de-excited by a 1056.5 keV $\gamma$-ray with positive angular distribution to the 3428.5 keV 9$^{-}$ state. The possible level assignments are 7$^{-}$,9 and 11$^{-}$ but since it does not de-excite to another different level than to the 3428.5 keV 9$^{-}$ state, then the latter is the most probable. Then, our level assignment is (11$^{-}$).  
\item[-] {\it Level at 4502.2 keV}. Level known from~\cite{BRODA79}. We see a level de-excitation of 1073.8 keV with negative angular distribution to the 3428.5 keV 9$^{-}$ state. The level assignment is 10 and we cannot determine its parity.
\item[-] {\it Level at 4520.4 keV}. Level seen for the first time and de-excited by a 1909 keV $\gamma$-ray to the 2611.5 keV 4$^{+}$ state. Since we do not have angular distribution or polarization information, we cannot make level assignment.
\item[-] {\it Level at 4529.1 keV}. Level seen for the first time in~\cite{TESINA}. We see the 1547.1 keV $\gamma$-ray de-excitation to the 2982.0 keV 7$^{-}$ yrast state. No level assignment can be made since no angular distribution or polarization information was extracted.
\item[-] {\it Level at 4532 keV}. There was a 0$^{+}$ level seen in the (p,t) reaction experiment~\cite{MANN}, but it could not be the same state since our level de-excites to a 4$^{+}$ state. We see a 1921 keV $\gamma$-ray with negative angular distribution to the 2611.5 keV 4$^{+}$ state. Thus, the possible level spins are 3 and 5.
\item[-] {\it Level at 4541.2 keV}. Level known from~\cite{BRODA79}. We see the known 1112.9 keV $\gamma$-ray with negative angular distribution and positive polarization to the 3428.5 keV 9$^{-}$ state and an additional de-excitation $\gamma$-ray  of 676.3 keV to the 3864.8 keV 10$^{+}$ state. Therefore, we confirm the 10$^{+}$ previous level assignment.
\item[-] {\it Level at 4580 keV}. Level seen for the first time and de-excited by a 1399 keV $\gamma$-ray with negative angular distribution to the 3182.4 keV 8$^{-}$ state and by a 1480 keV $\gamma$-ray with negative angular distribution to the 3098.9 keV 6$^{-}$ state. Our level-spin assignment is 7, but and no parity can be extracted. 
\item[-] {\it Level at 4608.3 keV}. Level seen for the first time and de-excited by a 1314.7 keV $\gamma$-ray with positive angular distribution to the 3293.7 keV 8$^{-}$ state. The possible level assignments are 8 and 10$^{-}$.
\item[-] {\it Level at 4666.8 keV}. Level known from~\cite{BRODA79}. We see the known 802.0 keV $\gamma$-ray with positive angular distribution to the 3864.8 keV 10$^{+}$ state and a new level de-excitation of 125.9 keV with negative angular distribution to the 4541.2 keV 10$^{+}$ state. Obviously, the angular distributions are in contradiction. In order to make a level assignment, we are more confident in the 802.0 keV transition. Then, our level assignment is (12$^{+}$).
\item[-] {\it Level at 4722 keV}. 4$^{-}$ level known previously from Tb(5$^{-}$) {\it $\beta$}-decay~\cite{STYCZEN}. We see a level de-excitation by a 3142 keV $\gamma$-ray to the 1579.4 keV 3$^{-}$ state.
\item[-] {\it Level at 4729.5 keV}. There was a (2$^{+}$,3$^{-}$) level seen in a (p,t) reaction experiment~\cite{MANN} but it could not be the same state since our level de-excites to an 8$^{-}$ level, which implies a level assignment too high to be seen in that experiment. As we noted, we see a 1435.9 keV $\gamma$-ray with negative angular distribution and positive polarization to the 3293.7 keV 8$^{-}$ state and a 1547 keV $\gamma$-ray to the 3182.4 keV 8$^{-}$ state. Then, the level assignment can be (9$^{+}$,7$^{+}$). 
\item[-] {\it Level at 4780.5 keV}. Level seen for the first time and de-excited by a 1391.8 keV $\gamma$-ray with negative angular distribution. Since we do not know the assignment for the fed level, we cannot say anything about the present one. 
\item[-] {\it Level at 4782 keV}. Level seen for the first time and de-excited by a 1800 keV $\gamma$-ray with negative angular distribution and without a Doppler shift to the 2982.0 keV 7$^{-}$ yrast state. Thus, it has to be an {\it M}1 transition and the level assignment is 8$^{-}$ or 6$^{-}$.
\item[-] {\it Level at 4802 keV}. Level seen for the first time and de-excited by a 1703 keV $\gamma$-ray without a Doppler shift to the 3098.9 keV 6$^{-}$ state. We cannot say anything about the level assignment.
\item[-] {\it Level at 4848 keV}. Level seen for the first time and de-excited by a 1554 keV $\gamma$-ray with negative angular distribution and without a Doppler shift to the 3293.7 keV 8$^{-}$ state. Thus, it has to be an {\it M}1 transition and the level assignment is 9$^{-}$ or 7$^{-}$.
\item[-] {\it Level at 4880 keV}. Level seen for the first time and de-excited by a 1451.8 keV $\gamma$-ray with negative angular distribution and without a Doppler shift to the 3428.5 keV 9$^{-}$ state. Thus, it has to be an {\it M}1 transition and the level assignment is 10$^{-}$ or 8$^{-}$. 
\item[-] {\it Level at 4898.3 keV}. Level seen for the first time. It is de-excited by a 1715.7 keV $\gamma$-ray (without Doppler shift) to the 3182.4 keV 8$^{-}$ state and by a 1604.7 keV $\gamma$-ray with negative angular distribution (without Doppler shift) to the 3293.7 keV 8$^{-}$ state. The level assignment can be 9$^{-}$ or 7$^{-}$.  
\item[-] {\it Level at 4943 keV}. There was a (2$^{+}$) level seen in a (p,t) reaction experiment~\cite{MANN}, but it could not be the same state since our level de-excites to an 8$^{-}$ state, which implies a level assignment too high to be seen in that experiment. The level is seen for the first time and de-excited by a 1760 keV $\gamma$-ray with a Doppler shift. We cannot say anything about the level assignment. 
\item[-] {\it Level at 5056.3 keV}. Level seen for the first time. It is de-excited by a 1191.5 keV $\gamma$-ray (apparently without a Doppler shift) to the 3864.8 keV 10$^{+}$ state. No angular distribution or polarization information was extracted, and we cannot make level assignment.  
\item[-] {\it Level at 5094.2 keV}. 11$^{+}$ level known from~\cite{BRODA79}. We see a level de-excitation by a 1229.4 keV $\gamma$-ray with negative angular distribution, without a Doppler shift, to the 3864.8 keV 10$^{+}$ state, confirming the {\it M}1 transition multipolarity and thus the previous level assignment.
\item[-] {\it Level at 5164.4 keV}. Level seen for the first time in~\cite{TESINA}. We see the 1299.7 keV $\gamma$-ray with negative angular distribution and without a Doppler shift to the 3864.8 keV 10$^{+}$ state. Thus, it has to be an {\it M}1 transition and the level assignment is 11$^{+}$ or 9$^{+}$. 
\end{itemize}




\renewcommand{\chaptermark}[1]{\markboth{Ap\'endice\ \thechapter}{}} 
\renewcommand{\sectionmark}[1]{\markright{\thesection\ #1}}
\include{bib}

\end{document}

%% file: titulo.tex
\begin{center}

{\Large\sc Universidad de Valencia - CSIC }
\vspace{0.2cm}

{\large\sc Departamento de F\'{\i}sica At\'omica, Molecular y Nuclear}
\vspace{0.1cm}

{\large\sc Instituto de F\'{\i}sica Corpuscular}
\vspace{1.5cm}

\mbox{\epsfig{file=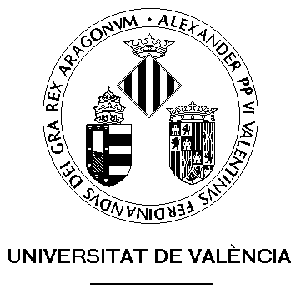,height=7.cm}}

\vspace{-0.5cm}
{\LARGE\bf Double Octupole States in $^{146}$Gd  } 

\vspace{4.cm}
{\large\sc Luis Caballero Ontanaya} 
\vspace{0.2cm}

{\large\sc Tesis Doctoral}
\vspace{0.2cm}

{\large\sc Noviembre de 2005} 
\vspace{0.2cm}

\end{center}

\vspace{4cm}

\newpage

\vspace*{4cm}

\newpage  

\begin{center}

{\Large\sc Universidad de Valencia - CSIC }
\vspace{0.2cm}


{\large\sc Instituto de F\'{\i}sica Corpuscular}
\vspace{1.2cm}

{\large\sc Departamento de F\'{\i}sica At\'omica, }
\vspace{0.1cm}

{\large\sc Molecular y Nuclear }
\vspace{3cm}

{\LARGE\bf Double Octupole States in $^{146}$Gd } 

\vspace{5cm}

\end{center}

 \hfill {\large\sc Luis Caballero Ontanaya} 
\vspace{0.3cm}

 \hfill {\large\sc Tesis Doctoral} 
\vspace{0.3cm}

 \hfill {\large\sc Noviembre de 2005} 

\newpage

\vspace*{4.cm}

\newpage
\vspace*{4.cm}
{\bf Berta Rubio Barroso}, Colaborador
Cient\'{\i}fico del Consejo
Superior de Investigaciones Cient\'{\i}ficas (CSIC)
\vspace*{1.5cm}

CERTIFICA: Que la presente memoria
{\bf ``Double Octupole States in $^{146}$Gd''}
ha sido realizada bajo su direcci\'on en el Instituto de F\'{\i}sica Corpuscular 
(Centro Mixto Universidad de Valencia - CSIC) por {\bf Luis Caballero Ontanaya} y 
constituye su Tesis Doctoral dentro del programa
de doctorado del Departamento de F\'{\i}sica At\'omica, Molecular y
Nuclear.
\vspace*{1.5cm}

Y para que as\'{\i} conste, en cumplimiento
con la legislaci\'on vigente,
presenta ante el Departamento de F\'{\i}sica At\'omica, Molecular y
Nuclear la
referida memoria, firmando el presente certificado en Burjassot (Valencia) a
07 de Diciembre de 2005.

\newpage
\vspace*{4.cm}
\newpage
\vspace*{5.cm}

\begin{flushright}
A te, principessa.
\end{flushright}

\newpage
\vspace*{5.cm}

\newpage
\vspace*{5.cm}

\begin{flushright}
{\it Y hoy Oscar dice simplemente: la mariposa tocaba el tambor.}\\
El tambor de hojalata,G\"unter Grass\\

\end{flushright}

\newpage
\vspace*{4.cm}

%% file: bib.tex
\bibliographystyle{alpha}